\documentclass[aps, prd, preprint, preprintnumbers, showpacs,
showkeys, amsfonts, nofootinbib, floatfix]{revtex4-1}
\usepackage{graphicx}
\usepackage{epsfig}
\usepackage{rotate}
\usepackage{rotating}
\usepackage{multirow}
\usepackage{longtable}
\usepackage[T1]{fontenc}
\usepackage{array}

\def\fo{\hbox{{1}\kern-.
25em\hbox{l}}}

\def\beq{\begin{equation}}
\def\eeq{\end{equation}}
\def\eq{\end{equation}}
\def\to{\rightarrow}

\def\bsg{\ifmmode B\to X_s\gamma\else $B\to X_s\gamma$\fi}
\def\bsll{\ifmmode B\to X_s\ell^+\ell^-\else $B\to X_s\ell^+\ell^-$\fi}
\def\bstt{\ifmmode B\to X_s\tau^+\tau^-\else $B\to X_s\tau^+\tau^-$\fi}
\def\shat{\ifmmode \hat{s}\else $\hat{s}$\fi}

\def\EmissT{\not \! \!  E_{T}}

\def\nolep{0\ell+\EmissT}
\def\twolep{2\ell+\EmissT}
\def\fourlep{4\ell+\EmissT}

\def\tetmulep{2e2\mu+\EmissT}
\def\ETsum{E_T^{\rm sum}}

\def\s2b{s_{2\beta}}

\def\u1{G_{SM}\otimes U(1)^{\prime}}

\newcommand{\newc}{\newcommand}

\newc{\lcal}{\int {\cal L}dt}

\newc{\lsp}{{\widetilde{\chi}^0_1}}
\newc{\niki}{{\widetilde{\chi}^0_2}}
\newc{\nuc}{{\widetilde{\chi}^0_3}}
\newc{\ndort}{{\widetilde{\chi}^0_4}}
\newc{\nbes}{{\widetilde{\chi}^0_5}}
\newc{\nalti}{{\widetilde{\chi}^0_6}}
\newc{\mnbir}{{M_{\widetilde{\chi}^0_1}}}
\newc{\mniki}{{M_{\widetilde{\chi}^0_2}}}
\newc{\mnuc}{{M_{\widetilde{\chi}^0_3}}}
\newc{\mndort}{{M_{\widetilde{\chi}^0_4}}}
\newc{\mnbes}{{M_{\widetilde{\chi}^0_5}}}
\newc{\mnalti}{{M_{\widetilde{\chi}^0_6}}}
\newc{\stauR}{{\widetilde{\tau}_R}}
\newc{\stau}{{\widetilde{\tau}_1}}
\newc{\staut}{{\widetilde{\tau}_2}}
\newc{\mstop}{m_{\widetilde{t}}}
\newc{\mHpm}{m_{H^\pm}}
\newc{\simgt}{\lower.7ex\hbox{$\;\stackrel{\textstyle>}{\sim}\;$}}
\newc{\simlt}{\lower.7ex\hbox{$\;\stackrel{\textstyle<}{\sim}\;$}}
\newc{\ie}{{\it i.e.}}
\newc{\etal}{{\it et al.}}
\newc{\eg}{{\it e.g.}}
\newc{\kev}{\hbox{\rm\,keV}}
\newc{\mev}{\hbox{\rm\,MeV}}
\newc{\gev}{\hbox{\rm\,GeV}}
\newc{\tev}{\hbox{\rm\,TeV}}
\newc{\xpb}{\hbox{\rm\, pb}}
\newc{\xfb}{\hbox{\rm\, fb}}

\newc{\mtop}{m_t}
\newc{\mbot}{m_b}
\newc{\mz}{m_Z}
\newc{\mw}{M_W}
\newc{\alphasmz}{\alpha_s(m_Z^2)}
\newc{\swsq}{\sin^2\theta_W}
\newc{\tw}{\tan\theta_W}
\newc{\cw}{\cos\theta_W}
\newc{\sw}{\sin\theta_W}
\newc{\BR}{\hbox{\rm BR}}
\newc{\zbb}{Z\to b\bar}
\newc{\Gb}{\Gamma (Z\to b\bar b)}
\newc{\Gh}{\Gamma (Z\to \hbox{\rm hadrons})}
\newc{\rbsm}{R_b^\hbox{\rm sm}}
\newc{\rbsusy}{R_b^\hbox{\rm susy}}
\newc{\drb}{\delta R_b}

\newc{\sgn}{\mbox{sgn}}
\newc{\tbeta}{\tan\beta}
\newc{\uL}{{\widetilde{u}_L}}
\newc{\uR}{{\widetilde{u}_R}}
\newc{\cL}{{\widetilde{c}_L}}
\newc{\cR}{{\widetilde{c}_R}}
\newc{\tL}{{\widetilde{t}_L}}
\newc{\tR}{{\widetilde{t}_R}}
\newc{\dL}{{\widetilde{d}_L}}
\newc{\dR}{{\widetilde{d}_R}}
\newc{\sL}{{\widetilde{s}_L}}
\newc{\sR}{{\widetilde{s}_R}}
\newc{\bL}{{\widetilde{b}_L}}
\newc{\bR}{{\widetilde{b}_R}}
\newc{\eL}{{\widetilde{e}_L}}
\newc{\eR}{{\widetilde{e}_R}}
\newc{\mhp}{m_{H^\pm}}
\newc{\mhalf}{m_{1/2}}
\newc{\emt}{{e/\mu /\tau}}

\newc{\lR}{\widetilde{\ell}_R}
\newc{\lL}{\widetilde{\ell}_L}
\newc{\nL}{\widetilde{\nu}_{\ell_L}}
\newc{\nR}{\widetilde{\nu}_{\ell_R}}
\newc{\neL}{\widetilde{\nu}_{e_L}}
\newc{\nmL}{\widetilde{\nu}_{\mu_L}}
\newc{\nlL}{\widetilde{\nu}_{\tau_L}}
\newc{\neR}{\widetilde{\nu}_{e_R}}
\newc{\nmR}{\widetilde{\nu}_{\mu_R}}
\newc{\nlR}{\widetilde{\nu}_{\tau_R}}
\newc{\naa}{\widetilde{\chi}^0_1}
\newc{\nbb}{\widetilde{\chi}^0_2}
\newc{\ncc}{\widetilde{\chi}^0_3}
\newc{\ndd}{\widetilde{\chi}^0_4}
\newc{\nee}{\widetilde{\chi}^0_5}
\newc{\nff}{\widetilde{\chi}^0_6}
\newc{\caa}{\widetilde{\chi}^{\pm}_1}
\newc{\cbb}{\widetilde{\chi}^{\pm}_2}
\newc{\phit}{\phi_t}
\newc{\phib}{\phi_b}
\newc{\phiew}{\phi_{ew}}
\newc{\htz}{h^0_t}
\newc{\hbz}{h^0_b}
\newc{\hewz}{h^0_{ew}}
\newc{\hsmz}{h^0_{sm}}
\newc{\huz}{h^0_u}
\newc{\hsusyz}{h^0_{susy}}
\newc{\lmop}{\rm LM1^\prime}
\newc{\lmtp}{\rm LM2^\prime}
\newc{\lmsp}{\rm LM6^\prime}
\newc{\smin}{\hat{s}_{\rm min}^{1/2}}

\def\slashchar#1{\setbox0=\hbox{$#1$}           % set a box for #1
   \dimen0=\wd0                                 % and get its size
   \setbox1=\hbox{/} \dimen1=\wd1               % get size of /
   \ifdim\dimen0>\dimen1                        % #1 is bigger
      \rlap{\hbox to \dimen0{\hfil/\hfil}}      % so center / in box
      #1                                        % and print #1
   \else                                        % / is bigger
      \rlap{\hbox to \dimen1{\hfil$#1$\hfil}}   % so center #1
      /                                         % and print /
   \fi}                                         %
\catcode`@=11
\long\def\@caption#1[#2]#3{\par\addcontentsline{\csname
  ext@#1\endcsname}{#1}{\protect\numberline{\csname
  the#1\endcsname}{\ignorespaces #2}}\begingroup
    \small
    \@parboxrestore
    \@makecaption{\csname fnum@#1\endcsname}{\ignorespaces #3}\par
  \endgroup}
\catcode`@=12

\begin{document}

\preprint{ CUMQ/HEP 156, IZTECH-P-07/2010}
\vspace{1.2cm}
\title{\Large Scalar Neutrinos at the LHC}
\author{Durmu\c{s} A. Demir$^{a}$}
\author{Mariana Frank$^b$}
\author{Levent Selbuz$^{a,c}$}
\author{Ismail Turan$^{d}$}
\affiliation{$^a$Department of Physics, {\.I}zmir Institute of
Technology, IZTECH, TR35430 {\.I}zmir, Turkey,}
\affiliation{$^b$Department of Physics, Concordia University, 7141
Sherbrooke St. West, Montreal, Quebec, Canada H4B 1R6,}
\affiliation{$^c$Department of Engineering Physics, Ankara
University, TR06100 Ankara, Turkey,}\affiliation{$^d$Ottawa-Carleton
Institute of Physics, Carleton University, 1125 Colonel By Drive
Ottawa, Ontario, Canada, K1S 5B6.}

%\date{\today}

\begin{abstract}
We study a softly-broken supersymmetric model whose gauge symmetry is
that of the standard model (SM) gauge group times an extra Abelian
symmetry $U(1)^{\prime}$. We call this gauge-extended model $U(1)^{\prime}$
model, and we study a $U(1)^\prime$ model with a secluded sector such that
neutrinos acquire Dirac masses via higher-dimensional terms allowed
by the $U(1)^\prime$ invariance. In this model the $\mu$ term of
the minimal supersymmetric model (MSSM) is dynamically induced by the vacuum expectation
value of a singlet scalar. In addition, the model contains exotic
particles necessary for anomaly cancellation, and extra singlet bosons for
achieving correct $Z^\prime/Z$ mass hierarchy. The neutrinos
are charged under $U(1)^{\prime}$, and thus, their production and
decay channels differ from those in the MSSM in strength and topology. We
implement the model into standard packages and  perform a detailed analysis of
sneutrino production and decay at the Large Hadron Collider, for various mass
scenarios, concentrating on three types of signals:  {\it (1)} $\rm 0\ell+ MET$,
{\it (2)}
$\rm 2\ell+MET$, and {\it (3)} $\rm 4\ell + MET$. We  compare the results with
those of
the MSSM whenever possible, and analyze the SM background for each
signal. The sneutrino production and decays provide clear signatures
enabling distinction of the $U(1)^\prime$ model from the MSSM at the LHC.
\end{abstract}

%%%%%%%%%%%%%%
%%%%%%%%%%%%%%
%\vspace*{-0.9cm}
\maketitle
%%%%%%%%%%%%%%

%%%%%%%%%%%%%%%%%%%%%%%%%%%%%%%%%%%%%%%%%%%%%%%%%%%%%%%%%%%%%%%

\tableofcontents

%%%%%%%%%%%%%%%%%%%%%%%%%%%%%%%%%%%%%%%%%%%%%%%%%%%%%%%
\section{Introduction and Motivation}
%%%%%%%%%%%%%%%%%%%%%%%%%%%%%%%%%%%%%%%%%%%%%%%%%%%%%%%

The MSSM is arguably the most popular `new physics' scenario referring to
a perturbative completion of the SM beyond Fermi energies. Motivated by the resolution of such
long standing problems of the SM as the gauge hierarchy problem, the existence
of dark matter and the added attraction of gauge unification, nevertheless, it
still has some outstanding problems. One of these is the so-called $\mu$ problem
\cite{muprob}.  Supersymmetric models which extend the MSSM via an extra gauge
group generally intend to solve $\mu$ problem and incorporate an  extra singlet
field, whose coupling to the Higgs fields and VEV generate dynamically the $\mu$
term.  These models extend the $SU(2)_L \otimes U(1)_Y$  MSSM electroweak
symmetry by an extra $U(1)$ gauge symmetry. Such an extension is minimal, and it
is well motivated in superstring theories \cite{Cvetic:1995rj},  grand unified
theories \cite{Hewett:1988xc} and  in dynamical electroweak breaking theories
\cite{Hill:2002ap}. The simplest versions contain a singlet field and an extra
neutral gauge boson. Other versions also allow right-handed neutrinos into the
spectrum. In a non-minimal version of the $U(1)$ extended MSSM,  which includes
several singlet ($S$) fields,  the tension between the electroweak scale and
developing a large enough $Z^\prime$ mass is resolved. We call this version of
the model   \emph{secluded sector} $U(1)^\prime$, a shorthand notation for
$SU(3)_c \otimes SU(2)_L \otimes U(1)_Y \otimes U(1)^\prime$,  the gauge
symmetry underlying the model, and describe it in the next section.  In the
MSSM, as in the  SM, neutrinos are massless. The fact that neutrino oscillation
imply non-vanishing neutrino masses is a strong motivation to consider an
extended form of the MSSM.  Small neutrino masses consistent with neutrino
oscillation phenomenology are usually explained by the see-saw mechanism
\cite{Minkowski:1977sc}. In the see-saw mechanism, large Majorana masses for
right-handed neutrinos induce small Majorana masses for left-handed neutrinos.
In the scalar sector,  right-handed sneutrinos  mix with the left handed
sneutrinos and give potentially new signals for extended symmetry. The choice of
$U(1)$ symmetry would determine the magnitude and type of neutrino masses. In
this paper, we consider a $U(1)^\prime$ extended form of the MSSM that contains
Dirac-type neutrino masses.

Direct or indirect detection of the superpartners of the Standard Model
particles, the definitive signal for supersymmetry, is one of the major aims of
the LHC experiments. Except for the LSP in the $R-$parity conserving
supersymmetry, the superpartners are expected to decay instantaneously into SM
particles, plus the LSP, detected as  missing energy. The common methodology
for detection is to analyze the production and cascade decays of the
supersymmetric particles.  As the right sneutrinos, which can mix with the left
sneutrinos are  a feature of the $U(1)^\prime$ model that distinguishes it from
MSSM, studying sneutrino signals would be an important test for this model.

 Systematic analyses of sneutrino decays in the MSSM have been performed in
\cite{Bartl:1999bg}. The aim of this article is to perform a comparative study
of LHC signals of sneutrino production and decays in the MSSM and in a
supersymmetric model with a secluded $U(1)^{\prime}$ breaking sector
\cite{Erler:2002pr} via their decay chain topologies. Differences between MSSM
and the \emph{secluded sector} $U(1)^{\prime}$ model likely reveal themselves
via decay modes of the sneutrino. We analyze the signals, and, for completeness,
we also include possible Standard Model backgrounds.

In most variants of the MSSM consistent with relic density calculations, the LSP
is the lightest neutralino, typically a mixed state of bino (fermionic partner
of the $U(1)_Y$ gauge boson) and the higgsino. In a previous work
\cite{Demir:2009kc}, we showed that a minimal $U(1)^{\prime}$
model (one extra singlet boson) could be consistent with the excess positron
observed in satellite experiments,
choosing on of the right-handed sneutrinos as the LSP. However, for the purpose
of this work (dependent on parameter space chosen to compare our results with
those of MSSM), the  \emph{secluded sector} $U(1)^{\prime}$  lightest neutralino
appears consistently to be the lightest supersymmetric particle (LSP) and
therefore is a potentially viable dark matter (DM) candidate, although its
composition is likely to differ from the lightest neutralino in MSSM.

Here we perform a thorough analysis of sneutrino production and decay in the
\emph{secluded sector} $U(1)^{\prime}$ model. In order to compare with previous
signals, we establish a set of three mSUGRA-inspired benchmarks for our model.
Similar to the mSUGRA benchmark points analyzed in MSSM (LM1, LM2, LM6)
\cite{Ball:2007zza,Battaglia:2003ab,Battaglia:2001zp}, we analyze the
corresponding scenarios in  \emph{secluded sector} $U(1)^{\prime}$ model
($\lmop,\lmtp,\lmsp$). Here LM stands for Low Mass, a choice likely to yield
visible signals at the LHC.

Our paper is organized as follows. We briefly introduce the model in
Section~\ref{sec:model}, then define the parameters and physical masses of
supersymmetric particles in the  \emph{secluded sector} $U(1)^{\prime}$ model in
Section~\ref{sec:num}. For each benchmark point, we insure  that DM candidate of
the $U(1)^{\prime}$ model yields relic densities consistent with the WMAP range
of cold dark matter density \cite{Spergel:2006hy}. We then perform a comparative
analysis of the production, decays and detectability of sneutrinos within these
benchmark supersymmetric scenarios. During this analysis we focus on the
multilepton plus missing energy signatures of the supersymmetric scenarios. We
present the results of our simulation analysis for the LHC. In
Section~\ref{sec:conc} we conclude the work. We leave the extensive details of
the model for the Appendices.

\section{The $U(1)^{\prime}$ Model \label{sec:model}}

The MSSM suffers from a naturalness problem due to the presence of $\mu$
parameter, responsible for giving masses to the Higgs bosons and Higgsino in the
superpotential. From a purely theoretical point of view, the value of this
parameter is expected to be either of the order of the GUT, Planck scale or
zero. For phenomenological aspects, however, it must be of the order of the
scale of electroweak symmetry breaking (EWSB) and it has to be non-zero to agree
with the experimental data. Seen from the low energy point of view, adding an
extra $U(1)$ is needed in order to solve the $\mu$ problem \cite{muprob} of the
MSSM. Basically the problem is  remedied by extending the matter and gauge
structure of the MSSM, {\it e.g.} within unified and/or string models by
introducing an additional singlet filed $S$, whose VEV generates the $\mu$ term
dynamically. Theories with an extra $U(1)^{\prime}$ broken at the electroweak-
to- TeV scale by SM singlets are known to be able to generate an appropriately
sized $\mu$ parameter (see {\it e.g.} \cite{muprob}).

The other success of the $U(1)^{\prime}$ symmetry is being able to generate
pertinent neutrino masses by introducing right-handed neutrinos into the
superpotential. The right-handed neutrino sector and the $\mu$ parameter can be
correlated for both Majorana \cite{biz2} and Dirac masses \cite{biz3}. We assume
here that lepton number is an accidental symmetry that is conserved at the
perturbative level. Hence, the neutrinos are Dirac fermions, requiring Yukawa
couplings of ${\cal{O}}\left(10^{-13}\right)$. These couplings are technically
natural, but an explanation for such a strong suppression is clearly desirable.
One way this can occur is if the $U(1)^{\prime}$ invariance suppresses leading
order contributions to Dirac neutrino masses and allows higher-dimensional
operators \cite{biz3}.

In this work, we extend the MSSM in the following ways. First, the gauge
structure of the MSSM, $SU(3)_C\otimes SU(2)_L\otimes U(1)_Y$, is enriched to
include an extra Abelian group factor $U(1)^\prime$. Second, we promote the
$\mu$ parameter into dynamical field, $S$, which is charged under the
$U(1)^\prime$. Third, exotics with Yukawa couplings to $S$ are included to make
the theory anomaly-free. Fourth, $Z^\prime/Z$ mass hierarchy in the model is
ensured by three additional $SU(2)$ singlet fields which are coming from
secluded sector of the model. The model also includes a term that provides
suppressed Dirac neutrino masses in accordance with observations. We present the
main relevant points in this section, leaving the details for the appendices.

In the minimal version of the model which contains only one singlet $S$, there
is some tension between the electroweak scale and the need to generate a large
enough $M_Z^\prime$. These two problems can be decoupled without fine tuning
when several additional fields are incorporated into the model. An example of
this kind of non-minimal model is \emph{secluded sector model}. The secluded
sector model involves an ordinary sector of symmetry breaking fields, which
includes two Higgs doublets, and an $SU(2)_L$ singlet $S$. After acquiring a
VEV, $S$ generates an effective $\mu$ parameter $\mu = h_{s} \langle S \rangle
$. The secluded sector of the model includes three $SU(2)_L$ singlet fields
$S_i, i=1,2,3$ which acquire large VEVs. All four VEVs of the singlet fields
$S,S_{1,2,3}$  contribute to $Z^\prime$ mass. Thus, in this model, $Z^\prime/Z$
mass hierarchy is implemented mainly through the secluded sector  of the model.

The superpotential of the model is given by
\begin{eqnarray}\label{eq:superpot}
\widehat{W}&=&h_u\widehat{Q}\cdot \widehat{H}_u \widehat{U}+
h_d\widehat{Q}\cdot \widehat{H}_d \widehat{D} + h_e\widehat{L}\cdot
\widehat{H}_d \widehat{E} + h_s \widehat{S}\widehat{H}_u \cdot
\widehat{H}_d +  \frac{1}{M_R}  \widehat{S}_1 \widehat{L}\cdot
\widehat{H}_u {\bf h_{\nu}} \widehat{N}+
\bar{h}_s \widehat{S}_1 \widehat{S}_2 \widehat{S}_3 \nonumber \\
&+& \sum_{i=1}^{n_{\cal{Q}}} {h}_Q^i \widehat{S} \widehat{\cal{Q}}_i
\widehat{\cal{\overline{Q}}}_i + \sum_{j=1}^{n_{\cal{L}}} {h}_L^j
\widehat{S} \widehat{\cal{L}}_j \widehat{\cal{\overline{L}}}_j
\end{eqnarray}
where the fields entering the equation, together with their quantum numbers are
listed in Table \ref{tab:charge}. Here, $M_R$ is a large mass scale and
$h_{\nu}$ is the Yukawa coupling responsible for generating neutrino
masses.

\begin{table}[t]
\addtolength{\tabcolsep}{0.7pt}
\begin{tabular*}{0.99\textwidth}{@{\extracolsep{\fill}} ccccccccccccccccc}
\hline \hline Field & $\widehat{Q}$ & $\widehat{U}$ &
$\widehat{D}$ & $\widehat{L}$ & $\widehat{N}$ & $\widehat{E}$ &
$\widehat{H}_u$ & $\widehat{H}_d$ & $\widehat{S}$ & $\widehat{S_1}$
& $\widehat{S_2}$ & $\widehat{S_3}$ & $\widehat{\cal{Q}}$ &
$\widehat{\cal{\overline{Q}}}$ & $\widehat{\cal{L}}$ &
$\widehat{\cal{\overline{L}}}$
  \\\hline
$\;$  $SU(3)_C$ & 3 & $\overline{3}$ &  $\overline{3}$ & 1 & 1 &  1& 1& 1 &1
& 1& 1 &1 &3 &$\overline{3}$
&1&1\\
$\;$  $SU(2)_L$ & 2 & 1 &  1 & 2 & 1 &  1& 2& 2 &1 & 1& 1 &1 &1 &1
&1 &1\\
$\;$  $U(1)_Y$ & 1/6 & -2/3 &  1/3 & -1/2 & 0 &  1&
1/2& -1/2 &0 & 0& 0 &0 &$Y_{\cal{Q}}$ &$-Y_{\cal{Q}}$ &$Y_{\cal{L}}$
& $-Y_{\cal{L}}$\\
$\;$  $U(1)^{\prime}$ & $Q_{Q}^{\prime}$ &
$Q_{U}^{\prime}$ & $Q_{D}^{\prime}$& $Q_{L}^{\prime}$ &
$Q_{N}^{\prime}$& $Q_{E}^{\prime}$& $Q_{H_u}^{\prime}$&
$Q_{H_d}^{\prime}$& $Q_{S}^{\prime}$ & $Q_{S_1}^{\prime}$&
$Q_{S_2}^{\prime}$ &$Q_{S_3}^{\prime}$ &$Q_{\cal{Q}}^{\prime}$
&$Q_{\cal{\overline{Q}}}^{\prime}$ &$Q_{\cal{L}}^{\prime}$
&$Q_{\cal{\overline{L }}}^{\prime}$
\\\hline\hline
\end{tabular*}
\caption{\sl\small Gauge quantum numbers of quark ($\widehat{Q},
\widehat{U}, \widehat{D}$), lepton ($\widehat{L}, \widehat{N},
\widehat{E}$), Higgs ($\widehat{H}_u, \widehat{H}_d$), SM-singlet
($\widehat{S}, \widehat{S}_1, \widehat{S}_2, \widehat{S}_3$), exotic
quark ($\widehat{\cal{Q}}, \widehat{\cal{\overline{Q}}}$) and exotic
lepton ($\widehat{\cal{L}}, \widehat{\cal{\overline{L}}}$)
superfields.} \label{tab:charge}
\end{table}

The $U(1)^{\prime}$ charges of the fields satisfy a number of conditions arising
from phenomenological constraints, as well as from gauge invariance of the model
and from the requirement of cancellation of gauge and gravitational anomalies.
They are as follows.

The $U(1)^{\prime}$ charges satisfy $Q^{\prime}_{H_u}+Q^{\prime}_{H_d}\neq 0$ to
forbid the bare $\mu$ term, $Q^{\prime}_{L}+Q^{\prime}_{H_u}+Q^{\prime}_{N}\neq
0$ to induce neutrino masses correctly, and
$Q^{\prime}_{S_1}+Q^{\prime}_{S_2}+Q^{\prime}_{S_3}=0$ to correctly generate the
$Z-Z^{\prime}$ mass hierarchy. Gauge invariance of the superpotential implies
\begin{eqnarray}
 \label{eq:gauge_cond}
0&=&Q^{\prime}_{S}+Q^{\prime}_{H_u}+Q^{\prime}_{H_d},
\nonumber \\
0&=&Q^{\prime}_{Q}+Q^{\prime}_{H_u}+Q^{\prime}_{U},
\nonumber \\
0&=&Q^{\prime}_{Q}+Q^{\prime}_{H_d}+Q^{\prime}_{D},
\nonumber \\
0&=&Q^{\prime}_{L}+Q^{\prime}_{H_d}+Q^{\prime}_{E},
\nonumber \\
0&=&Q^{\prime}_{\cal{Q}}+Q^{\prime}_{\cal{\overline{Q}}}+Q^{\prime}_{S},
\nonumber \\
0&=&Q^{\prime}_{\cal{L}}+Q^{\prime}_{\cal{\overline{L}}}+Q^{\prime}_{S},
\nonumber \\
0&=&Q^{\prime}_{S_1}+Q^{\prime}_{L}+Q^{\prime}_{H_u}+Q^{\prime}_{N}.
\end{eqnarray}

For the model to be anomaly-free the $U(1)^{\prime}$ charges of fields must
satisfy
\begin{eqnarray}
0&=&3(2Q^{\prime}_{Q}+Q^{\prime}_{U}+Q^{\prime}_{D})+
n_{\cal{Q}}(Q^{\prime}_{\cal{Q}}+Q^{\prime}_{\cal{\overline{Q}}}),
\\
0&=&3(3Q^{\prime}_{Q}+Q^{\prime}_{L})+Q^{\prime}_{H_d}+Q^{\prime}_{H_u},
\\
0&=&3(\frac{1}{6}Q^{\prime}_{Q}+\frac{1}{3}Q^{\prime}_{D}+
\frac{4}{3}Q^{\prime}_{U}+
\frac{1}{2}Q^{\prime}_{L}+Q^{\prime}_{E})
+\frac{1}{2}(Q^{\prime}_{H_d}+Q^{\prime}_{H_u})\nonumber \\
&+&3n_{\cal{Q}} Y^2_{\cal{Q}} (Q^{\prime}_{\cal{Q}}+
Q^{\prime}_{\cal{\overline{Q}}})+ n_{\cal{L}}
Y^2_{\cal{L}} (Q^{\prime}_{\cal{L}}+
Q^{\prime}_{\cal{\overline{L}}}),
 \\
0&=&3(6Q^{\prime}_{Q}+3Q^{\prime}_{U}+3Q^{\prime}_{D}+2Q^{\prime}_{L}+
Q^{\prime}_{E}+Q^{\prime}_{N})
+2Q^{\prime}_{H_d}+2Q^{\prime}_{H_u}\nonumber \\
&+&Q^{\prime}_{S}+Q^{\prime}_{S_1}+Q^{\prime}_{S_2}+Q^{\prime}_{S_3}+
3 n_{\cal{Q}}
(Q^{\prime}_{\cal{Q}}+Q^{\prime}_{\cal{\overline{Q}}})+
n_{\cal{L}}(Q^{\prime}_{\cal{L}}+Q^{\prime}_{\cal{\overline{L}}}),
\\
0&=&3(Q^{\prime\ 2}_{Q}+Q'^2_{D}-2Q^{\prime\
2}_{U}-Q^{\prime\ 2}_{L}+Q^{\prime\ 2}_{E})-Q^{\prime\
2}_{H_d}+Q^{\prime\ 2}_{H_u}+ 3n_{\cal{Q}} Y_{\cal{Q}}
(Q^{\prime\ 2}_{\cal{Q}}- Q^{\prime\
2}_{\cal{\overline{Q}}})\nonumber \\
&+&n_{\cal{L}} Y_{\cal{L}}
(Q^{\prime\ 2}_{\cal{L}}-Q^{\prime\
2}_{\cal{\overline{L}}}),
 \\
0&=&3(6Q^{\prime\ 3}_{Q}+3Q^{\prime\ 3}_{D}+3Q^{\prime\
3}_{U}+2Q^{\prime\ 3}_{L}+Q^{\prime\ 3}_{E}+Q^{\prime\
3}_{N})+ 2Q^{\prime\ 3}_{H_d}+2Q^{\prime\
3}_{H_u}+Q^{\prime\ 3}_{S}\nonumber \\
&+&Q^{\prime\
3}_{S_1}+Q^{\prime\ 3}_{S_2}+Q^{\prime\ 3}_{S_3}+
3n_{\cal{Q}}(Q^{\prime\ 3}_{\cal{Q}}+Q^{\prime\
3}_{\cal{\overline{Q}}})+n_{\cal{L}}(Q^{\prime\
3}_{\cal{L}}+Q^{\prime\ 3}_{\cal{\overline{L}}}),
\end{eqnarray}
which correspond to vanishing of
$U(1)^{\prime}$-$SU(3)_C$-$SU(3)_C$,
$U(1)^{\prime}$-$SU(2)_L$-$SU(2)_L$,
$U(1)^{\prime}$-$U(1)_Y$-$U(1)_Y$,
$U(1)^{\prime}$-graviton-graviton,
$U(1)^{\prime}$-$U(1)^{\prime}$-$U(1)_Y$, and
$U(1)^{\prime}$-$U(1)^{\prime}$-$U(1)^{\prime}$ anomalies,
respectively. All these anomaly cancellation conditions are
satisfied for a particular pattern of charges and parameters. It is
found that the solution to the mixed anomaly constraints requires
$n_{\cal{Q}} = 3$ color triplet pairs with hypercharge $Y_{\cal{Q}}
= -1/3$, and $n_{\cal{L}} = 5$ singlet pairs with $Y_{\cal{L}} = -
\sqrt{2/5}$. With these parameter values one obtains the
$U(1)^{\prime}$ model displayed in Table \ref{tab:charge-sol}. The
$U(1)^\prime$ charges for Higgs fields in the model are chosen as
\begin{eqnarray}
Q^\prime_{S}=-Q^\prime_{S_1}=-Q^\prime_{S_2}=\frac{1}{2}Q^\prime_{S_3},
~~~Q^\prime_{H_u}+Q^\prime_{H_d}+Q^\prime_{S}=0.
\end {eqnarray}
Under the conditions above, the supersymmetry breaking soft terms for the
secluded sector model are
\begin{eqnarray}
 V_{soft}=V_{soft}^I+V_{soft}^o
\end{eqnarray}
where $ V_{soft}^I$ are the allowed $U(1)^\prime$ dimension-2 operators
\begin{eqnarray}
V_{soft}^I=(m^2_{SS_1}SS_1+m^2_{SS_2}SS_2+m^2_{S_1S_2}S_1^\dag
S_2+h.c.)
\end {eqnarray}
and $V_{soft}^o$ term is defined as
\begin{eqnarray}
V_{soft}^o&=m^2_{H_u}|H_u|^2+m^2_{H_d}|H_d|^2+m^2_S|S|^2+\sum_{i=1}^3m^2_{S_i}
|S_i|^2\\
\nonumber
 &-(A_sh_sSH_uH_d+A_{\bar s}\bar h_sS_1S_2S_3+h.c.)
\end{eqnarray}
We set $m^2_{S_1S_2}=0$ as only two of the $S_i$ fields are needed to break
the global $U(1)$ symmetries. To insure that the potential is not bounded
from below, we require
\begin{eqnarray}
&&m^2_S+m^2_{S_1}+2m^2_{SS_1}>0\,,\nonumber\\
&&m^2_S+m^2_{S_2}+2m^2_{SS_2}>0.
\end {eqnarray}
In the model, the charge of the quark doublet $\widehat{Q}$ is kept
as a free parameter after the  normalization
$Q^\prime_{H_u}=-2$, $Q^\prime_{H_d}=1$, $Q^\prime_{S}=1$,
$Q^\prime_{S_1}=-1$, $Q^\prime_{S_2}=-1$, $Q^\prime_{S_3}=2$.
\begin{table}[t]
\begin{tabular*}{0.99\textwidth}{@{\extracolsep{\fill}} cccc}
\hline\hline $\displaystyle
\begin{array}{l} Q^\prime_{H_u}=-2\\
Q^\prime_{H_d}=1\\
Q^\prime_{S}=1\\
Q^\prime_{S_1}=-1
\end{array}$&$\begin{array}{l} Q^\prime_{S_2}=-1\\
Q^\prime_{S_3}=2\\
Q^\prime_{Q}=x\\
Q^\prime_{U}=2-x
\end{array}$&$\begin{array}{l} Q^\prime_{D}=-1-x\\
Q^\prime_{L}=\frac{1}{3}-3x\\
Q^\prime_{E}=-\frac{4}{3}+3x\\
Q^\prime_{N}=\frac{8}{3}+3x
\end{array}$&$\begin{array}{l} Q^\prime_{\cal{Q}}=
\frac{4-12x-\sqrt{2}\Omega}{18}\\
Q^\prime_{\cal{\overline{Q}}}=\frac{-22+12x+\sqrt{2}\Omega}{18}\\
Q^\prime_{\cal{L}}=\frac{-15+13\sqrt{10}-12\sqrt{10}x+\sqrt{5}\Omega}{30}\\
Q^\prime_{\cal{\overline{L}}}=\frac{-15-13\sqrt{10}+
12\sqrt{10}x-\sqrt{5}\Omega}{30}
\end{array}$\\ \hline\hline
\end{tabular*}
\caption{\label{tab:charge-sol}\sl\small A set of $U(1)^{\prime}$ charges
satisfying all gauge invariance and anomaly cancellation
conditions. The charge of the quark doublet $\widehat{Q}$ is left
free, and for simplicity $\Omega(x)=\sqrt{241+708x+612x^2}$ is
introduced.}
\end{table}

In this model the left and right sneutrinos mix, and the mixing matrix can in
general be expressed as
\begin{eqnarray}
\label{eq:sneutrino_mix}
 {\cal L}_m^{\widetilde \nu }= -\sum_{i,j=1}^3
({\widetilde \nu}_L^{i*} {\widetilde \nu}_R^{j*})\left(
 \begin{array}{cc}
 m_{\widetilde \nu_{L L}^i}^{2}& m_{\widetilde \nu_{L R}^{ij}}^{2}\\[1.ex]
 m_{\widetilde \nu_{R L}^{ij}}^{2} & m_{\widetilde \nu_{R R}^j}^{2}
 \end{array}
 \right)
 \left(
\begin{array}{c}
\widetilde \nu_L^i\\ [1.ex] \widetilde \nu_R^j
\end{array}
\right),
\end{eqnarray}
where $i,j$ are the flavor indices and the matrix elements are given
by
\begin{eqnarray}
\label{eq:mix_comp}
m^2_{\widetilde \nu_{LL}^i}&=&M^2_{L_i}+({\bf m}_\nu^{ii})^2+
\frac{1}{4}(g_Y^2Y_L-\frac{g^2}{2})(\langle H_u^0\rangle^2-
\langle H_d^0\rangle^2)\nonumber \\
&&+\frac{1}{2}g_{Y'}^2Q'_L(Q'_{H_u}\langle
H_u^0\rangle^2+Q'_{H_d}\langle H_d^0\rangle^2+Q'_{S}\langle
S\rangle^2\rho_s)
\nonumber \\
m^2_{\widetilde \nu_{RR}^j}&=&M^2_{N_j}+({\bf m}_\nu^{ii})^2+
\frac{1}{4}g_Y^2Y_N(\langle H_u^0\rangle^2-\langle H_d^0\rangle^2)\nonumber \\
&&+\frac{1}{2}g_{Y'}^2Q'_N(Q'_{H_u}\langle
H_u^0\rangle^2+Q'_{H_d}\langle H_d^0\rangle^2+Q'_{S}\langle
S\rangle^2\rho_s)
\nonumber \\
m^2_{\widetilde\nu_{LR}^{ij}}&=&(m^2_{\widetilde \nu_{RL}^{ij}})^* =
{\bf m}_\nu^{ij}\left[A^*_{\nu_i} +
\frac{\mu}{\tan\beta}+\frac{\bar{h}_s\langle S_2 \rangle\langle S_3
\rangle}{\sqrt{2}\langle S_1 \rangle}\right].
\end{eqnarray}
Here $ M^2_{L_i}$ and $ M^2_{N_i}$ are soft mass terms and
$A_{\nu_i}$ are trilinear couplings (assumed flavor-diagonal). Dirac
neutrino masses ${\bf m}_\nu$, the $\mu$ parameter and $\rho_s$ in the
equations above are expressed as
\begin{eqnarray}
\hspace*{-0.8cm} {\bf m_{\nu}} = \frac{1}{M_R} \langle S_1 \rangle
\langle H_u^0\rangle {\bf h_{\nu}}\equiv {\bf Y}_{\nu} \left(\langle
H_u^0\rangle/\sin\beta\right), \hspace{0.5cm} \mu =\frac{h_{s}
\langle S \rangle}{\sqrt{2}}, \hspace{0.5cm}
\rho_s=1+\frac{\sum_{i=1}^3Q'_{S_i}v^2_{s_i}}{Q'_{S}v_s^2}.
\label{eq:rho}
\end{eqnarray}

In this model, neutrino masses are chosen to be Dirac-type.
The effective neutrino Yukawa coupling ${\bf Y}_{\nu}$ leads to neutrino masses
in agreement with the experiment. Numerically, we obtain \cite{Demir:2009kc}
\begin{eqnarray}
\label{eq:hnu-eq}
\left|{\bf Y}_{\nu}\right|
%\equiv \frac{\langle S \rangle}{M_R} \left|{\bf Y_{\nu}}\right| \sin \beta
\simeq 3 \times 10^{-13} \left(\frac{|m_{\nu}|^2}{2.8\times 10^{-3}\
{\rm eV}^2}\right)^{1/2}.
\end{eqnarray}

%%%%%%%%%%%%%%%%%%%%%%%%%%%%%%%%%%%%%%%%%%%%%%%%%
\section{MSSM vs. $U(1)^{\prime}$ at the LHC Energies \label{sec:num} }
%%%%%%%%%%%%%%%%%%%%%%%%%%%%%%%%%%%%%%%%%%%%%%%%%

\subsection{Parameter Space and Relic Density}

Motivated by the fact that the scalar neutrino LSP
can in principle explain the WMAP data as well as excess
positron flux measured by various satellite experiments
\cite{Demir:2009kc}, we  analyze the model further
by investigating the production and decay mechanism of the
scalar neutrinos at LHC.
%In addition to the MSSM fields, the
%model has four extra scalar fields, a new gauge boson $Z^\prime$
%and their corresponding super fields.

The model we consider here, the secluded-$U(1)^\prime$ with
right-handed neutrinos, has some advantages over the so-called the
minimal $U(1)^\prime$ where only one additional scalar field is
introduced. The squark phenomenology in this minimal $U(1)^\prime$
model has been explored in Ref.~\cite{Ali:2009md} where there is difficulty
with inducing a small $\mu_{eff}$ while
satisfying the $Z^\prime$ mass bound, which is around 1 TeV. This is
because both  $\mu_{eff}$ and $m_{Z^\prime}$ are proportional to the
vacuum expectation value of the additional scalar field $S$. One needs
three additional scalars to ameliorate the picture  the
VEVs of the new scalars are kept large large. This is  one motivation for the
secluded $U(1)^\prime$ model.  For further details of the model, see
\cite{Erler:2002pr}.

%%%%%%%%%%%%%%%%%%%%%%%%%%%%%%%%%%%%%%%%%%%%%%%%%%%%%%%%%%%%%%%%%%%%%%%%%%%%%%%
\begin{figure}[htb]
%\vskip 0.2in
\begin{center}
	\includegraphics[width=6.5in,height=1.5in]{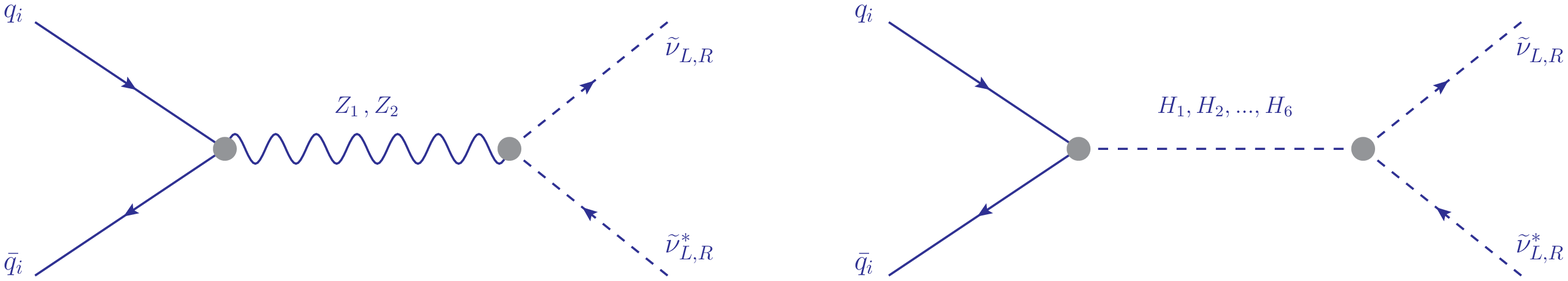}
\end{center}
\vskip -0.2in
      \caption{\sl \small The Feynman diagrams for the production of the scalar
neutrinos
in the
secluded-$U(1)^\prime$ model. $H_i, i=1,...,6$ are the CP-even physical
Higgs bosons.}
\label{fig:prod}
\end{figure}
%%%%%%%%%%%%%%%%%%%%%%%%%%%%%%%%%%%%%%%%%%%%%%%%%%%%%%%%%%%%%%%%%%%%%%%%%%%%%%%

The Feynman diagrams contributing to the hard production of scalar neutrinos are
given in Fig~\ref{fig:prod}. For simplicity we neglect the mixing between $Z$ (
the $Z$ boson of the SM) and $Z^{\prime}$ in the numerical analysis so that $Z_1\equiv Z$ and $Z_2\equiv
Z^\prime$. In addition to $Z^\prime$ exchange (left-handed diagram), all CP-even
Higgs bosons contribute to the process in the $s$-channel (right-handed
diagram).

 \setlength{\voffset}{-0.5in}
\begin{table}[htbp]
 \begin{center}
\setlength{\extrarowheight}{-5.9pt}
\small
\begin{tabular*}{0.99\textwidth}{@{\extracolsep{\fill}} ccccccc}
\hline\hline
$\rm Parameters$&\multicolumn{3}{c}{$\rm MSSM$}&\multicolumn{3}{c}{$\rm U(1)'$}
\\ \hline
& $\rm LM1$ & $\rm LM2$ & $\rm LM6$  & $\rm LM1^\prime$  &$\rm LM2^\prime$ &
$\rm LM6^\prime$
 \\ \cline{2-7}\cline{2-7}
$\rm sign(\mu)$ & $+$ & $+$& $+$ & $+$ & $+$ & $+$ \\
 $\tan\beta$&10&35&10&10&35&10\\
 $Q_{Q}^{\prime}$&--&--&--&-2&-2&-2\\
 $\mu\,(\mu_{eff})$&373&506&583&373&506&583\\
 $h_{\nu}$&--&--&--&1&1&1\\
 $h_s$&--&--&--&0.5&0.7&0.7\\
 $\bar{h}_s$&--&--&--&0.75&0.75&0.70\\
 $A_s$&--&--&--&200&200&200\\
 $A_{\bar {s}}$&--&--&--&100&100&100\\
 $v_{s_1}$&--&--&--&1450&1350&1600\\
 $v_{s_2}$&--&--&--&1250&1250&1450\\
 $v_{s_3}$&--&--&--&1150&1100&1300\\
 $R_{Y^\prime}$&--&--&--&49.4&45&42\\
 $M_{\tilde \nu{_e{_R}}}$&--&--&--&400&500&600\\
 $M_{\tilde \nu{_\mu{_R}}}$&--&--&--&450&550&650\\
 $M_{\tilde \nu{_\tau{_R}}}$&--&--&--&500&600&700\\
 $M_1$&98&139&159&98&139&159\\
 $M_2$&189&266&303&189&266&303\\
 $M_3$&630&871&989&630&871&989\\
 $M_{L_1}$&181&295&284&199&295&284\\
 $M_{E_1}$&110&218&171&121&218&171\\
 $M_{Q_1}$&586&821&916&586&821&916\\
 $M_{U_1}$&569&797&888&569&797&888\\
 $M_{D_1}$&567&795&885&567&795&885\\
 $M_{L_2}$&181&295&284&199&295&284\\
 $M_{E_2}$&110&218&171&121&218&171\\
 $M_{Q_2}$&586&821&916&586&821&916\\
 $M_{U_2}$&569&797&888&569&797&888\\
 $M_{D_2}$&567&795&885&567&795&885\\
 $M_{L_3}$&180&283&284&198&283&284\\
 $M_{E_3}$&108&182&168&121&182&168\\
 $M_{Q_3}$&538&731&842&538&731&842\\
 $M_{U_3}$&467&652&729&467&652&729\\
 $M_{D_3}$&563&748&879&563&748&879\\
 $M^2_{SS_{1,2}}$&--&--&--&$-2\times10^6$&$-2\times10^6$&$-2\times10^6$\\
 $A_t$&-517&-698&-806&-517&-698&-806\\
 $A_b$&-791&-960&-1224&-791&-960&-1224\\
 $A_\tau$&-159&-139&-251&-159&-139&-251\\
\hline\hline
\end{tabular*}
\caption{\label{tab:inputs}\sl\small The scenarios (benchmark points) 
LM1, LM2, and LM6  (for the MSSM  {\it i. e.} minimal supergravity), and
$LM1^\prime,~LM2^\prime$ and $LM6^\prime$ (for the $U(1)^{\prime}$ model). 
The unprimed $LMX$ and primed $LMX^{\prime}$ benchmark points similar
mass spectra. Parameter $R_{Y^\prime}$ is defined in Appendix~\ref{app:C}.}
\end{center}
 \end{table}

\begin{table}[htbp]
  \begin{center}
 \setlength{\extrarowheight}{-5.0pt}
 \small
 \begin{tabular*}{0.99\textwidth}{@{\extracolsep{\fill}} ccccccc}
\hline\hline
$\rm Masses $&\multicolumn{3}{c}{$\rm MSSM$}&\multicolumn{3}{c}{$\rm
U(1)^\prime$}
\\ \hline
& $\rm LM1$ & $\rm LM2$ & $\rm LM6$  & $\rm LM1^\prime$  &$\rm LM2^\prime$ &
$\rm LM6^\prime$
 \\ \cline{2-7}\cline{2-7}
$m_{Z^\prime}$&--&--&--&1476&1418&1661  \\
 $m_{\tilde\chi^0_1}$&96&141&161&96&63&79\\
 $m_{\tilde\chi^0_2}$&178&264&302&99&138&158\\
 $m_{\tilde\chi^0_3}$&340&448&513&177&258&295\\
 $m_{\tilde\chi^0_4}$&360&462&529&356&443&425\\
 $m_{\tilde\chi^0_5}$&--&--&--&392&527&603\\
 $m_{\tilde\chi^0_6}$&--&--&--&412&536&609\\
 $m_{\tilde\chi^0_7}$&--&--&--&633&593&657\\
 $m_{\tilde\chi^0_8}$&--&--&--&1364&1311&1438\\
 $m_{\tilde\chi^0_9}$&--&--&--&5312&6592&7110\\
 $m_{\tilde\chi^\pm_1}$&177&264&303&174&256&293\\
 $m_{\tilde\chi^\pm_2}$&362&466&532&397&523&598\\
 $m_{\tilde e_L}$&186&298&287&155&248&271\\
 $m_{\tilde e_R}$&120&223&178&193&285&206\\
 $m_{\tilde \mu_L}$&186&298&287&155&248&271\\
 $m_{\tilde \mu_R}$&120&223&178&193&285&206\\
 $m_{\tilde \tau_1}$&111&146&171&144&168&195\\
 $m_{\tilde \tau_2}$&190&309&289&200&305&276\\
 $m_{\tilde\nu_e}$&168&287&276&133&235&259\\
 $m_{\tilde\nu_{\mu}}$&168&287&276&133&235&259\\
 $m_{\tilde\nu_{\tau}}$&168&274&275&132&219&258\\
 $m_{\tilde\nu_{e_R}}$&--&--&--&412&514&604\\
 $m_{\tilde\nu_{\mu_R}}$&--&--&--&460&563&654\\
 $m_{\tilde\nu_{\tau_R}}$&--&--&--&509&612&704\\
 $m_{H_1^0}$&109&112&112&218&252&238\\
 $m_{H_2^0}$&371&423&576&780&807&735\\
 $m_{H_3^0}$&--&--&--&852&870&942\\
 $m_{H_4^0}$&--&--&--&884&1198&1089\\
 $m_{H_5^0}$&--&--&--&1251&1883&1339\\
 $m_{H_6^0}$&--&--&--&2789&2770&2844\\
 $m_{A_1^0}$&371&423&576&418&412&431\\
 $m_{A_2^0}$&--&--&--&868&1256&1085\\
 $m_{A_3^0}$&--&--&--&1257&1883&1246\\
 $m_{A_4^0}$&--&--&--&2591&2586&2599\\
 $m_{H^\pm}$&380&431&581&867&1881&1081\\
\hline\hline
\end{tabular*}
\caption{\label{tab:spec}\sl\small The complete mass spectra of the benchmark 
points (scenarios) given in Table~\ref{tab:inputs} for both MSSM and the secluded $U(1)^\prime$.}
\end{center}
 \end{table}

Once the scalar neutrinos are produced, they will decay. The decay pattern
strictly depends on scenario chosen for the free parameters. Since we are
interested
in rather light scalar neutrinos (assuming low-energy SUSY exits), we prefer to
choose MSSM-like Low Mass (LM) scenarios
\cite{Ball:2007zza}.  Battaglia {\it et. al} have proposed updated
post-WMAP
benchmark points
for the constrained MSSM  \cite{Battaglia:2003ab} modifying earlier proposal
\cite{Battaglia:2001zp}, and we include these points in Table {\ref{tab:inputs}.

To compare our results with MSSM predictions,  we choose three low-mass MSSM scenarios (benchmark points),
namely LM1, LM2 and LM6, from the low mass scenarios of MSUGRA and use {\tt
Softsusy} package \cite{Allanach:2001kg} to generate the MSSM spectrum. In the
secluded $U(1)^\prime$ we choose LM-like scenarios, denoted as $\rm LM1^\prime$,
$\rm LM2^\prime$ and $\rm LM6^\prime$  by keeping the overlapping parameters the
same and fixing the additional parameters to agree with phenomenological
constraints on masses. The input parameters for LM1, LM2 and LM6
for MSSM as well as their corresponding prime versions for the secluded
$U(1)^\prime$ are given in Table~\ref{tab:inputs}. As seen from
Table~\ref{tab:inputs}, the VEVs of the additional scalars ($S_1,S_2$ and $S_3$)
$v_{s_i}, i=1,2,3$ are taken above the TeV scale so that the $Z^\prime$ mass
bound is satisfied no matter what the VEV of the scalar field $S$ is chosen. In
fact, for convenience, the parameters $\mu_{eff}$ and $h_s$ are taken as free
parameters and the VEV of $S$ are determined accordingly using the relation
given in Eq.~(\ref{eq:rho}). From Table~\ref{tab:spec} it is seen that the
scalar neutrino masses are rather light. The left-handed sneutrinos masses are
varying in the 168 GeV-287 GeV range while the right-handed ones are in the 412
GeV-704 GeV depending on the LM scenario as well as on the flavor of the scalar
neutrino.  The right-handed scalar neutrinos are heavier, showing the same
pattern as in the neutrino sector. With these chosen masses  we can foresee that
the production cross section for the left-handed sneutrinos will dominate the
one for the right-handed ones.

The validity of the MSSM scenarios LM1, LM2 and LM6 has been confronted with
both the LEP and Tevatron data. There will be no contributions to the LEP
observables from our $\rm LMX^\prime,~X=1,\, 2,\, 6$ scenarios since the
lightest Higgs boson mass in the model is 218 GeV, which is already above the
LEP energy. For the Tevatron case, however, one needs do a more careful analysis.
Nevertheless, as the  $\rm LMX^\prime$ scenarios aim to be consistent with the
corresponding MSSM scenarios, in the limit where the extra $U(1)^\prime$
particles decouple, we expect consistency with the Tevatron data. To verify
this point, we used the package {\tt HiggsBounds} \cite{Bechtle:2008jh}, which
yields results for any arbitrary Higgs sector.

\begin{table}[htbp]
  \begin{center}
 \begin{tabular*}{0.99\textwidth}{@{\extracolsep{\fill}} ccccccc}
\hline\hline
$\rm Observables $&\multicolumn{3}{c}{$\rm MSSM$}&\multicolumn{3}{c}{$\rm
U(1)^\prime$}
\\ \hline
& $\rm LM1$ & $\rm LM2$ & $\rm LM6$  & $\rm LM1^\prime$  &$\rm LM2^\prime$ &
$\rm LM6^\prime$
 \\ \cline{2-7}\cline{2-7}
$\rm \sigma(pp\rightarrow \tilde\nu_{e_R}^{} \tilde\nu_{e_R}
^*)/fb$ & -  & -  & - & 80.5 & 67.8 & 29.1 \\
$\rm \sigma(pp\rightarrow \tilde\nu_{\mu_R}^{} \tilde\nu_{\mu_R}
^*)/fb$ & -  & -  & - & 66.7 & 55.1 & 24.0 \\
$\rm \sigma(pp\rightarrow \tilde\nu_{\tau_R}^{} \tilde\nu_{\tau_R}
^*)/fb$ & -  & -  & - & 54.9 & 44.6 & 19.7 \\
$\rm \sigma(pp\rightarrow \tilde\nu_{\ell_L}^{} \tilde\nu_{\ell_L}
^*)/fb$ & 36.7  & 4.1  & 5.3 & 887.6 & 734.0 & 371.9 \\
$\rm \sigma(pp\rightarrow \tilde\nu_{\tau_L}^{} \tilde\nu_{\tau_L}
^*)/fb$ & 37.2  & 4.9  & 5.3 & 890.7 & 778.7 & 373.1 \\
%\cline{2-7}
$\rm \sigma_{TOT}(pp\rightarrow \tilde\nu_i^{}  \tilde\nu_i
^*)/fb$ & 110.6 & 13.1 & 15.9 & 2868.0 & 2414.2 & 1189 \\
\cline{2-7}
  $\Omega_{\rm DM}h^2$ & 0.120 & 0.120 & 0.120 & 0.115 & 0.109 &0.100\\
\hline\hline
\end{tabular*}
\caption{\sl \small The production cross section and the relic density
$\Omega_{\rm DM}$ values for the
$LM$ scenarios considered in the paper.} \label{tab:cs}
\end{center}
\end{table}

The production cross sections for the scattering $pp \to
\tilde\nu_{\ell_{L,R}}^{}\tilde\nu_{\ell_{L,R}}^*$ processes are listed in
Table~\ref{tab:cs}, for both MSSM and the secluded $U(1)^\prime$ model. The
values were obtained  implementing the secluded $U(1)^\prime$ model into  {\tt
CalcHEP} \cite{calchep} with the help of {\tt LanHEP} \cite{Semenov:2008jy}. The
parton distributions in the proton have been parametrized by using {\tt CTEQ6M}
of {\tt LHAPDF} \cite{Whalley:2005nh}. The MSSM total cross sections (including
the three scalar neutrino flavors) are in the range of 4 to 110 fb while in the
secluded $U(1)^\prime$ model they are varying between 1.1 pb to 2.6 pb. The new
right-handed sneutrino cross sections in the secluded $U(1)^\prime$ model are
about  10 times smaller than  the cross sections for their left-handed
counterparts, and are in the range of 20 fb to 80 fb.

%%%%%%%%%%%%%%%%%%%%%%%%%%%%%%%%%%%%%%%%%%%%%%%%%%%%%%%%%%%%%%%%%%%%%%%%%%%%%%%
\begin{figure}[htbp]
%\vskip 0.2in
\begin{center}
	\includegraphics[width=6.5in,height=4.8in]{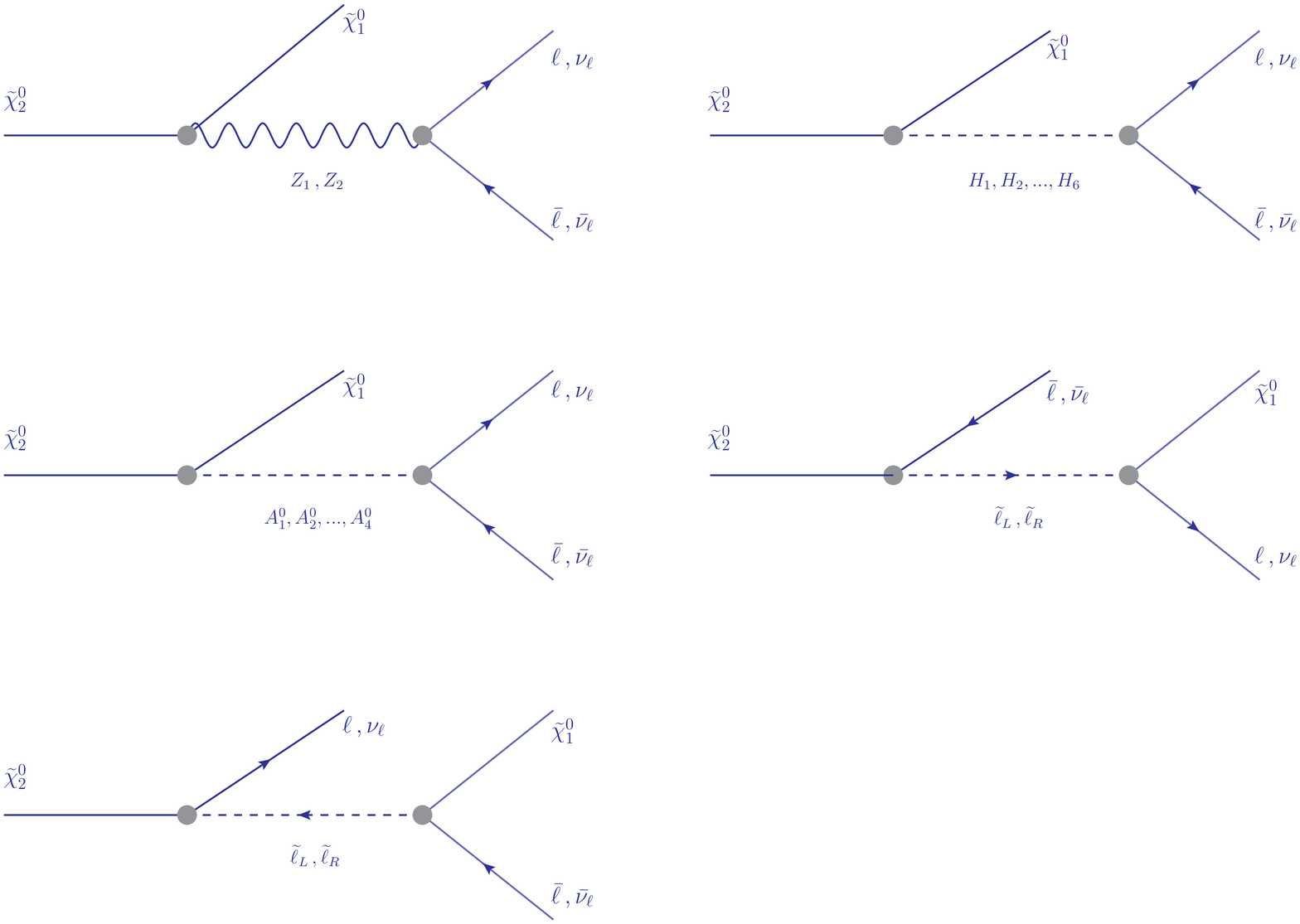}
\end{center}
\vskip -0.3in
      \caption{\sl\small The Feynman diagrams for the three-body decay channels
of the
 next-to-LSP $\niki$ in the secluded-$U(1)^\prime$ model. Here $H_i, i=1,...,6$ are
the CP-even physical Higgs bosons while $A^0_i, i=1,...,4$ are CP-odd
ones.}
\label{fig:dec}
\end{figure}
%%%%%%%%%%%%%%%%%%%%%%%%%%%%%%%%%%%%%%%%%%%%%%%%%%%%%%%%%%%%%%%%%%%%%%%%%%%%%%%

In Table~\ref{tab:cs}, we also included  the relic density of the dark matter
for all six scenarios. This calculation is straightforward using the {\tt
Micromegas} package \cite{Belanger:2008sj}, once we include the model files from
 {\tt CalcHEP}. All the numbers are within the $1\sigma$ range of the WMAP
result \cite{wmap} which can be given with those from the Sloan Digital Sky
Survey \cite{Spergel:2006hy} \begin{eqnarray} \Omega_{DM} h^2 =
0.111^{+0.011}_{-0.015}\,. \end{eqnarray} We note that the relic density of the
dark matter $\Omega_{\rm DM} h^2$ is very sensitive to the free parameter
$R_{Y^\prime}$ listed in Table~\ref{tab:inputs} which varies between 42 to 50.
It's defined (see also Appendix~\ref{app:C}) as the ratio between bare $U(1)$
gaugino masses $$R_{Y^\prime}\equiv
M_{\widetilde{Y}^{\prime}}/M_{\widetilde{Y}}$$ where $M_{\widetilde{Y}}$ and
$M_{\widetilde{Y}^{\prime}}$ are the Bino and Bino$^\prime$ mass parameters
appearing in the $9\times9$ neutralino mixing matrix. More details are given in
Appendix~\ref{app:C}. In Table~\ref{tab:spec} the Lightest Supersymmetric
Particle (LSP) is the lightest neutralino $\lsp$ with masses 96 GeV, 63 GeV and
79 GeV for the $\lmop, \lmtp$ and $\lmsp$ scenarios, respectively. The
next-to-lightest supersymmetric particle is $\niki$ with masses 99 GeV, 138 GeV
and 158 GeV, respectively. For such a spectrum,  there will be no kinematically
available two-body decays for the $\niki$, so that three-body channels need to
be considered. The three-body decay modes relevant to the analysis here are
given in Fig.~\ref{fig:dec}. These decay modes will be considered in the LHC
simulation but not the relic density calculation, as they give negligible
contributions. The sizable contributions to the relic density are for the
$\lmop$ scenario
\begin{itemize}
 \item $\lsp\, \niki \to \tau^- \tau^+\;\; ( 15 \%)$
 \item $\niki\, \niki \to \tau^- \tau^+\;\; ( 13 \%)$
 \item $\lsp\, \niki \to e^- e^+ / \mu^- \mu^+\;\;$ (8\% for each channel)
 \item $\niki\, \niki \to e^- e^+ / \mu^- \mu^+\;\;$ (8\% for each channel)
 \item $\niki\, \niki \to \nu_l \bar{\nu}_l\,, l=e,\mu,\tau\;\;$ (5\% for
each channel)
 \item $\lsp\, \lsp \to \tau^- \tau^+\;\; (5\%)$
 \item $\lsp\, \lsp \to e^- e^+ /\mu^- \mu^+\;\;$ ( 3\% for each channel)
 \item $\lsp\, \niki \to \nu_l \bar{\nu}_l\,, l=e,\mu,\tau\;\;$ (3\% for
each channel)
 \item $\lsp\, \niki \to W^- W^+\;\; (2\% )$
\end{itemize}
In the $\lmtp (\lmsp)$ only $\lsp\, \lsp$ annihilation
contributes to the relic density of the dark matter as follows
\begin{itemize}
 \item $\lsp\, \lsp \to \tau^- \tau^+\;\; (75\%~ (38 \%))$
 \item $\lsp\, \lsp \to \mu^- \mu^+\;\; (8 \%~( 26 \% ))$
 \item $\lsp\, \lsp \to e^- e^+\;\; (8\%~(26\% ))$
 \item $\lsp\, \lsp \to b \bar{b}\;\; (3\% ~ (\% 1))$
 \item $\lsp\, \lsp \to \nu_\tau \bar{\nu}_\tau\;\; (1 \%~ (3 \% ))$
 \item $\lsp\, \lsp \to \nu_l \bar{\nu}_l\,, l=e,\mu\;\; (0 \%~(3 \% ))$
 \item $\lsp\, \lsp \to d \bar{d} /s \bar{s}\;\; (1 \% ~(0 \% ))$
\end{itemize}
Contributions from the $\niki\ \niki$ or $\lsp\ \niki$ annihilations for the
$\lsp$ scenario are due to the fact that $\lsp$ and $\niki$ are almost
degenerate in mass and since the two-body decay channel limit is used in $\tt
Micromegas$, $\niki$ acts very similar to $\lsp$. There is no sizable
contributions from $\niki$ in the other two scenario since $\niki$ is much
heavier. The bino, wino, higgsino and singlino compositions of the neutralinos
for the scenarios $\lmop,\lmtp$ and $\lmsp$ are given  in
Table~\ref{tab:neut_comps} in Appendix~\ref{app:D}. The LSP $\lsp$ is mainly
bino for $\lmop$ but mostly singlino ($\widetilde{S}$) for $\lmtp$ and $\lmsp$
($94.2 \%$ and $93.6 \%$, respectively, for the two scenarios). The situation is
reversed for the next-to-LSP, $\niki$.

\subsection{The LHC Signals}

After discussing the chosen scenarios and the details of the relic density
calculation of the dark matter, we proceed to discuss the signals
at LHC from scalar neutrino production processes. To  determine
and classify all possible signals for the scenarios $\lmop, \lmtp$ and $\lmsp$,
we need to look at the decay topology of the scalar neutrinos.

Since we include MSSM scenarios LM1, LM2 and LM6 for comparison purposes, we
first outline the main decay modes governing the decay channels. The left-handed
scalar neutrinos $\nL$ decay to $\nu_\ell\, \lsp$ with about $100 \% $ branching
ratio for the LM1 and LM6, since all the other neutralinos are heavier than the
scalar neutrinos. The picture is a bit more complicated for the LM2 where $\nL,
\ell=e,\mu$ decay to  $\nu_\ell \, \lsp\; (71 \%),\; \ell\, \caa\; (20 \%)$ and
$\nu_\ell\, \niki\; (8.8 \%)$. For the $\nlL$, the branching decay ratios are
$W\, \stau\; (61.5\%),\;  \nu_\tau\, \lsp\; (34.2\%),\; \tau\, \caa\; (3\%)$ and
$\nu_\tau\, \niki\; (1.3\%)$. Further in the decay chain $\niki$ decays mainly
to $\tau \stau/\bar{\tau} \stau^*$ (48\% for each channel), and the chargino
$\caa to \nu_\tau \stau \;$ (with $95.4\%$ branching ratio) and $W\,\lsp\;$
($4.6\%$ branching ratio).

In the secluded $U(1)^\prime$ model, the decay modes of the scalar neutrinos
with more than 1\% branching ratio are, for the scenarios $\lmop/\lmtp/\lmsp$
\begin{itemize}
 \item $\nL(\nR)\to \nu_\ell\, \lsp\,,\;\;\
 8.6\%\;(0\%)\;\ /\;\
91.7\%\;(84.8\%)\;\ /\;\ 93.2\%\;(65.7\%)$
\item $\nL(\nR)\to \nu_\ell\, \niki\,,\;\;\ 91.4\%\;(90.6\%)\;\ /\;\
8.3\%\;(0\%)\;\ /\;\ 6.8\%\;(0\%)$
\item $\nL(\nR)\to \nu_\ell\, \ndort\,,\;\;\ 0\%\;(8.8\%)\;\ /\;\
0\%\;(14.8\%)\;\ /\;\ 0\%\;(34.1\%)$
\end{itemize}
There will be further decays of $\niki$ and $\ndort$ in the chain. It is better
to consider $\ndort$ first. Again in the same notation ($\lmop/\lmtp/\lmsp$) it
decays as
\begin{itemize}
 \item $\ndort \to \tau\, \stau^*\;(\tau\,\staut^*)\,,\;\;\ 6.5\%\;(4.9\%)\;\
/\;\ 9.1\%\;(3.7\%)\;\ /\;\ 8.6\%\;(3.9\%)$
 \item $\ndort \to \ell\, \lL^*\;(\ell\,\lR^*)\,,\;\;\ 6.0\%\;(5.3\%)\;\
/\;\ 5.2\%\;(5.0\%)\;\ /\;\ \ 4.0\%\;(8.3\%)$
 \item $\ndort \to \nu_\ell\, \nL^*\,(\nu_\tau\,\nlL^*)\,,\;\;\
5.2\%\;(5.2\%)\;\
/\;\ 5.4\%\;(5.9\%)\;\ /\;\ 4.2\%\;(4.3\%)$
\end{itemize}
where $\ell=e,\mu$ and the conjugated decay modes are not listed. Then the decay
modes of the scalar leptons for $\ell=e,\mu$\footnote{We discard $\ell=\tau$
case since such a pattern ends up with a
$\tau$ lepton in the final state. We concentrate only on the first two generations
of the charged leptons.} are
\begin{itemize}
 \item $\lL\to \ell\, \lsp\;(\ell\,\niki)\,,\;\;\ 6.7\%\;(93.3\%)\;\ /\;\
90.1\%\;(9.9\%)\;\ /\;\ 92.6\%\;(7.4\%)$
 \item $\lR\to \ell\, \lsp\;(\ell\,\niki)\,,\;\;\ 20\%\;(80\%)\;\ /\;\
71.6\%\;(28.4\%)\;\ /\;\ 89.7\%\;(10.3\%)$
\end{itemize}

As can be seen from the above decay patterns, each decay ends up with either an
LSP $\lsp$ or next-to-LSP particle $\niki$. As mentioned earlier,  $\niki$
cannot decay into two-body but instead must undergo the one of the  three-body
decays given in Fig.~\ref{fig:dec}. The relative ratios are\footnote{ Note that
the $\tau^-\tau^-$ channel is not kinematically open for the $\lmop$ scenario.}
given in the ($\lmop/\lmtp/\lmsp$) order as
\begin{itemize}
 \item $\niki \to \nu_\ell\, \bar{\nu}_\ell\,(\nu_\tau\, \bar{\nu}_\tau)\,
\lsp\,,\;\;\
24\%\;(24\%)\;\ /\;\ 4.5\%\;(6.3\%)\;\ /\;\ 1.8\%\;(1.8\%)$
 \item $\niki \to \ell^+\, \ell^-\,(\tau^+\, \tau^-)\,
\lsp\,,\;\;\
14\%\;(0\%)\;\ /\;\ 10.8\%\;(63\%)\;\ /\;\ 28.7\%\;(37\%)$
\end{itemize}

In the light of these decay patterns, there are mainly three types of signal:
(1) $0\ell+\EmissT$, (2) $2\ell+\EmissT$ and (3) $4\ell+\EmissT$. It is
in fact also possible to produce signals with six or eight leptons but the
probability is very suppressed thus we ignore such signals. Therefore, in the
rest of this section we discuss these three signals at LHC. Predictions from MSSM
will be included as well. In MSSM there is no the $4\ell+\EmissT$ type of
signal in MSSM for the LM1 and LM2 and LM2 scenarios. The $2\ell+\EmissT$
signal is possible through chargino $\caa$ decay.

The usual concern is the possible background for the signals from the SM. For
the $0\ell+\EmissT$ mode, the background will come from the Drell-Yan (D-Y),
$pp\to \nu_\ell \bar{\nu}_\ell$, and $pp\to ZZ$ where each of $Z$ decays
invisibly. Since the D-Y has a huge cross section, some cuts need to be
implemented. In the $2\ell+\EmissT$ case, in addition to the D-Y and $ZZ$
production (where one of the $Z$ decays leptonicaly), there is $W^+W^-$
production. In principle there could be contributions from the $t\bar{t}$ with
jet veto, but we ignore such possibility since the $b$-jets are going to be
quite energetic and can be tagged. The process $pp\to ZZ\to 4\ell$ can be the
background for the $4\ell+\EmissT$ decay mode. However, a simple $\EmissT$ cut
would eliminate events from the SM process $pp\to ZZ\to 4\ell$. We confirmed
this with our event simulation.

At the first stage, the following basic cuts are applied to suppress the SM
background. It is required that, whenever relevant,
\begin{itemize}
 \item Each isolated charged lepton (electron or muon) has a transverse momentum
$p_T(\ell)>15 \gev$.
 \item The missing transverse energy satisfies $\EmissT> 100\gev$.
 \item The leptons are constrained to be in the central barrel region of the
detector by forcing the pseudorapidity $|\eta|<2$.
 \item The cone size between two charged lepton $\Delta R_{\ell\ell}$ is at
least 0.4. Here $\Delta R_{\ell\ell}=\left(\Delta \eta^2 + \Delta
\phi^2 \right)^{1/2}$ defined in the pseudorapidity-azimuthal angle plane.
\end{itemize}

As mention above, a missing transverse energy cut $\EmissT> 100\gev$ practically
gets rid of the SM background for the $\fourlep$ signal, which is now considered
background free. For the $\fourlep$ signal, in order to get enough statistics
after the cuts (as much less number of events pass the cuts as compared of the
other two signals), we relaxed the some of the above cuts. We use $p_T(\ell)>5
\gev$ and $\Delta R_{\ell\ell}>0.2$ for the analysis of this signal.

With the above cuts,  the SM background is still larger than the signals
$\nolep$ and $\twolep$. The D-Y and $W^+W^-$ dominate the $ZZ$ cross section and
they are all well above the signal for $\EmissT<500 \gev$. Such background
domination happens in various other distributions in most part of the region.
There is no point to present these figures. Instead we need to find a better way
to handle the background. After examining the results at the first stage, we
decided to use $\ETsum$,  also known as the effective mass $m_{\rm eff}$ in
literature. This variable could be helpful in reducing the backgrounds while
keeping most of the signal events especially if we use a suitable value for the
cuts. $\ETsum$ is defined as the scalar sum of the lepton transverse momenta and
the missing transverse energy
\begin{eqnarray}
 \ETsum\equiv m_{\rm eff}=\sum_\ell |p_T(\ell)|\ + \ \EmissT.
\end{eqnarray}
where the missing transverse energy $\EmissT$ is the sum of the total $x$ and
 $y$ components of the momenta in quadratures. Since it has been observed
that the signal processes lead to mostly high $m_{\rm eff}$ (or $\ETsum$)
distributions, a cut on $m_{\rm eff}$ would substantially reduce the
background. Thus, as a second stage for the cuts $m_{\rm eff}=\ETsum > 750
\gev$ has been employed (but only for the $\nolep$ and $\twolep$ cases.)

Global inclusive variables like $\EmissT$ and $\ETsum$ are used to estimate the
mass scalar of the parent particles produced in the hard scattering (thus
estimating the scale of the new physics). In a recent paper by Konar \etal\
\cite{Konar:2008ei}, a new global inclusive variable, called $\smin$, is
proposed as an alternative. For SUSY models with $R$ parity conservation, the
decay chain always ends with an LSP, which is left undetected at the collider.
This makes mass reconstruction procedure almost impossible, especially if there
are more than one LSP (there are at least have two LSPs in the final state).
Without going into extensive details of the signal, there is an easy way to
approach guessing the scale of the new physics through the parameter $\smin$. It
is defined as \cite{Konar:2008ei}
\begin{eqnarray}
\smin = \sqrt{E^2-P_z^2}+\sqrt{\EmissT^{\,2}+M_{\rm invisible}^2}
\end{eqnarray}
where $E$ is the total calorimeter energy, $\vec{P}$ is the total {\it visible}
momentum and $M_{\rm invisible}$ is the total mass of all invisible particles
produced in each event, which is the only unknown. All  the others variables can
be measured at the detector. Hence $\smin(M_{\rm invisible})$ is the variable to
consider. The {\it peak} of the $\smin$ distribution is associated with the mass
threshold of the parent particles originated from the hard scattering. Of
course, an estimation needs to be done for the total invisible mass  $M_{\rm
invisible}$. In most of the cases the $\smin(0)$ gives a pretty good idea about
the masses of the parent particles. It is shown that the method works better for
signals with fewer invisible particles and/or more visible particles. It also
works better with higher SUSY scales where Initial State Radiations (ISR) are
less significant. We include some figures for $\smin(0)$ in the $\nolep$ as well
as $\fourlep$ signal.

The events are generated at the partonic level with {\tt CalcHEP} \cite{calchep}
and passed to {\tt Pythia} \cite{pythia} with the use of {\tt CalcHEP-Pythia}
interface for hadronization and cuts. We simulated $4\times 10^{6}$ events for
the  $\nolep$, $\twolep$ and $\fourlep$ signals. Since the relative number of
events in each signal turns out to be proportional to the relevant branching ratio
combination, the number of events can be simply weighed by
$w=\sigma(pp\rightarrow \tilde\nu_\ell^{} \tilde\nu_\ell^*)\times {\cal L} /
N_{\rm tot}$ where ${\cal L}$ is the integrated luminosity and $N_{\rm tot}$ is
the total number of event generated. We set ${\cal L}=100 \xfb^{-1}$,  the
ultimate goal that is expected at the LHC.  Even though the current reach of LHC
center of mass energy is  $7 \tev$, we use $14 \tev$ in the numerical study,
which maximizes the reach in the parameter space.
%%%%%%%%%%%%%%%%%%%%%%%%%%%%%%%%%%%%%%%%%%%%%%%%%%%%%%%%%%%%%%%%%%%%%%%%%%%%%%%
\begin{figure}[htbp]
%\vskip -0.3in
\begin{center}$
	\begin{array}{cc}
\hspace*{-1.7cm}
	\includegraphics[width=3.8in,height=3.2in]{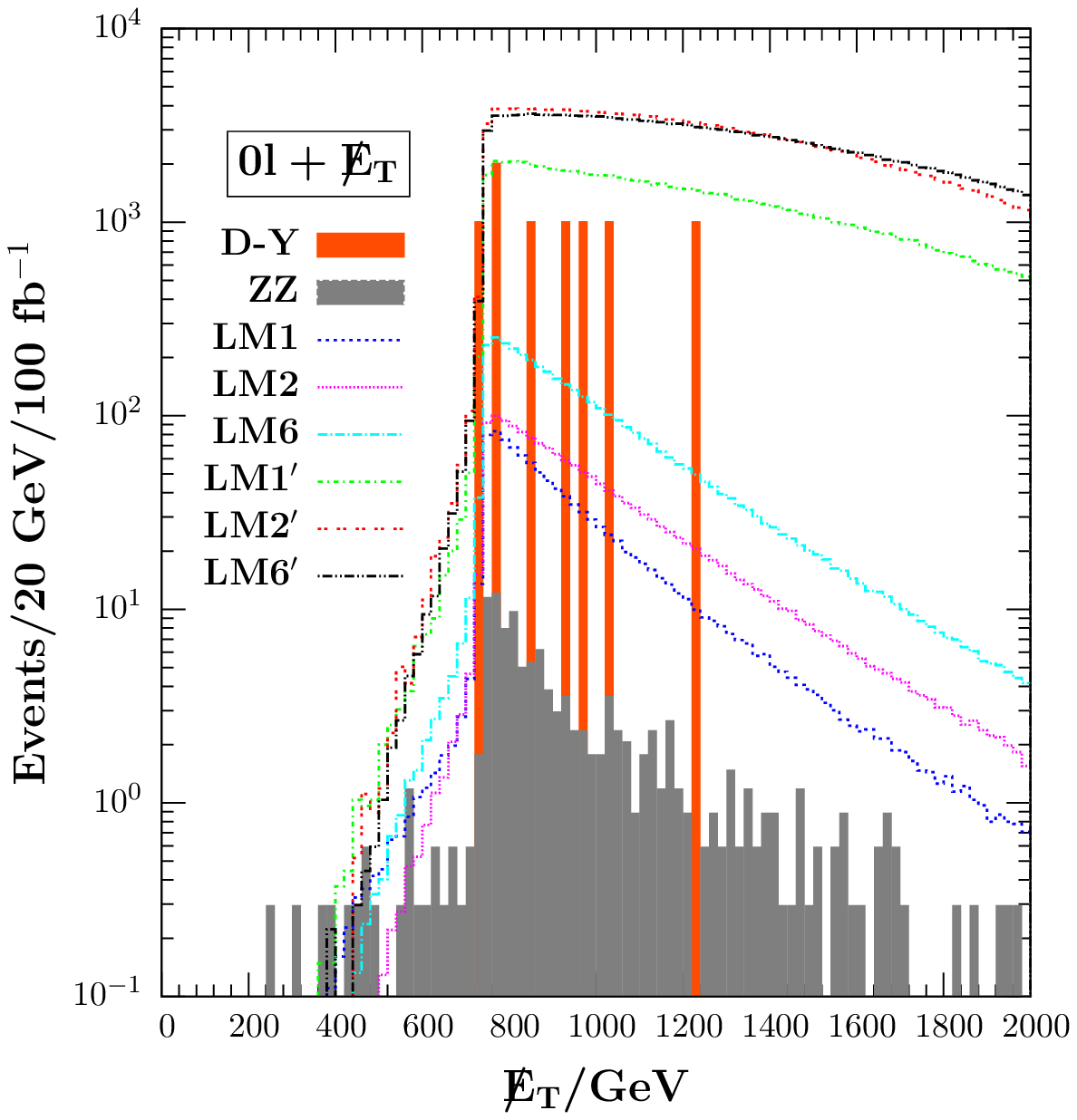}
&\hspace*{-1.5cm}
	\includegraphics[width=3.8in,height=3.2in]{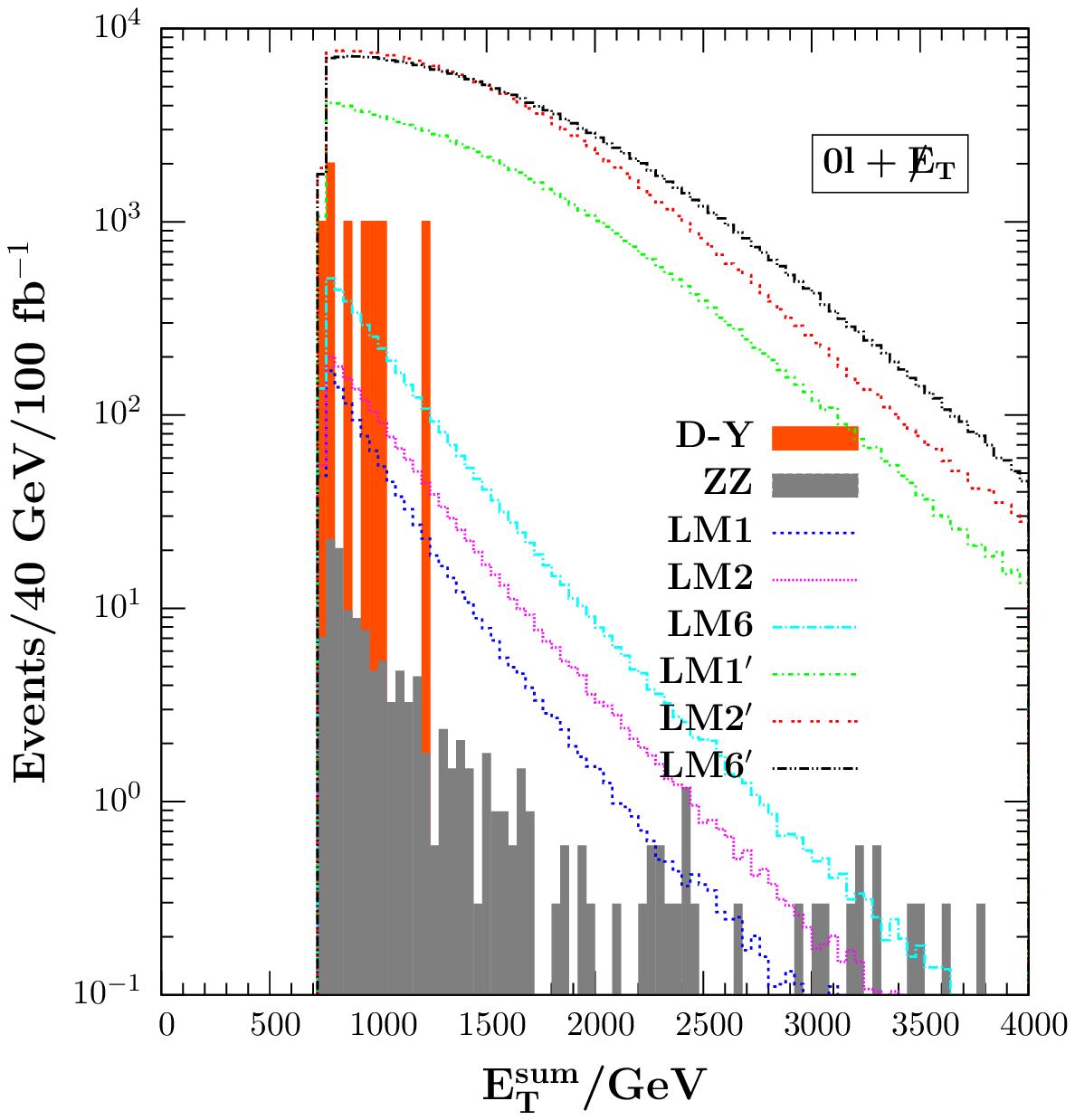} \\
\hspace*{-1.7cm}
	\includegraphics[width=3.8in,height=3.2in]{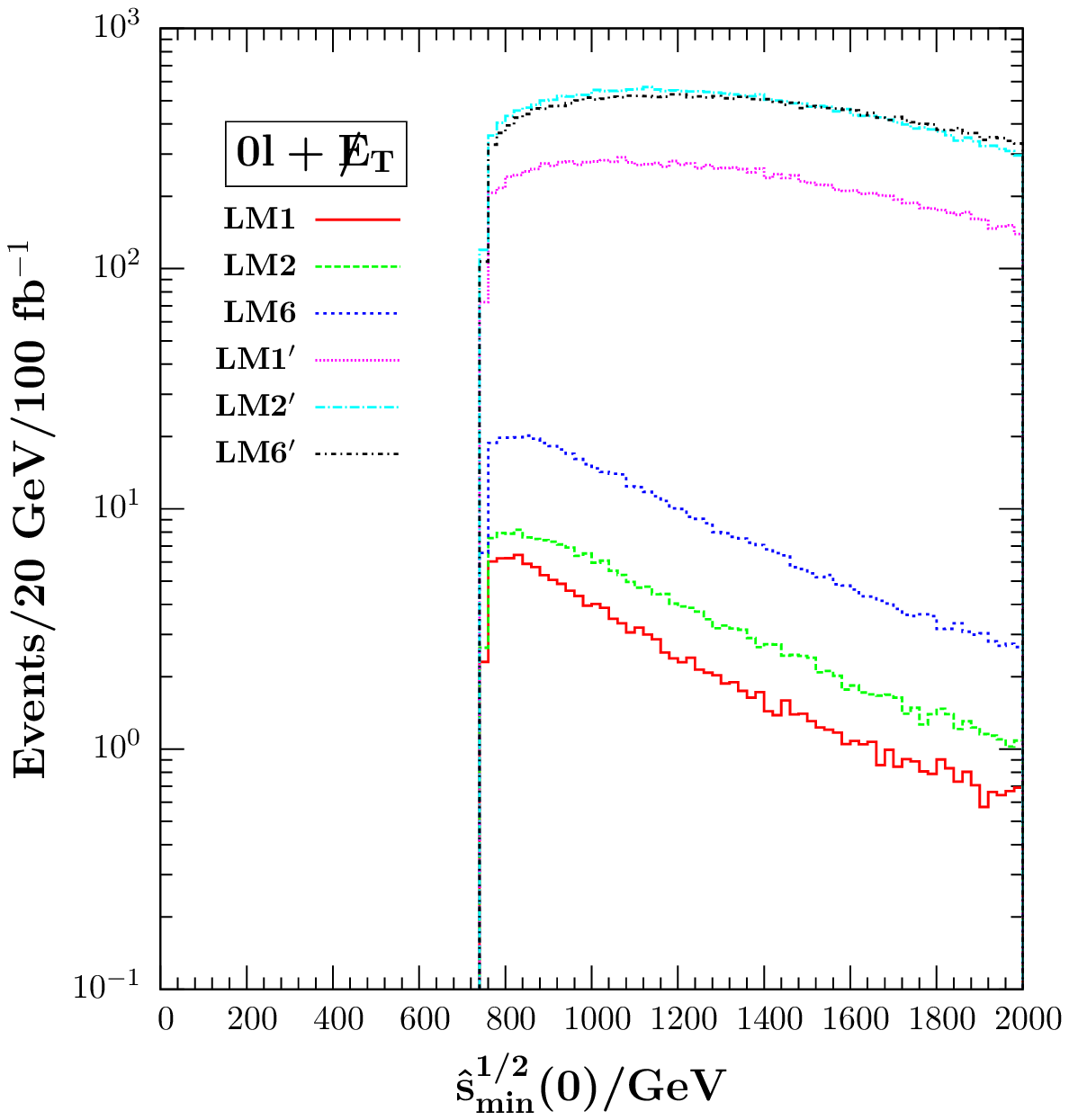}
&\hspace*{-1.5cm}
	\includegraphics[width=3.8in,height=3.2in]{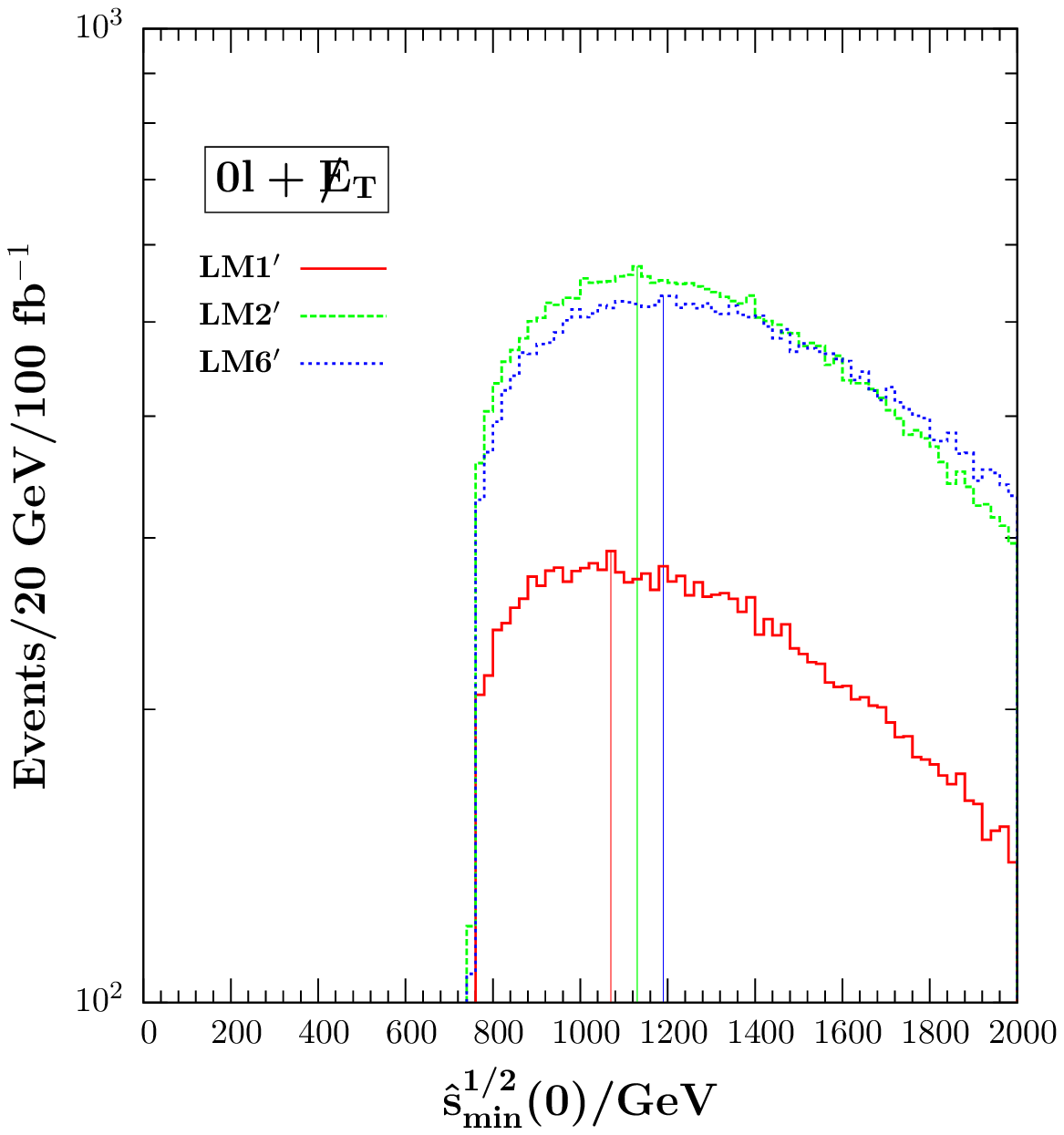} \\
	\end{array}$
\end{center}
\vskip -0.1in
      \caption{\sl\small The $\EmissT$, $\ETsum$, and $\smin(0)$ distributions
of the $\nolep$ signal at $14 \tev$ with integrated luminosity ${\cal
L}=100\xfb^{-1}$, for the three scenarios in both MSSM and the secluded
$U(1)^\prime$ model.}
\label{fig:0lep}
\end{figure}
%%%%%%%%%%%%%%%%%%%%%%%%%%%%%%%%%%%%%%%%%%%%%%%%%%%%%%%%%%%%%%%%%%%%%%%%%%%%%%%

\subsubsection{The Missing Energy Signal: $\nolep$}
%%%%%%%%%%%%%%%%%%%%%%%%%%%%%%%%%%%%%%%%%%%%%%%%%%%%%%%

The distributions of $\EmissT$, $\ETsum$, and $\smin(0)$ are depicted in
Fig.~\ref{fig:0lep} for the three scenarios $\lmop, \lmtp$ and $\lmsp$ as well
as the three benchmarks for the MSSM. In general the $\lmsp$ scenario has the
largest event pass the cuts, with similar results for the $\lmtp$, while the
$\lmop$ has the lowest. In fact about $80\%$ of the events pass the cuts in
$\lmtp$ and $\lmsp$ but only $60\%$ do so for the $\lmop$. In all three
scenarios, $100\%$ of the events pass the $\EmissT$ cut so that we are only
losing $20\%$ to $40\%$ of them by employing the $m_{eff}$ cut. This is because
the direct LSP decay modes of the scalar neutrinos (both the left handed and
right handed one) are either not available or suppressed for the $\lmop$, so
that the $\nolep$ signal would emerge from indirect decay channels through
$\niki$ or $\ndort$ decays, with smaller branching ratio combinations. This can
be understood from details provided earlier. The distributions for the secluded
$U(1)^\prime$ model dominate the ones for the MSSM since basically the total
production cross section in the secluded $U(1)^\prime$ model is much bigger.

The background distributions for the D-Y and $ZZ$ processes are also included in
the $\EmissT$ and $\ETsum$ graphs. The rate of success for the D-Y events
passing both of the cuts are only about eight in $10^6$. To give an idea how
effective the $m_{\rm eff}$ cut is, the success rate of events was about a bit
more than $3\%$ before implementing the $m_{\rm eff}$ cut. The situation is even
more drastic for the $ZZ$ case. While the almost $100\%$ of them passed the
$\EmissT$ cut, this number goes down to $0.3\%$ with the $m_{\rm eff}$ cut.

We included the $\smin(0)$ graphs in the second row of Fig.~\ref{fig:0lep} to
estimate the mass scale of the parent particles, i.e., the left handed and right
handed scalar neutrinos. The graph on the bottom right panel  is nothing but the
zoom-in version of the one left handed side for the secluded $U(1)^\prime$
model. We cannot say anything about the MSSM case since the sum of the parent
particle masses are varying in the $300\gev$ to $600\gev$ range, so that the
$\smin(0)$ peak is washed out due to the $m_{\rm eff}$ cut at $750\gev$. Indeed,
for the secluded model we should expect two different peaks, one for the
production of the left handed scalar neutrinos and the other one for the right
handed ones. The peak for the left-handed sneutrinos which are much lighter are
also washed out. We will see the picture clearer for the $\fourlep$ signal.

In the right panel, we also indicated the positions of the peaks, which are
correlated to the masses of the right handed scalar neutrinos. Of course, in
reality to determine the peak position by fitting the data, a better job is
needed . We just want to prove a point here. The peak position is related to the
mass of $\nR$ (since we produce them in pair)
\begin{eqnarray}
 m_{\nR}\approx \frac{1}{2} \left(\smin(0)\right)_{\rm peak}\,.
\end{eqnarray}
From the peak positions in the graph we can estimate the average right handed
sneutrino masses $m_{\nR}\sim (530,565,600)\gev$ for the $(\lmop,\lmtp,\lmsp)$,
respectively, while the real average values should read $(460,563,654)\gev$ from
Table~\ref{tab:spec}. One source of error is not knowing the mass of the LSP
(though we find out that this is not significant here since the LSP mass is
rather light and around $100\gev$) and the other is lack of a real fitting to
the data to pin down the location of the peaks. The estimated values are still
fairly good. We should also note that the method works better for signals with
more visible particles.
%%%%%%%%%%%%%%%%%%%%%%%%%%%%%%%%%%%%%%%%%%%%%%%%%%%%%%%%%%%%%%%%%%%%%%%%%%%%%%%
\begin{figure}[htb]
%\vskip -0.3in
\begin{center}$
	\begin{array}{cc}
\hspace*{-1.7cm}
	\includegraphics[width=3.8in,height=3.2in]{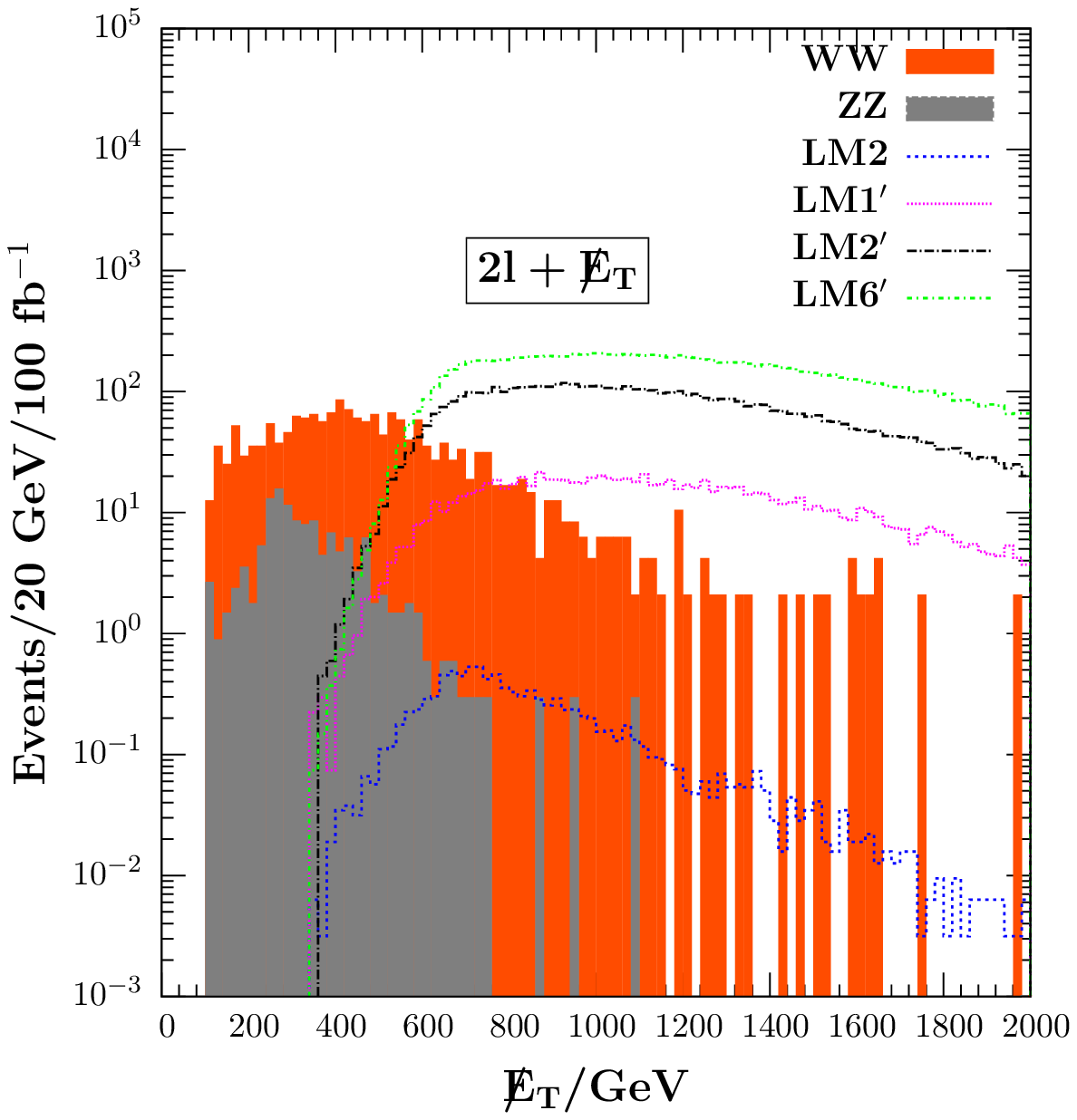}
&\hspace*{-1.5cm}
	\includegraphics[width=3.8in,height=3.2in]{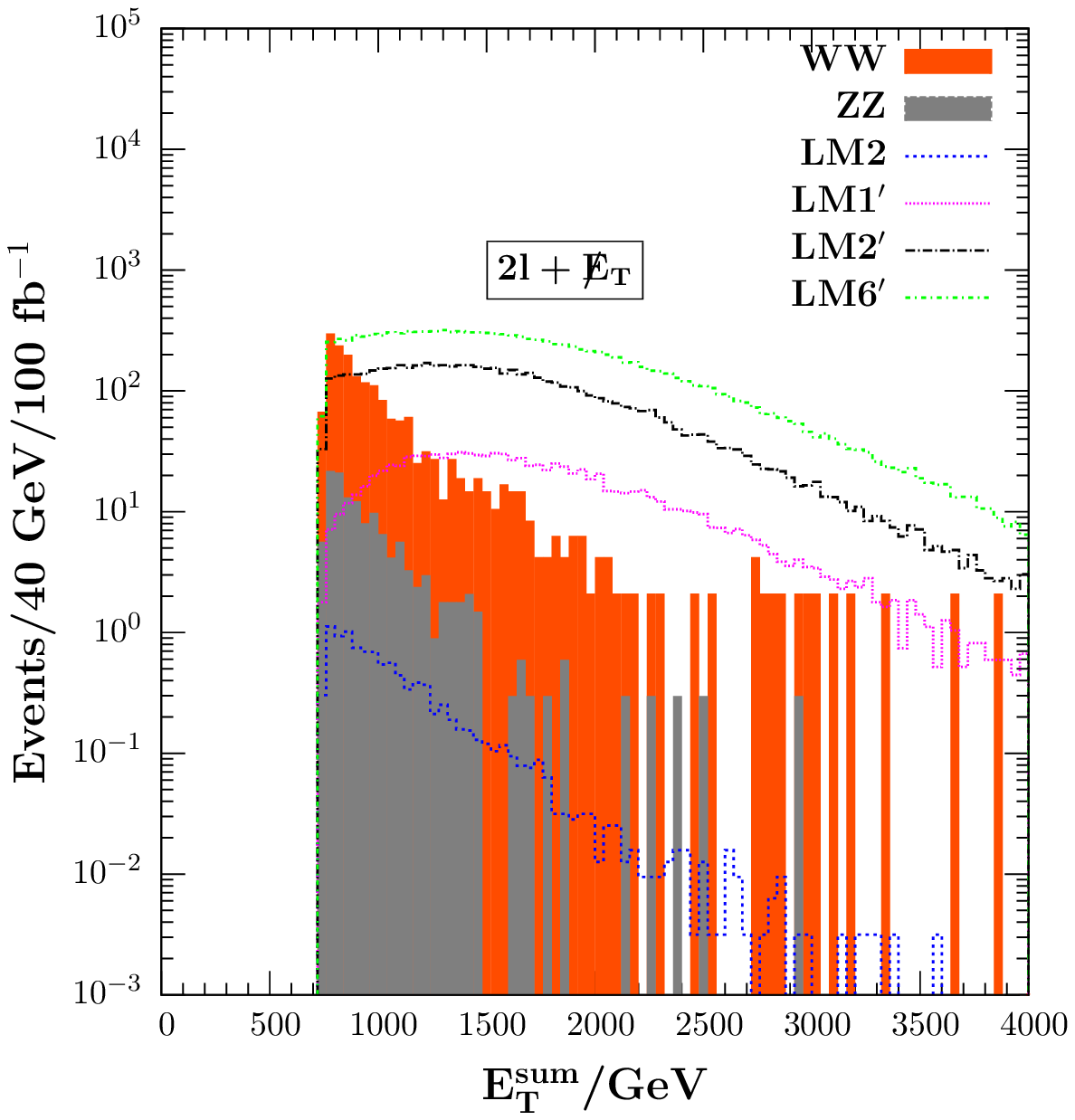} \\
	\end{array}$
\end{center}
\vskip -0.1in
      \caption{\sl\small The $\EmissT$ and $\ETsum$ distributions
of the $\twolep$ signal at $14 \tev$ with integrated luminosity ${\cal
L}=100\xfb^{-1}$ for all three scenarios in the MSSM and secluded
$U(1)^\prime$ model.}
\label{fig:2lep_et}
\end{figure}
%%%%%%%%%%%%%%%%%%%%%%%%%%%%%%%%%%%%%%%%%%%%%%%%%%%%%%%%%%%%%%%%%%%%%%%%%%%%%%%
%%%%%%%%%%%%%%%%%%%%%%%%%%%%%%%%%%%%%%%%%%%%%%%%%%%%%%%%%%%%%%%%%%%%%%%%%%%%%%%
\begin{figure}[htb]
%\vskip -0.3in
\begin{center}$
	\begin{array}{cc}
\hspace*{-1.7cm}
	\includegraphics[width=3.8in,height=3.2in]{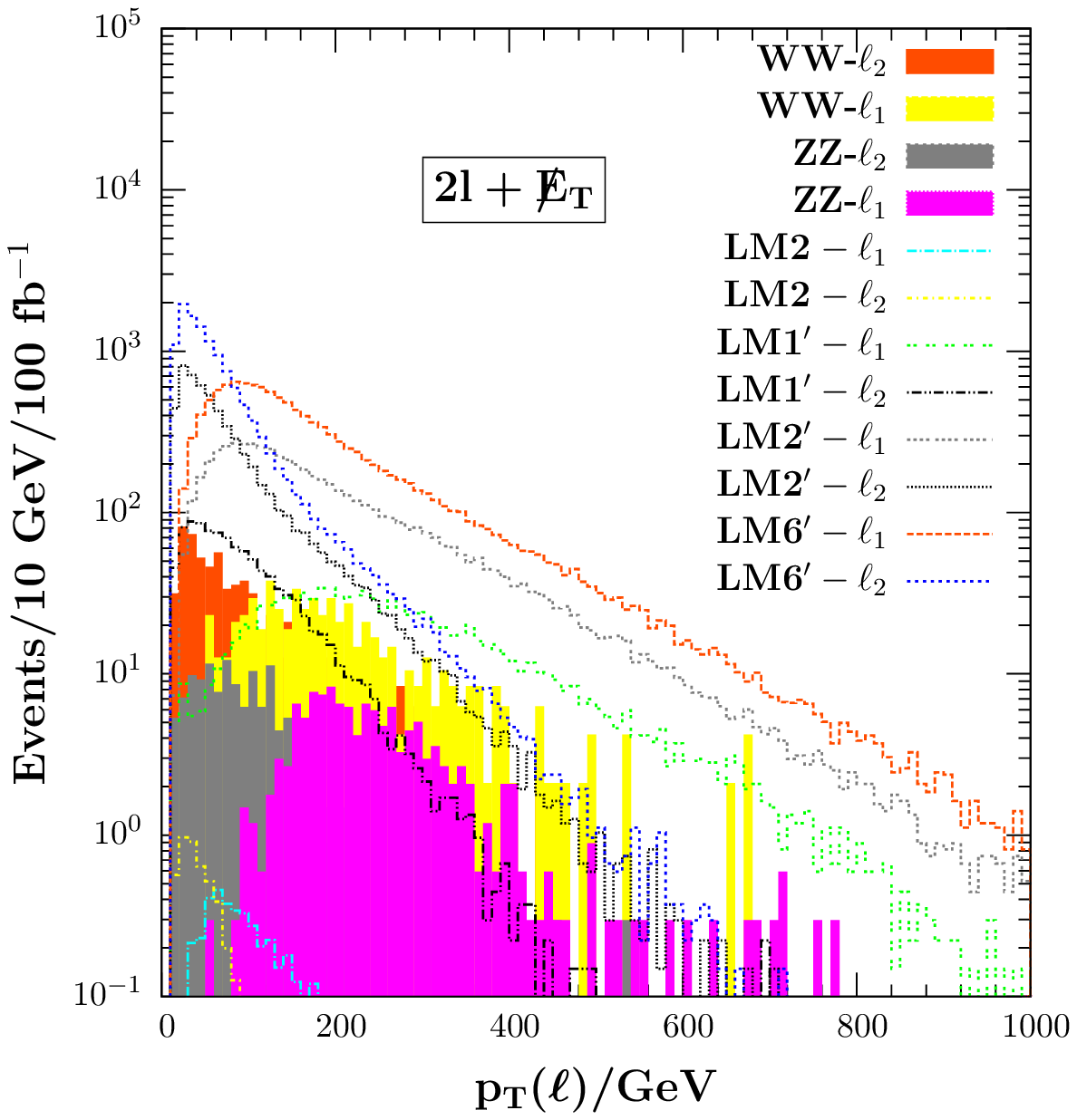}
&\hspace*{-1.5cm}
	\includegraphics[width=3.8in,height=3.2in]{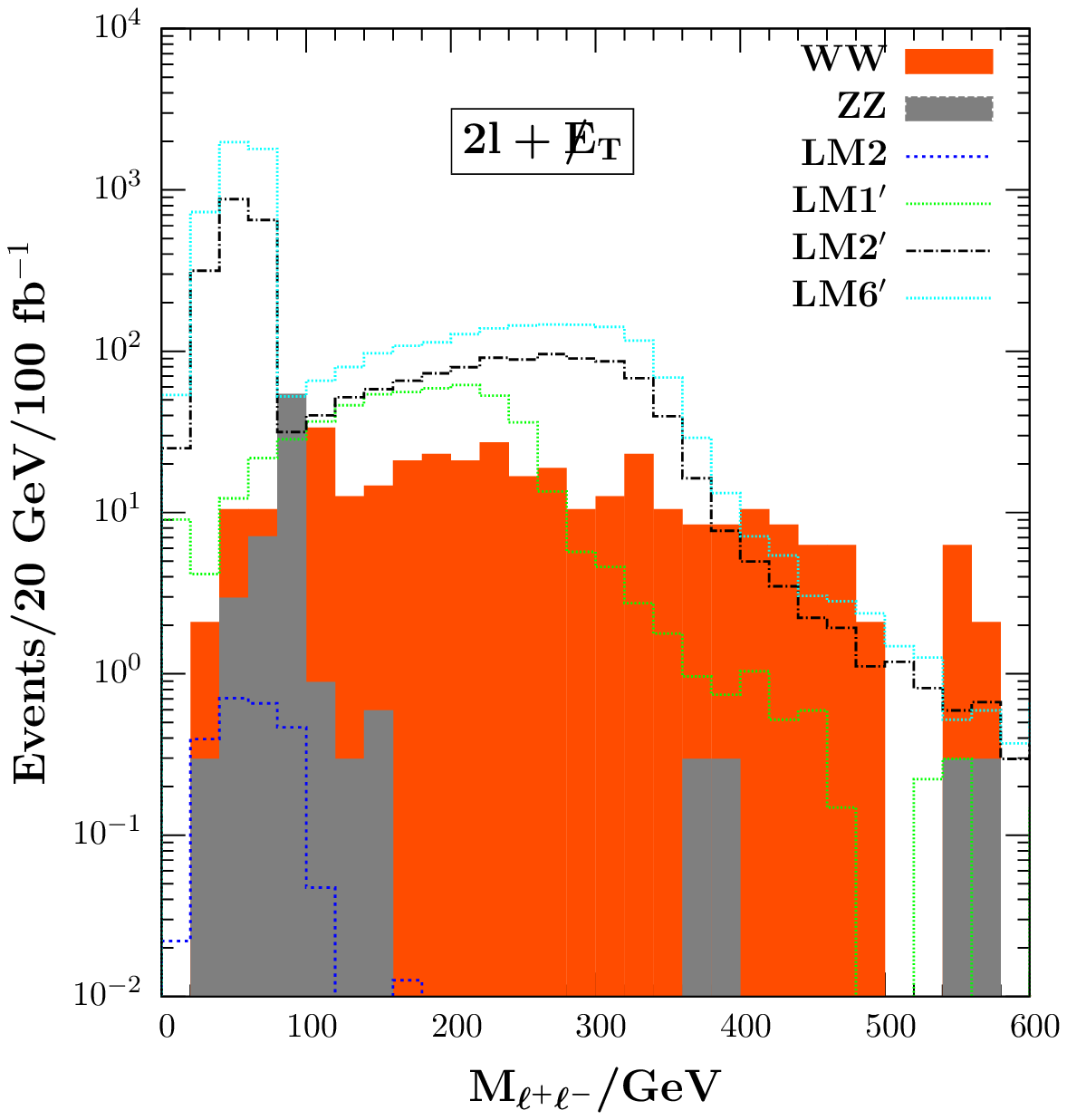} \\
\hspace*{-1.7cm}
	\includegraphics[width=3.8in,height=3.2in]{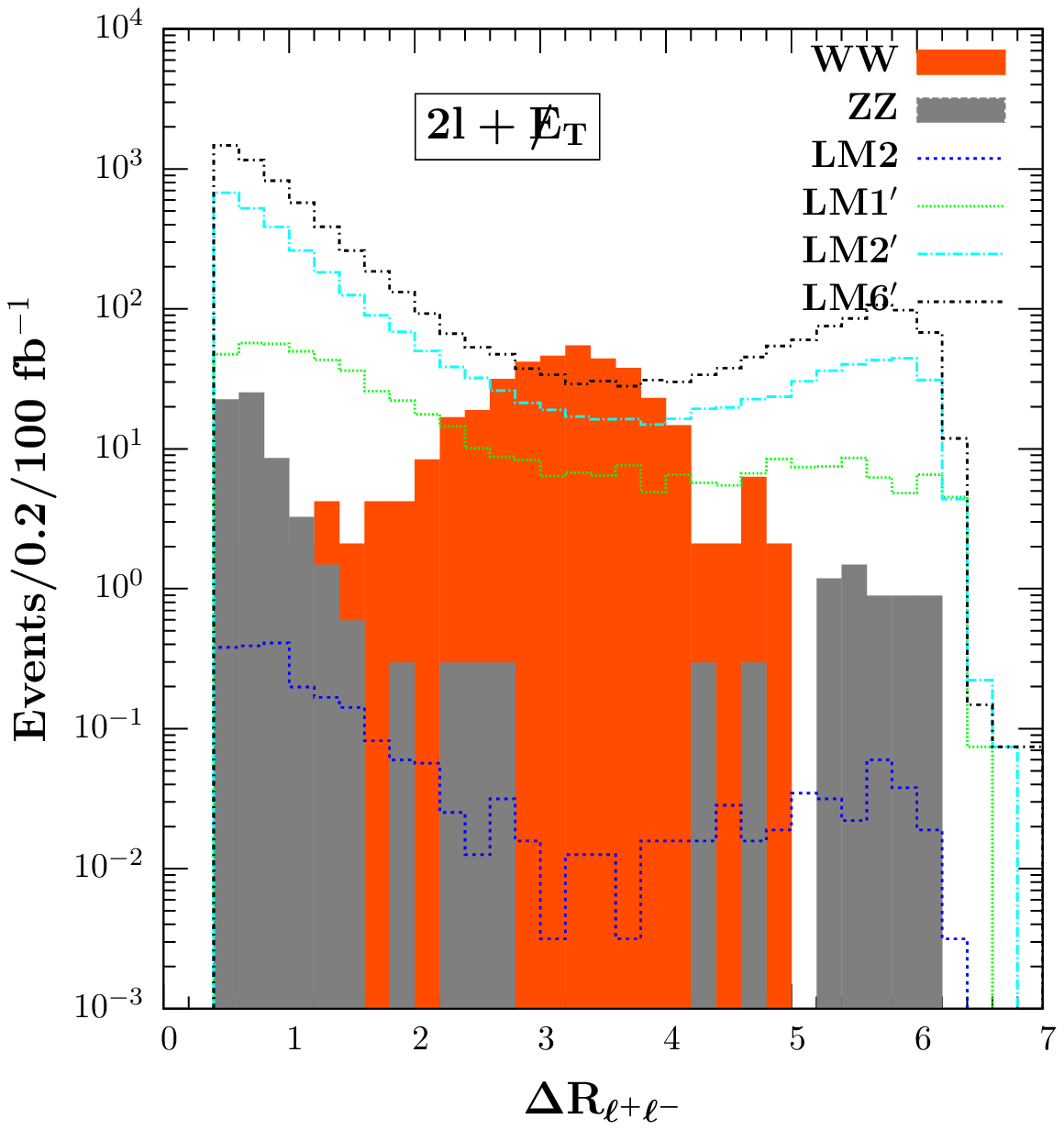}
&\hspace*{-1.5cm}
	\includegraphics[width=3.8in,height=3.2in]{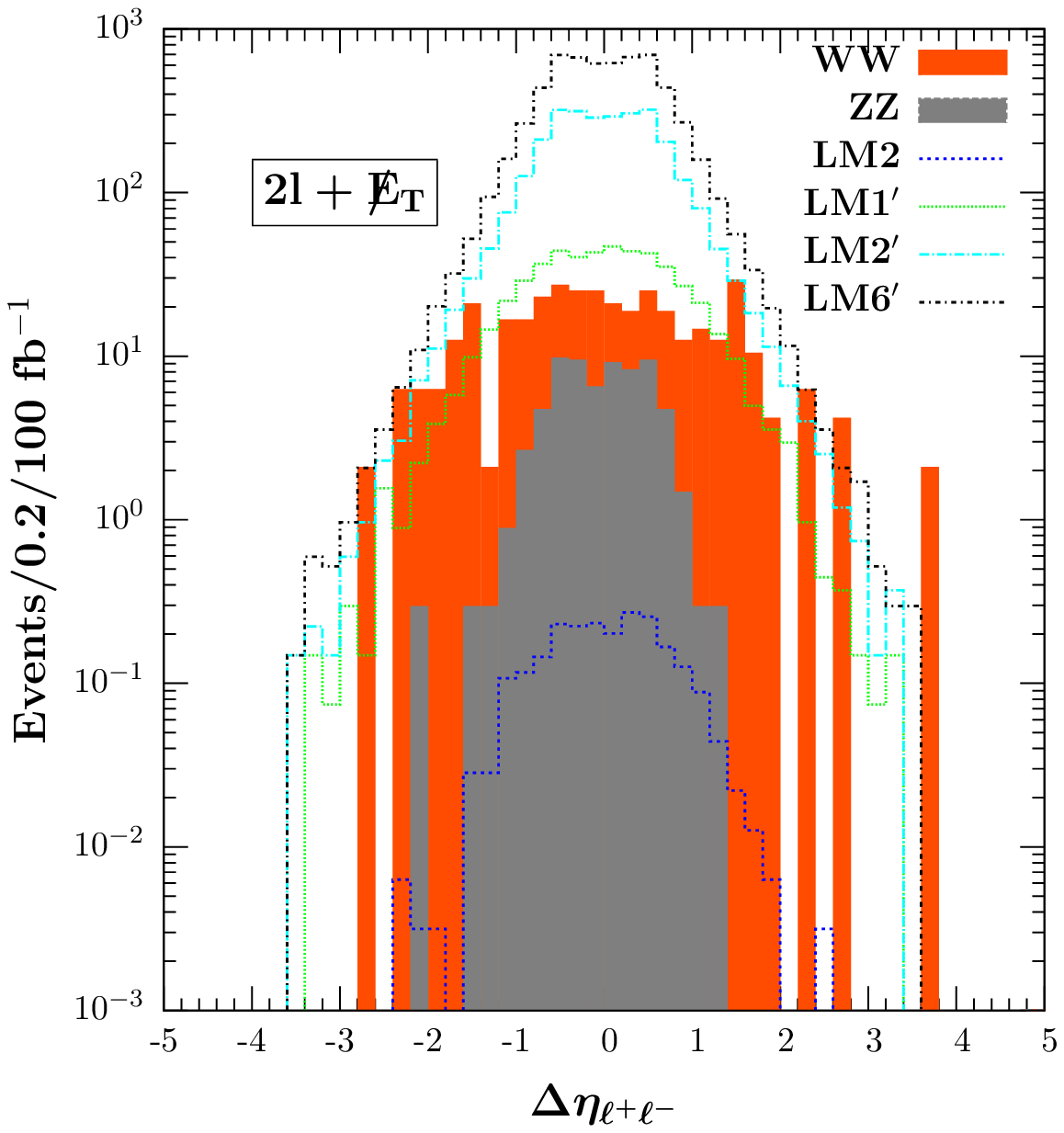} \\
	\end{array}$
\end{center}
\vskip -0.1in
      \caption{\sl\small The $p_T(\ell), M_{\ell^+\ell^-},
\Delta R_{\ell^+\ell^-}$ and $\Delta \eta_{\ell^+\ell^-}$ distributions
of the $\twolep$ signal at $14 \tev$ with integrated luminosity ${\cal
L}=100\xfb^{-1}$ for all three scenarios in both MSSM and secluded
$U(1)^\prime$ model. Here,  $\ell_1$ represents the hardest lepton.}
\label{fig:2lep_rest}
\end{figure}
%%%%%%%%%%%%%%%%%%%%%%%%%%%%%%%%%%%%%%%%%%%%%%%%%%%%%%%%%%%%%%%%%%%%%%%%%%%%%%%

\subsubsection{The Dilepton Signal: $\twolep$}
%%%%%%%%%%%%%%%%%%%%%%%%%%%%%%%%%%%%%%%%%%%%%%%%%%%%%%%

We analyze the $\twolep$ signal  in a similar fashion to the $\nolep$ one in the
previous subsection.  The results are shown in Fig.~\ref{fig:2lep_et} and
Fig.~\ref{fig:2lep_rest}. The main background is from $WW$ and $ZZ$. The D-Y
does not contribute due to the transverse missing energy cut. After all the
cuts, about $0.2\%$ and $0.4\%$ of the events pass for the $WW$ and $ZZ$
backgrounds, respectively. The rates were about $7\%$ and $19\%$, respectively,
before the $m_{\rm eff}$ cut. The situation for the $(\lmop,\lmtp,\lmsp)$
scenarios after all the cuts signal is $(0.9\%,50\%,50\%)$  survival, but
$100\%$ in each cases before the $m_{\rm eff}$ cut. For the MSSM, only the LM2
gives $\twolep$  signal since, for the other two scenarios, the $\nL\lsp$ are
the only final staes. The number of events past the cuts for the LM2 decreases
to $15\%$ from $100\%$ after inclusion of the $m_{\rm eff}$ cut.

We depicted the $p_T$ spectra of both leptons ordered with respect to their
hardness in Fig.~\ref{fig:2lep_rest}. As expected the MSSM leptons are softer,
and the distribution for ones from the $\lmtp$ and $\lmsp$ are very similar. The
$\lmop$ scenario is somewhere in between. In the invariant mass of the leptons,
the $\lmtp$ and $\lmsp$ curves peak at around $60\gev$ and from the mass spectra
in Table~\ref{tab:spec}, the mass difference $m_{\niki}-m_{\lsp}$ is between
$75\gev$ to $80\gev$. The $\twolep$ signal mainly goes through $\niki$. For the
$\lmop$ the mass difference is $3\gev$ and not visible. The $ZZ$ peaks at around
$Z$ boson mass as expected.

In Fig.~\ref{fig:2lep_rest}, we also include the $\Delta R_{\ell^+\ell^-}$ and
$\Delta \eta_{\ell^+\ell^-}$ distributions. It is seen that for both the MSSM
and the secluded $U(1)^\prime$ model more leptons emerge with smaller
separation, unlike the $WW$ case the peak is at the point where the others have
minimum. The background can be reduced further by adjusting the $M_{\rm eff}$
cut value. The leptons peak when they have the same pseudorapidity.
%%%%%%%%%%%%%%%%%%%%%%%%%%%%%%%%%%%%%%%%%%%%%%%%%%%%%%%%%%%%%%%%%%%%%%%%%%%%%%%
\begin{figure}[htb]
%\vskip -0.3in
\begin{center}$
	\begin{array}{cc}
\hspace*{-1.7cm}
	\includegraphics[width=3.8in,height=3.2in]{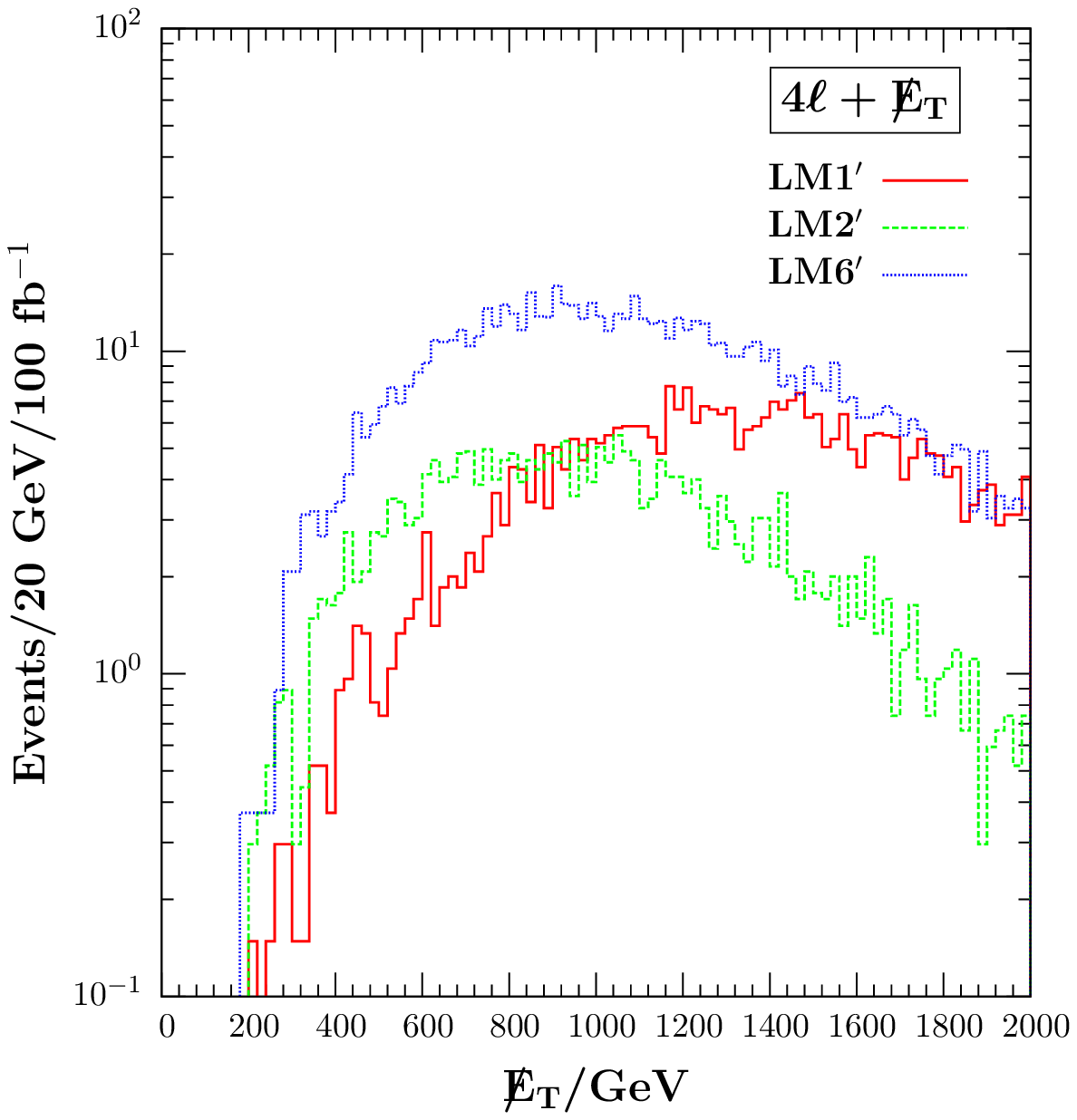}
&\hspace*{-1.5cm}
	\includegraphics[width=3.8in,height=3.2in]{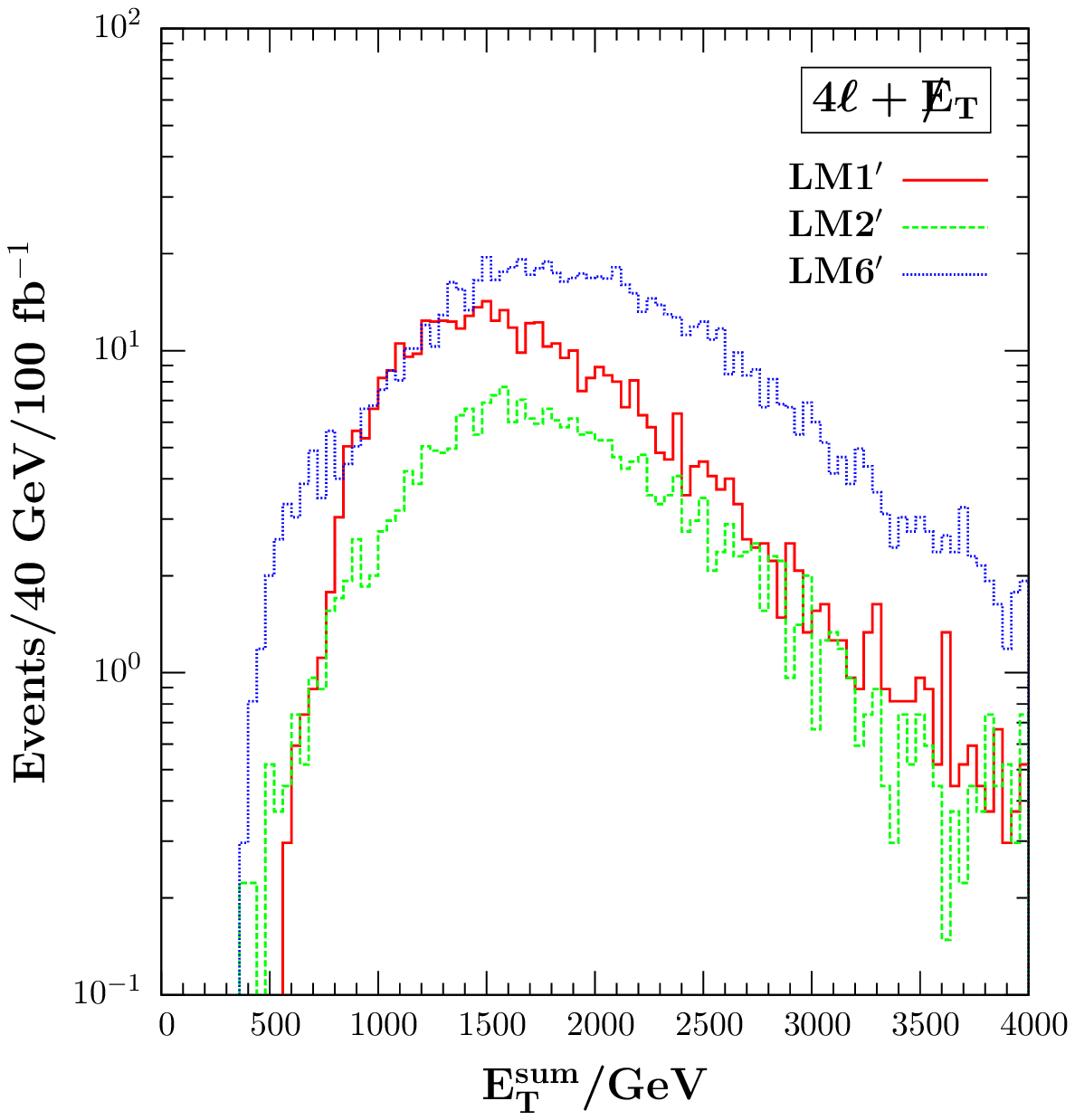} \\
\hspace*{-1.7cm}
	\includegraphics[width=3.8in,height=3.2in]{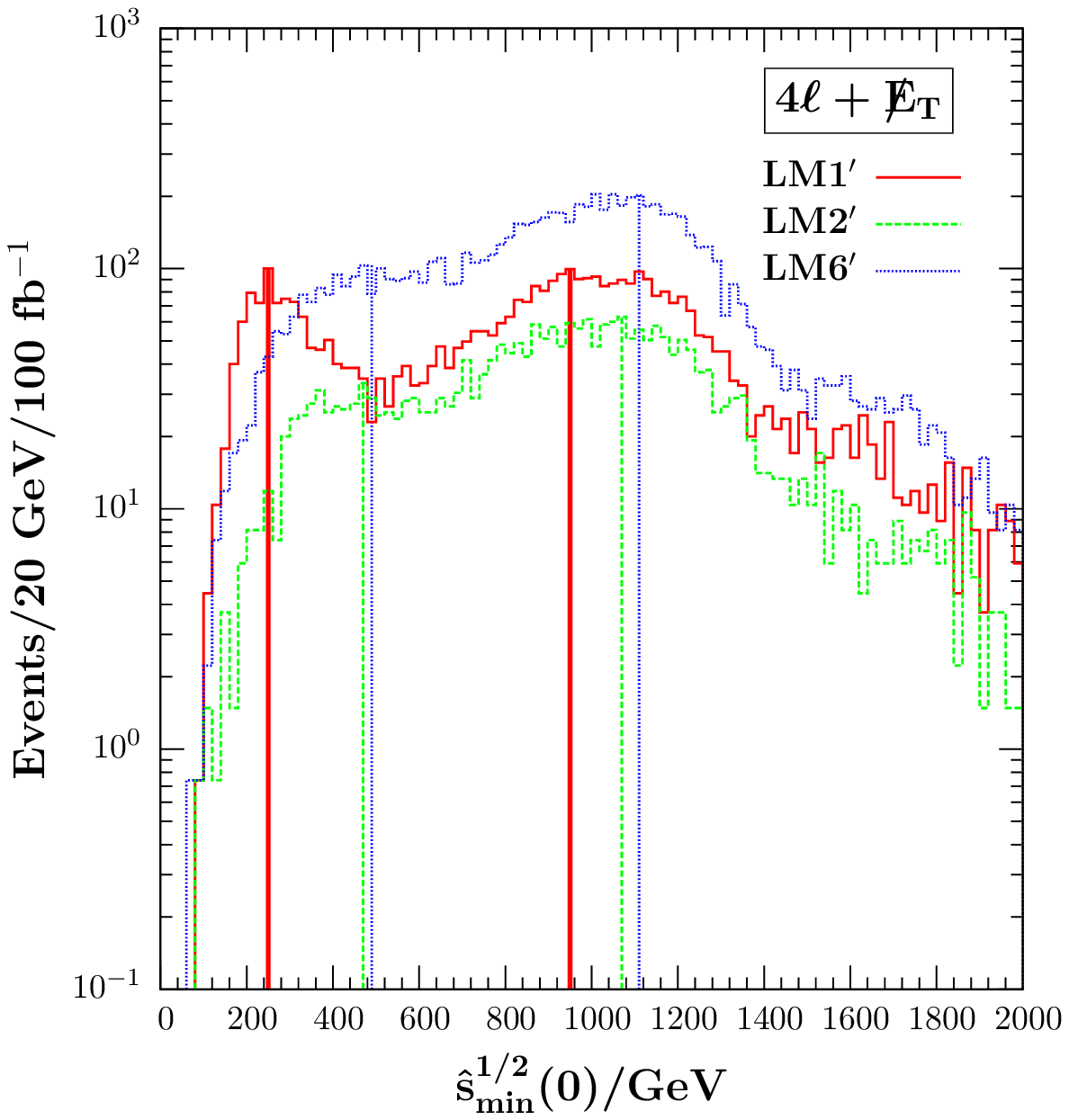}
&\hspace*{-1.5cm}
	\end{array}$
\end{center}
\vskip -0.1in
      \caption{\sl\small The $\EmissT$, $\ETsum$ and $\smin(0)$ distributions
of the tetralepton ($\fourlep$) signal at $14 \tev$ with integrated luminosity ${\cal
L}=100\xfb^{-1}$, for all three scenarios in the secluded $U(1)^\prime$ model.}
\label{fig:4lep_et}
\end{figure}
%%%%%%%%%%%%%%%%%%%%%%%%%%%%%%%%%%%%%%%%%%%%%%%%%%%%%%%%%%%%%%%%%%%%%%%%%%%%%%%

\subsubsection{The Tetralepton Signal: $\fourlep$}
%%%%%%%%%%%%%%%%%%%%%%%%%%%%%%%%%%%%%%%%%%%%%%%%%%%%%%%

As we mentioned earlier, the $\fourlep$ signal is practically background-free.
The $ZZ$ background disappears after the $\EmissT$ cut. Taking into account
having relatively few  $\fourlep$ events, we relaxed the $p_T$ and $\Delta R$
cut values. It is also true that MSSM scenarios LM1, LM2 and LM6 do not yield a
$\fourlep$ type of signal. For the $\lmop$, only $2\%$ of the events pass the
cuts and among them $1\%$ of these are $2e2\mu$, while the rest of the events
are shared between $4e$ and $4\mu$. The situation is different for the $\lmtp$
and $\lmsp$. The events which pass the cuts are around $68\%$ for both cases and
again half of them are the $2e2\mu$ type and the rest is shared equally between
$4e$ and $4\mu$. In fact, there are more $\fourlep$ events in the $\lmop$
scenario as compared to the other two scenarios (about $7.5\%$ of $N_{\rm
tot}=4\times 10^{6}$ for $\lmop$ but only $0.1\%$ and $0.4\%$ for $\lmtp$ and
$\lmsp$, respectively). The reason is that the signal goes through $\niki$
which is the dominant mode for the $\lmop$ but not for the $\lmtp$ or $\lmsp$.
However, the cuts reduce the $\lmop$ events very significantly. Again the reason
is the fact that $\niki$ and $\lsp$ are almost degenerate for $\lmop$, which
leads to very soft leptons.
%%%%%%%%%%%%%%%%%%%%%%%%%%%%%%%%%%%%%%%%%%%%%%%%%%%%%%%%%%%%%%%%%%%%%%%%%%%%%%%
\begin{figure}[htb]
%\vskip -0.3in
\begin{center}$
	\begin{array}{cc}
\hspace*{-1.7cm}
	\includegraphics[width=3.8in,height=3.2in]{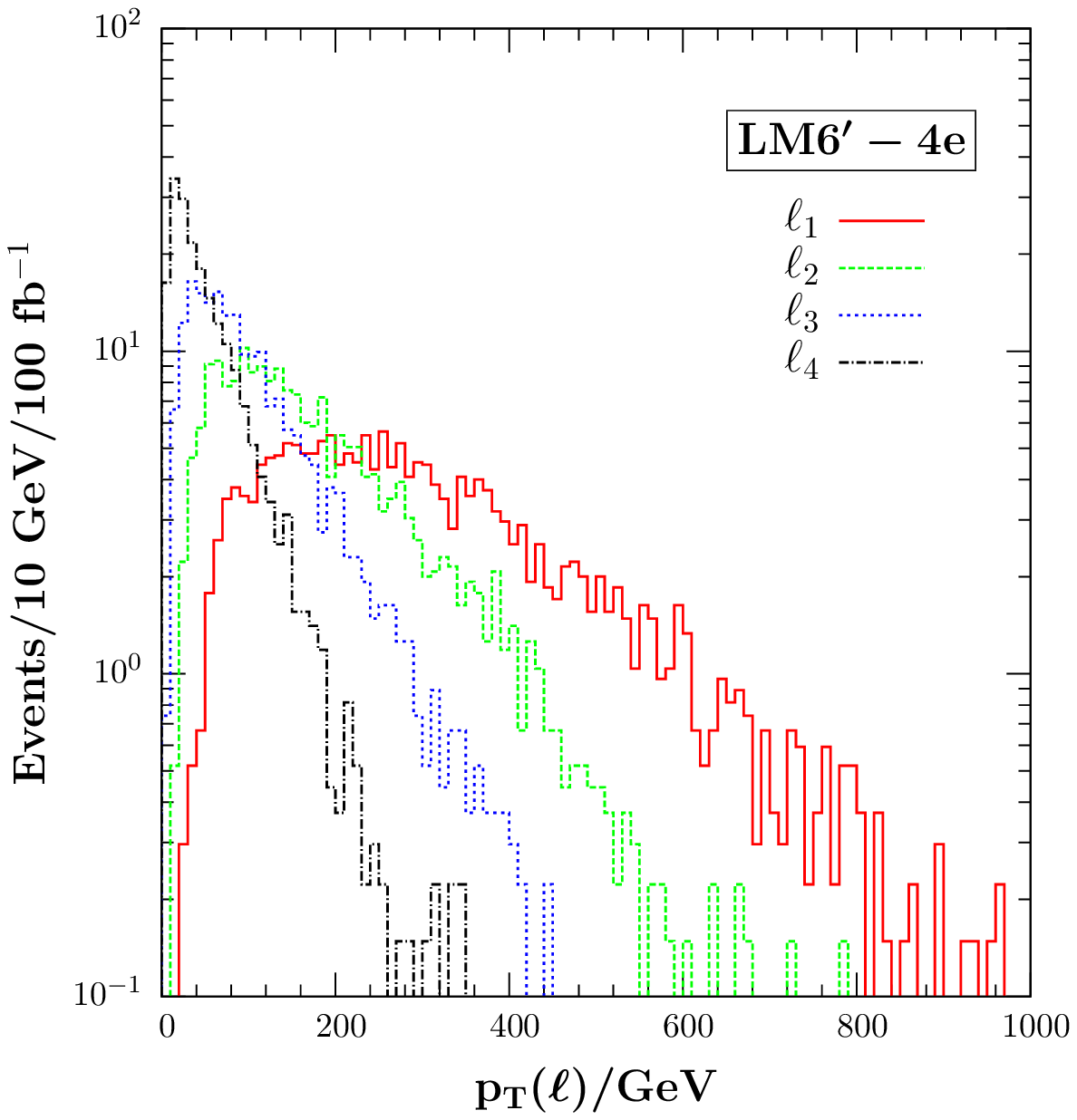}
&\hspace*{-1.5cm}
	\includegraphics[width=3.8in,height=3.2in]{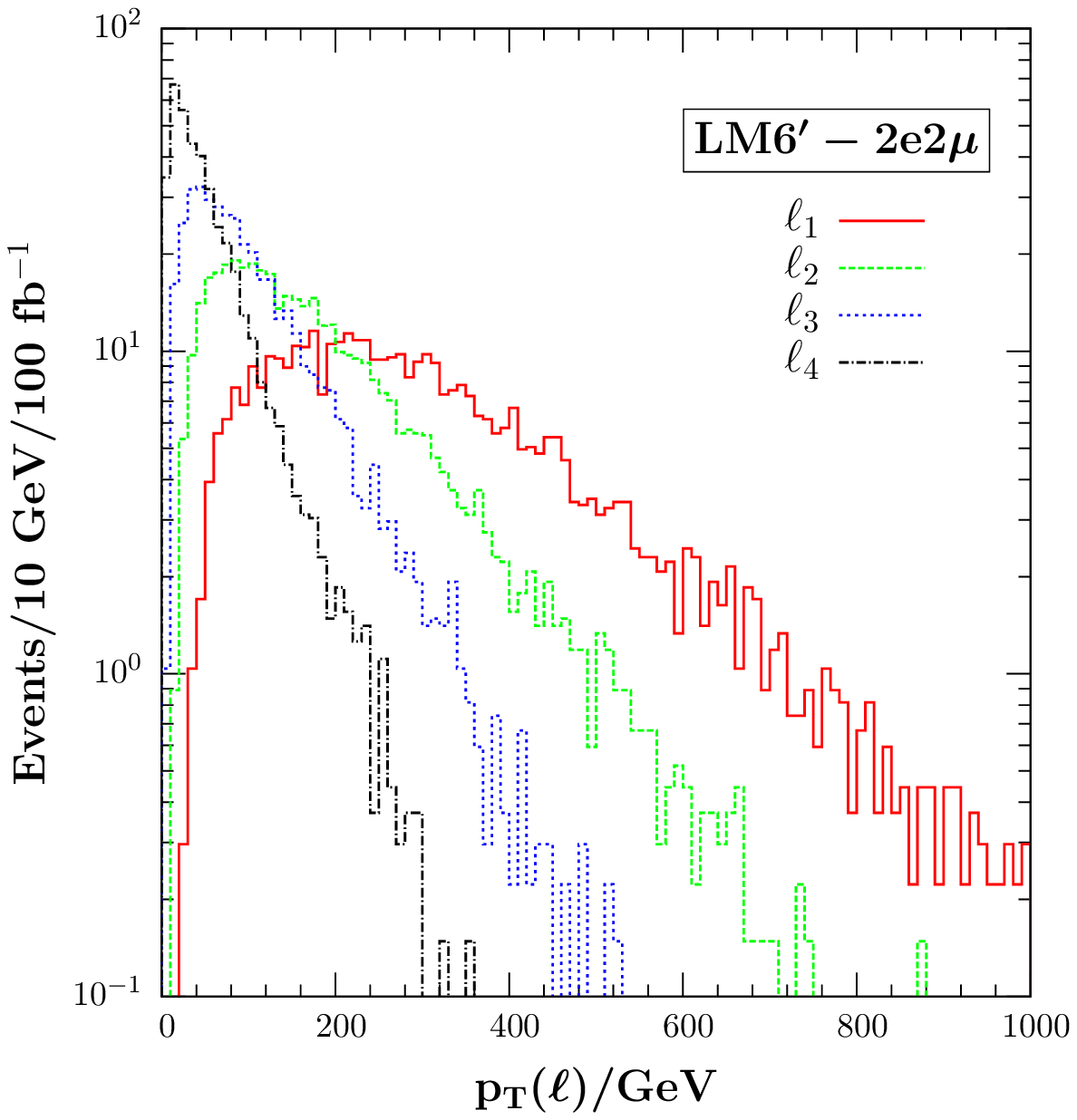}
	\end{array}$
\end{center}
\vskip -0.1in
      \caption{\sl\small The $p_T$ distribution of the $4e+\EmissT$ and
$\tetmulep$ signals at  $14 \tev$ with integrated
luminosity ${\cal L}=100\xfb^{-1}$ for all three scenarios in the secluded
$U(1)^\prime$ model. The $4\mu+\EmissT$ case is similar. Also the hardness of
the leptons are in decreasing order.}
\label{fig:4lep_pt}
\end{figure}

%%%%%%%%%%%%%%%%%%%%%%%%%%%%%%%%%%%%%%%%%%%%%%%%%%%%%%%%%%%%%%%%%%%%%%%%%%%%%%%
In Fig.~\ref{fig:4lep_et}, $\EmissT$, $\ETsum$, and $\smin(0)$ distributions
of the $\fourlep$ signal at $14 \tev$ with integrated luminosity ${\cal
L}=100\xfb^{-1}$ are shown for the three scenarios in the secluded $U(1)^\prime$
model. As promised, we include a $\smin(0)$ graph with the peak correlated
with the $\nL\nL^*$ production as well as the $\nR\nR^*$ production (no
$m_{\rm eff}$ cut). We can roughly tell the positions of the peak without doing
a serious fitting. For the $\lmop$, the first peak is around $250\gev$ and the
second one is around $950\gev$. For the $\lmtp$, they are at
$(470\gev,1070\gev)$ for the first and the second peaks, respectively. For the
$\lmsp$, the peak positions are close to the $\lmtp$ case, i.e., they are
at $(490\gev,1110\gev)$. Then we can estimate the masses for $\lmop/\lmtp/\lmsp$
\begin{itemize}
 \item $\left(m_{\nL},m_{\nR}\right)_{\rm est.}\approx\;\;
(125\gev,475\gev)\;\;/\;\;(235\gev,535\gev)\;\;/\;\;(245\gev,555\gev)$
\end{itemize}
while the theoretical average values obtained, including three flavors
\begin{itemize}
 \item $\left(m_{\nL},m_{\nR}\right)_{\rm theo.}\approx\;\;
(132\gev,460\gev)\;\;/\;\;(230\gev,563\gev)\;\;/\;\;(258\gev,654\gev)$
\end{itemize}
We suspect that the deviations are mainly responsible for not determining the
peak position after fitting the data to a curve. Also, simple averaging is not
quite right. One should include a relative weight based on the relative
contributions from different flavor channels.
%%%%%%%%%%%%%%%%%%%%%%%%%%%%%%%%%%%%%%%%%%%%%%%%%%%%%%%%%%%%%%%%%%%%%%%%%%%%%%%
\begin{figure}[htb]
%\vskip -0.3in
\begin{center}$
	\begin{array}{cc}
\hspace*{-1.7cm}
	\includegraphics[width=3.8in,height=3.2in]{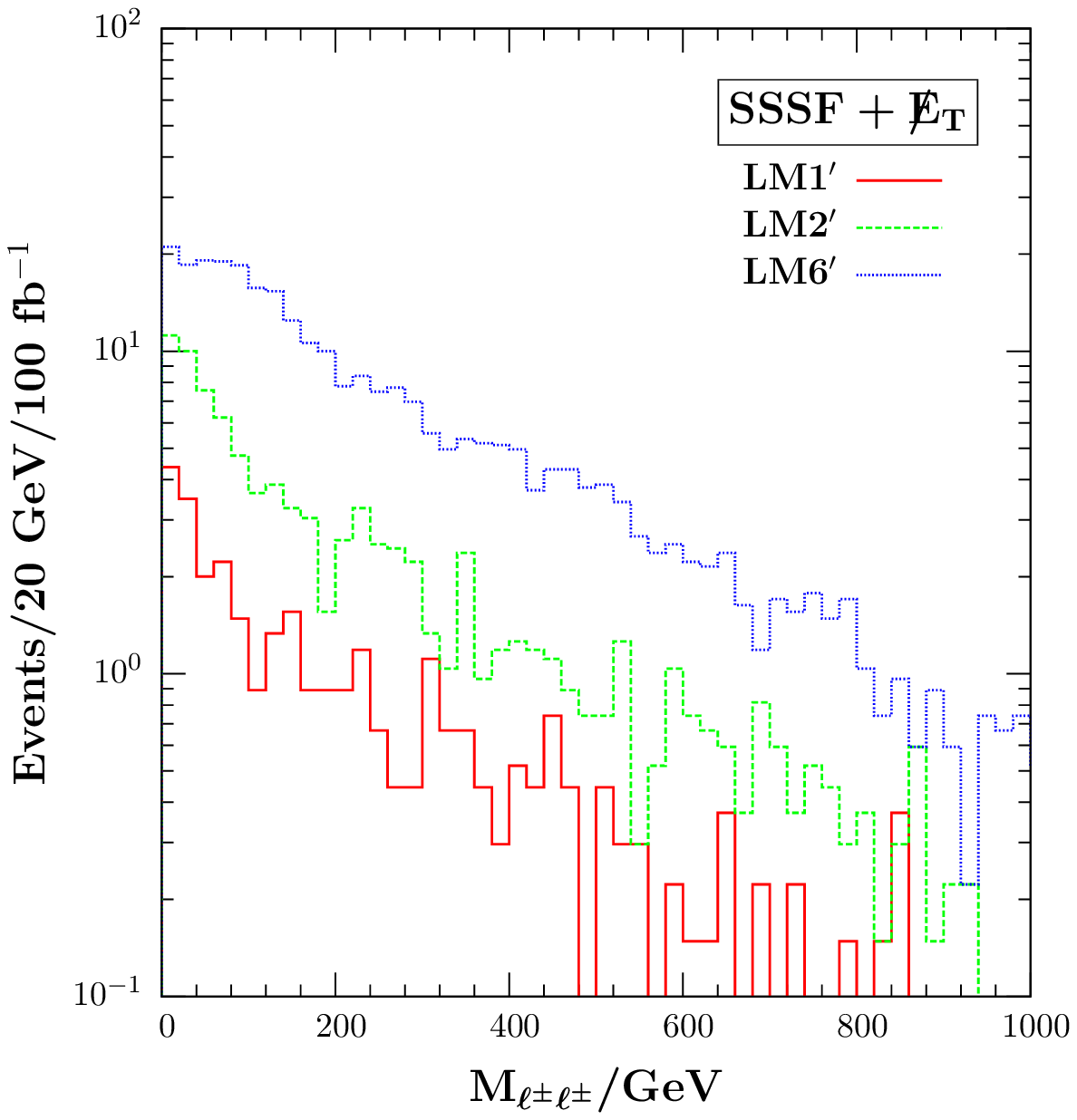}
&\hspace*{-1.5cm}
	\includegraphics[width=3.8in,height=3.2in]{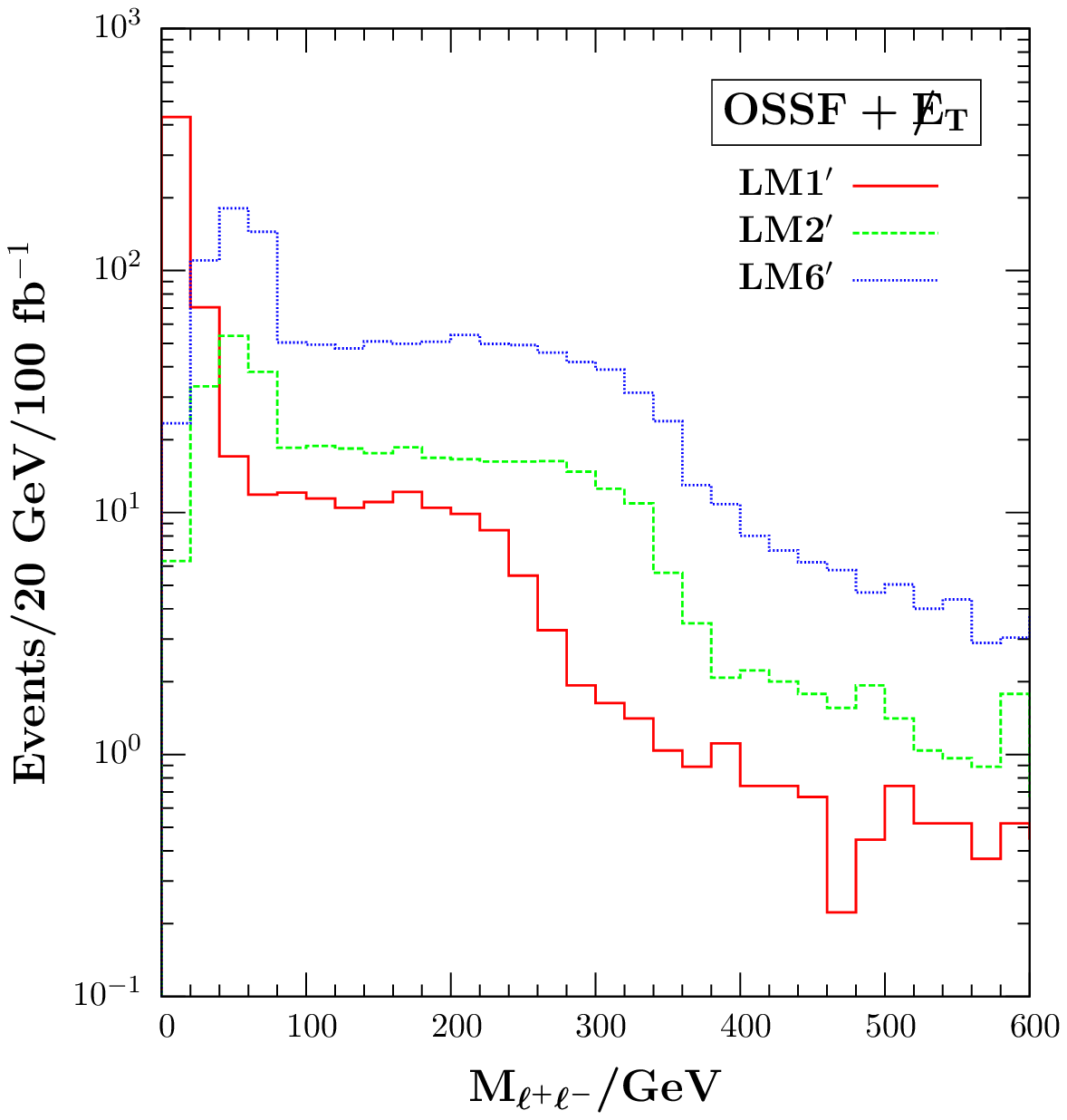}\\
\hspace*{-1.7cm}
	\includegraphics[width=3.8in,height=3.2in]{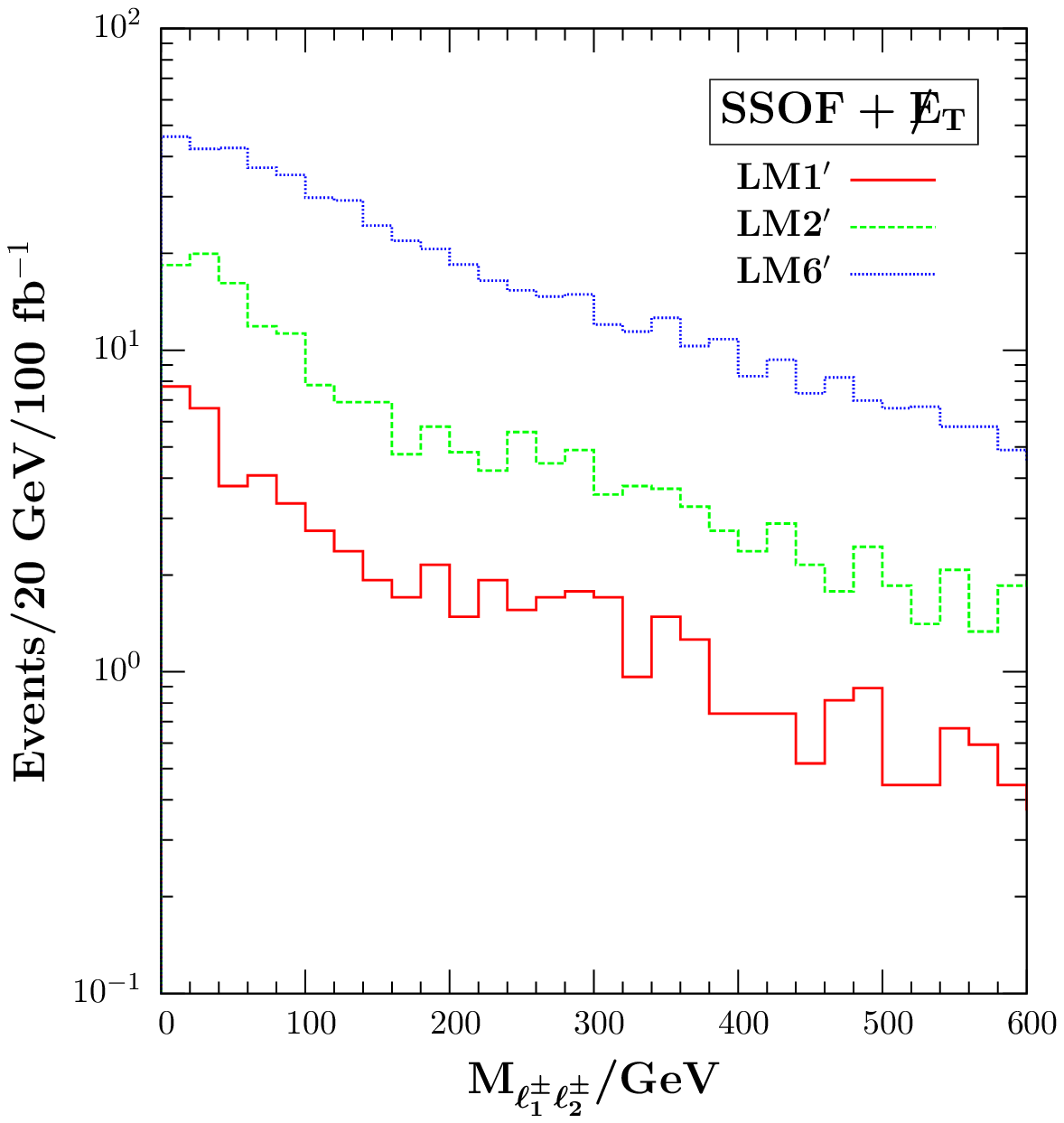}
&\hspace*{-1.5cm}
	\includegraphics[width=3.8in,height=3.2in]{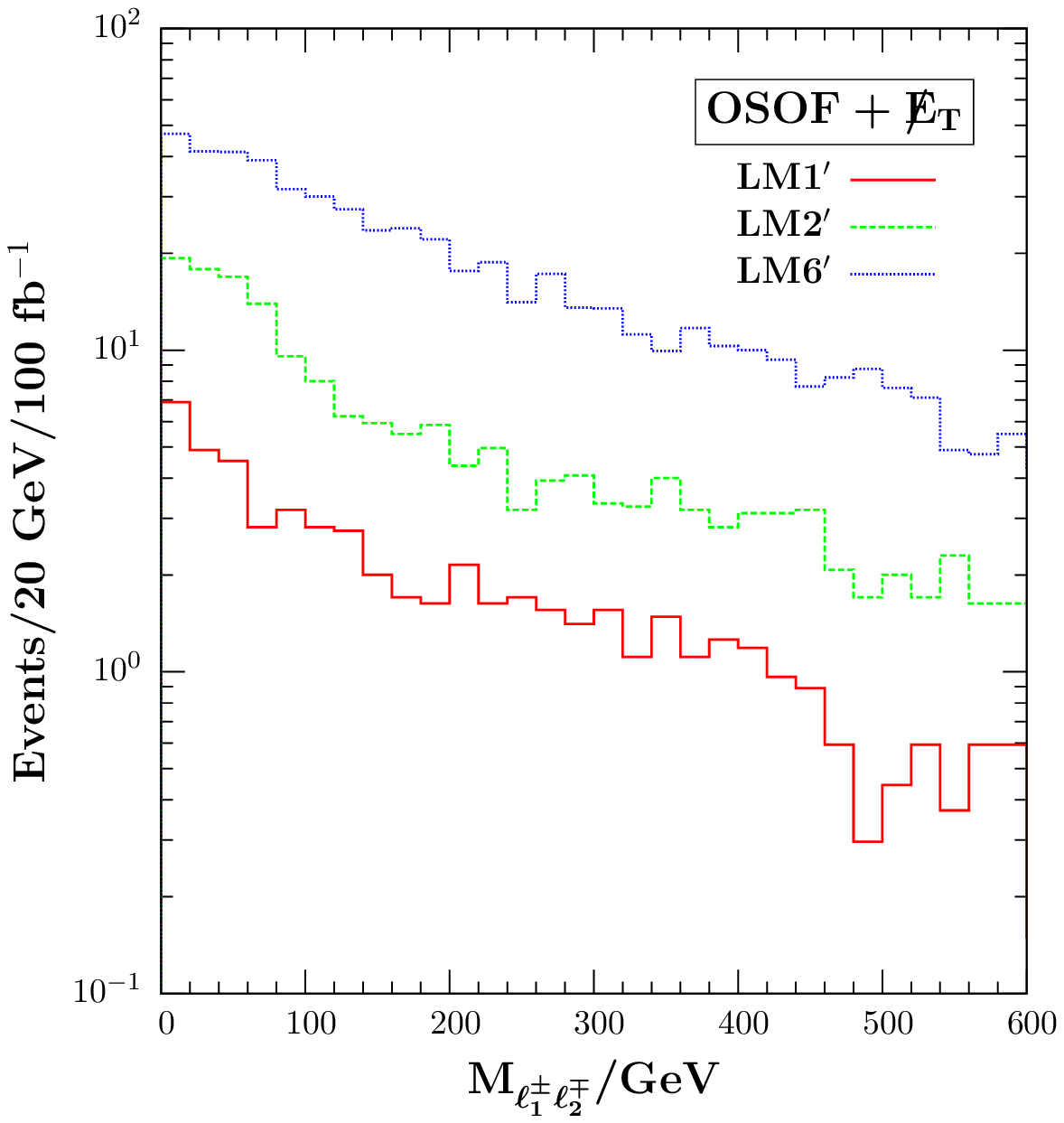}
	\end{array}$
\end{center}
\vskip -0.1in
      \caption{\sl\small Various invariant mass distributions of the $\fourlep$
signal at  $14 \tev$ with integrated luminosity ${\cal
L}=100\xfb^{-1}$ for all three scenarios in the secluded $U(1)^\prime$ model.}
\label{fig:4lep_minv}
\end{figure}
%%%%%%%%%%%%%%%%%%%%%%%%%%%%%%%%%%%%%%%%%%%%%%%%%%%%%%%%%%%%%%%%%%%%%%%%%%%%%%%
%%%%%%%%%%%%%%%%%%%%%%%%%%%%%%%%%%%%%%%%%%%%%%%%%%%%%%%%%%%%%%%%%%%%%%%%%%%%%%%
\begin{figure}[htb]
%\vskip -0.3in
\begin{center}$
	\begin{array}{cc}
\hspace*{-1.7cm}
	\includegraphics[width=3.8in,height=3.2in]{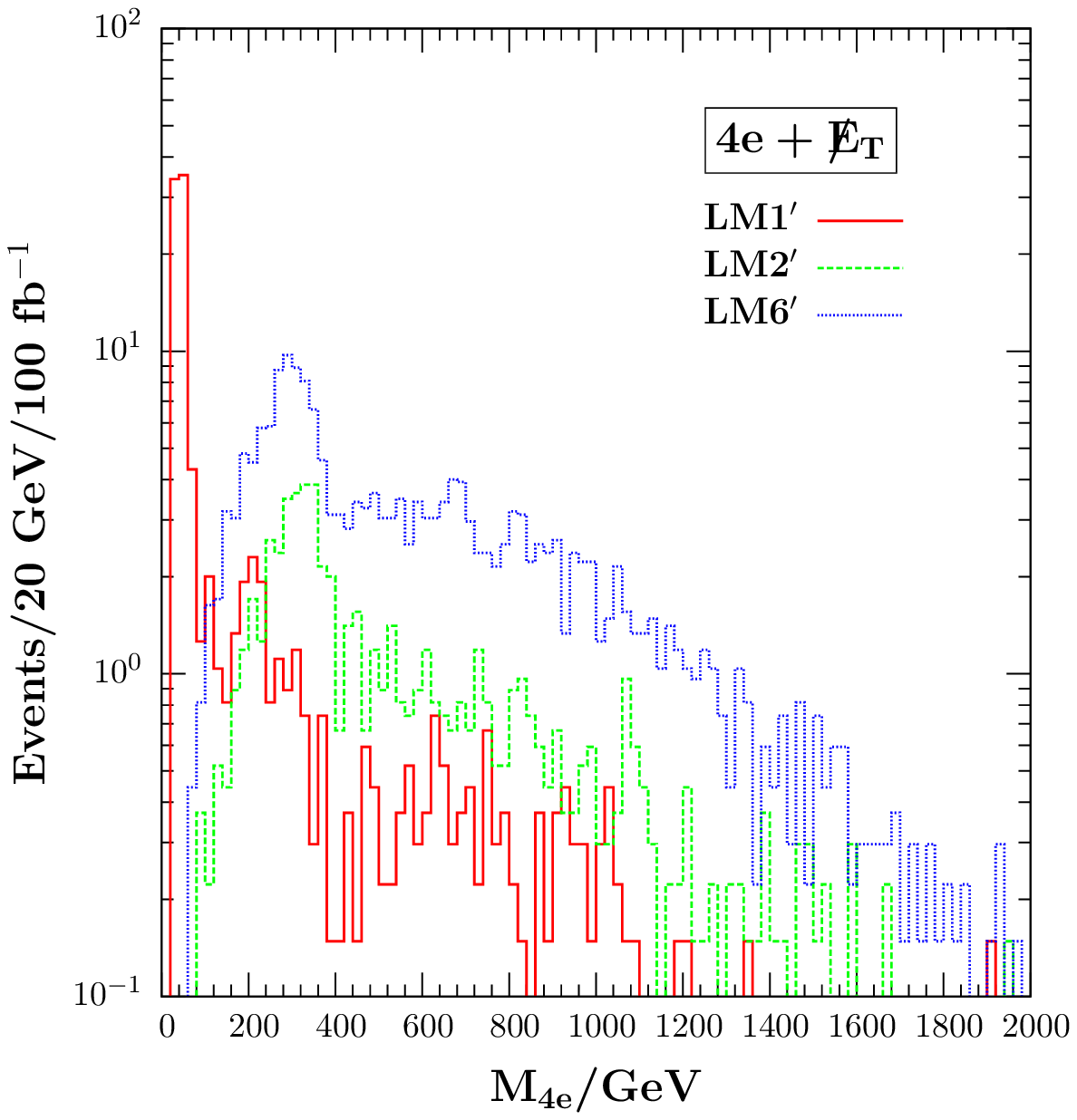}
&\hspace*{-1.5cm}
	\includegraphics[width=3.8in,height=3.2in]{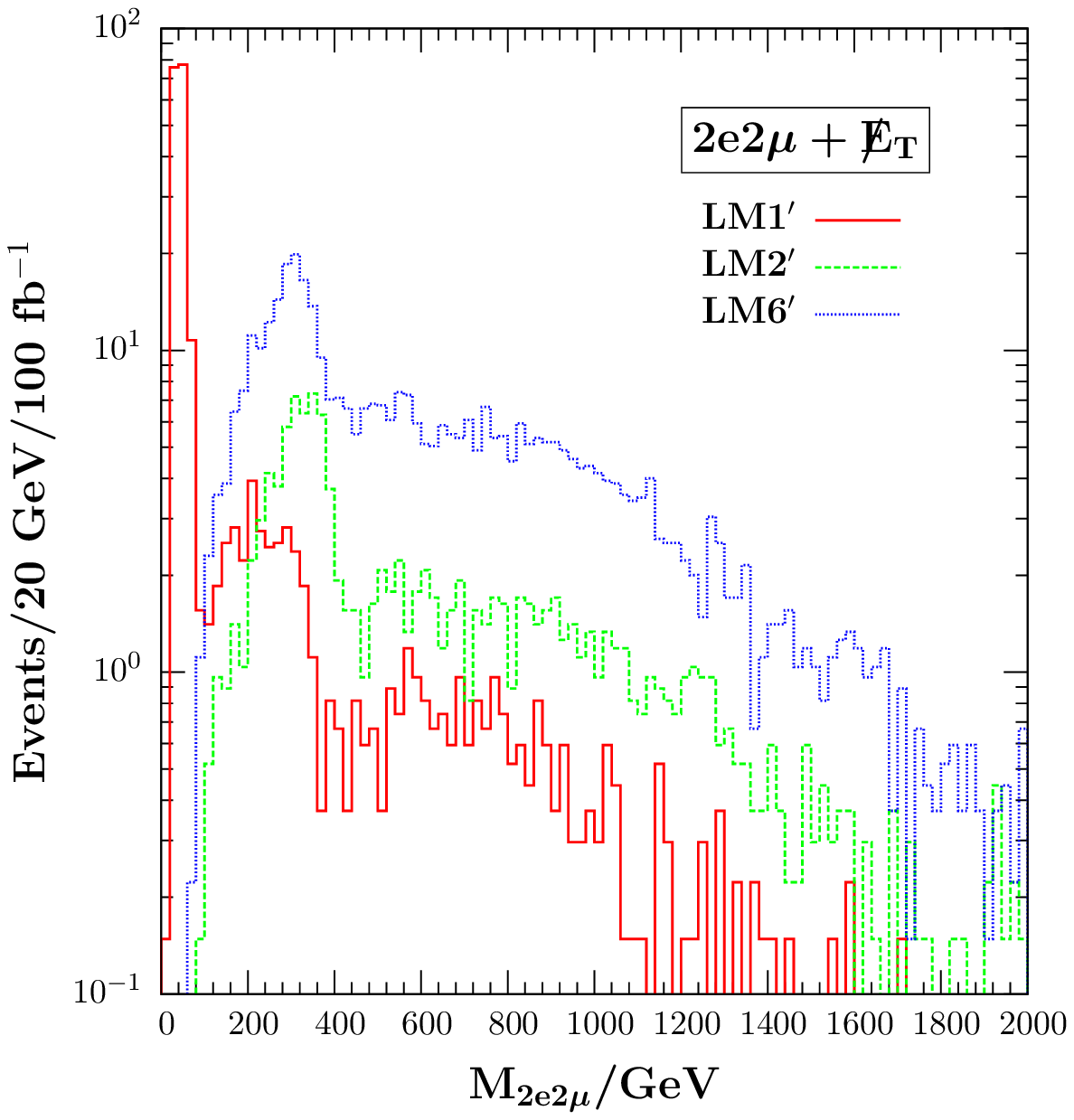}
	\end{array}$
\end{center}
\vskip -0.1in
      \caption{\sl\small Four lepton invariant mass distributions of 
$\fourlep,\ell=e,\mu$ and $\tetmulep$ signals at  $14 \tev$ with integrated
luminosity ${\cal L}=100\xfb^{-1}$ for all three scenarios in the secluded
$U(1)^\prime$ model.}
\label{fig:4lep_minv4l}
\end{figure}
%%%%%%%%%%%%%%%%%%%%%%%%%%%%%%%%%%%%%%%%%%%%%%%%%%%%%%%%%%%%%%%%%%%%%%%%%%%%%%%
%%%%%%%%%%%%%%%%%%%%%%%%%%%%%%%%%%%%%%%%%%%%%%%%%%%%%%%%%%%%%%%%%%%%%%%%%%%%%%%
\begin{figure}[htb]
%\vskip -0.3in
\begin{center}$
	\begin{array}{cc}
\hspace*{-1.7cm}
	\includegraphics[width=3.8in,height=3.2in]{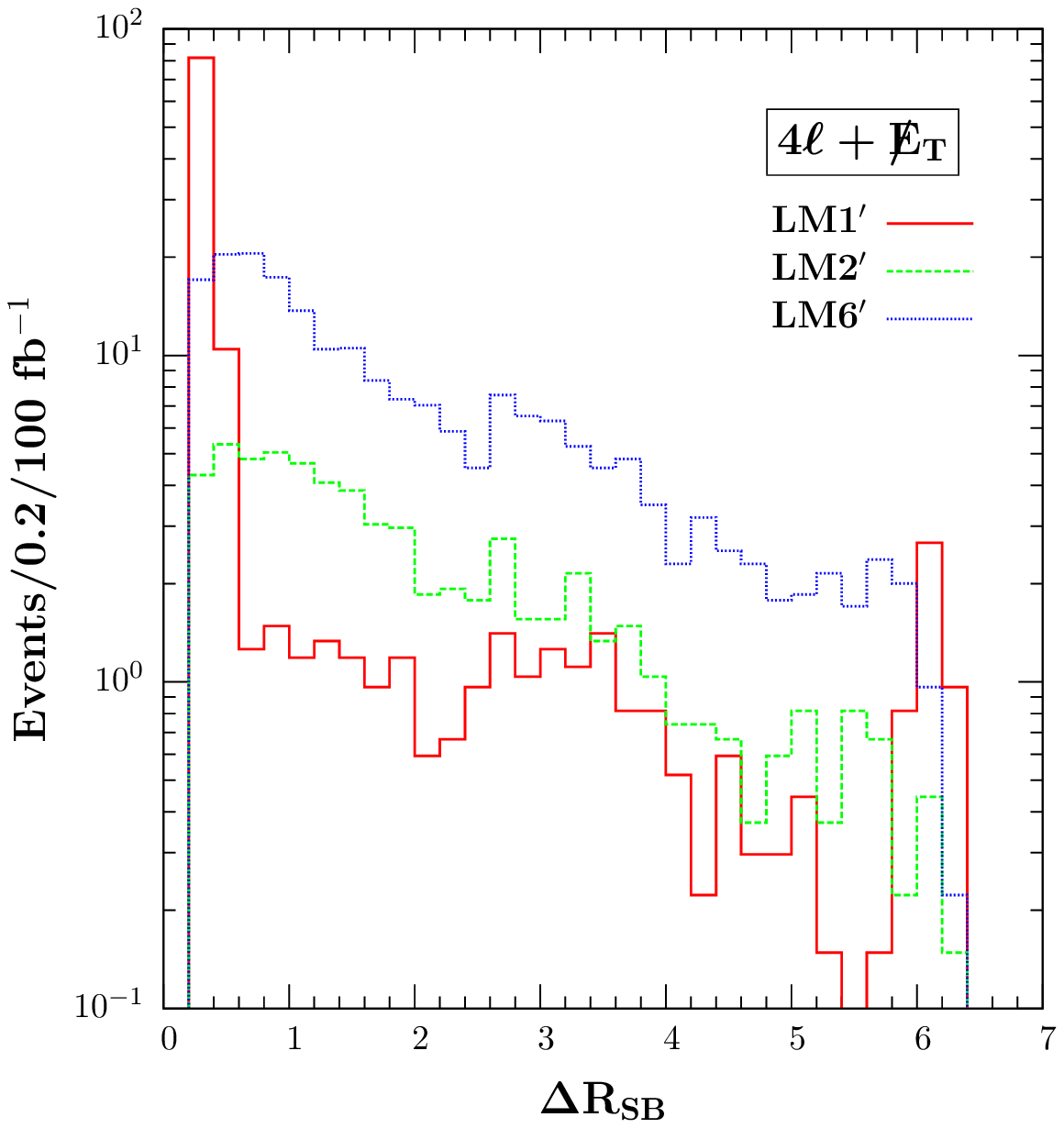}
&\hspace*{-1.5cm}
	\includegraphics[width=3.8in,height=3.2in]{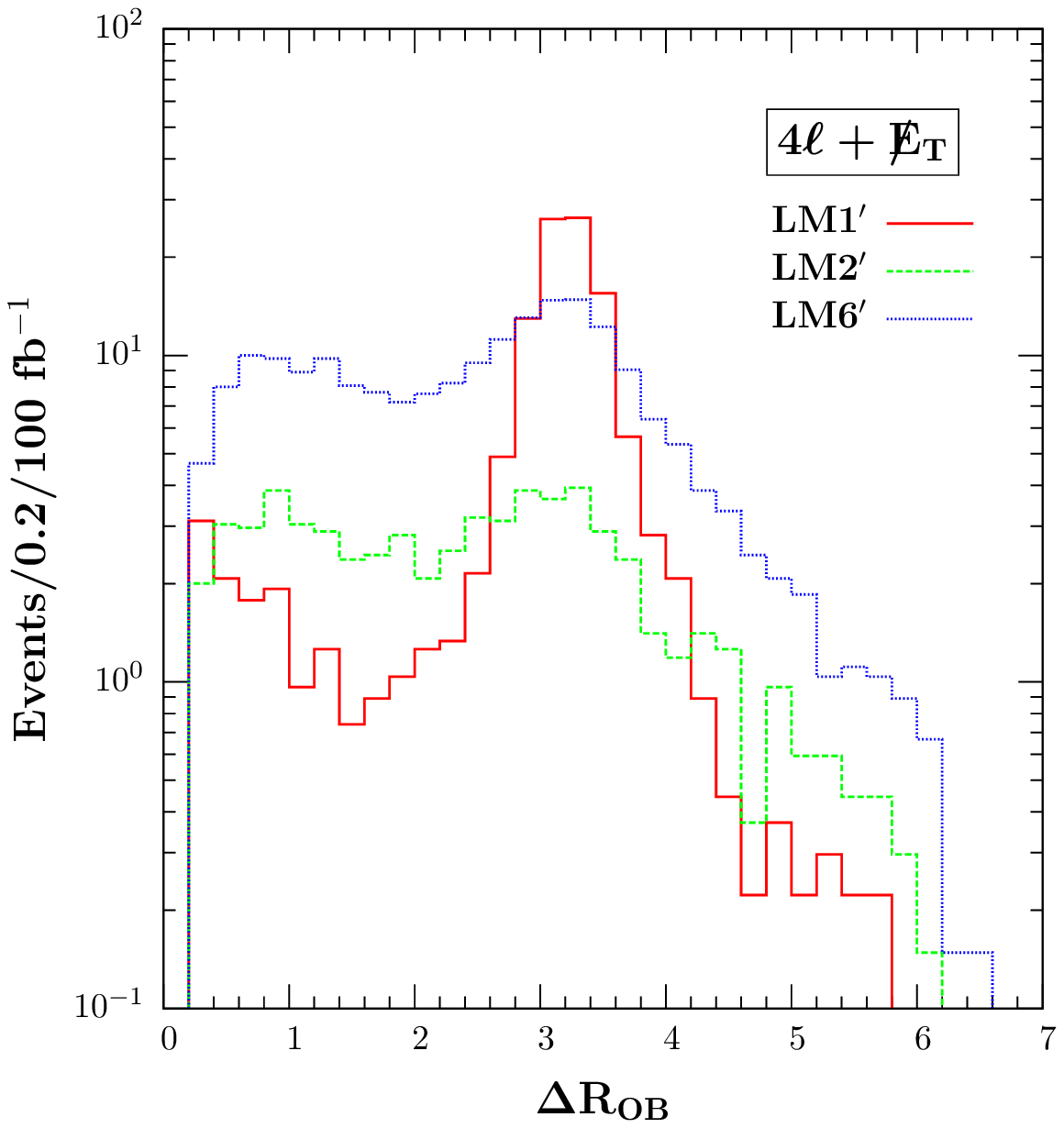}\\
\hspace*{-1.7cm}
	\includegraphics[width=3.8in,height=3.2in]{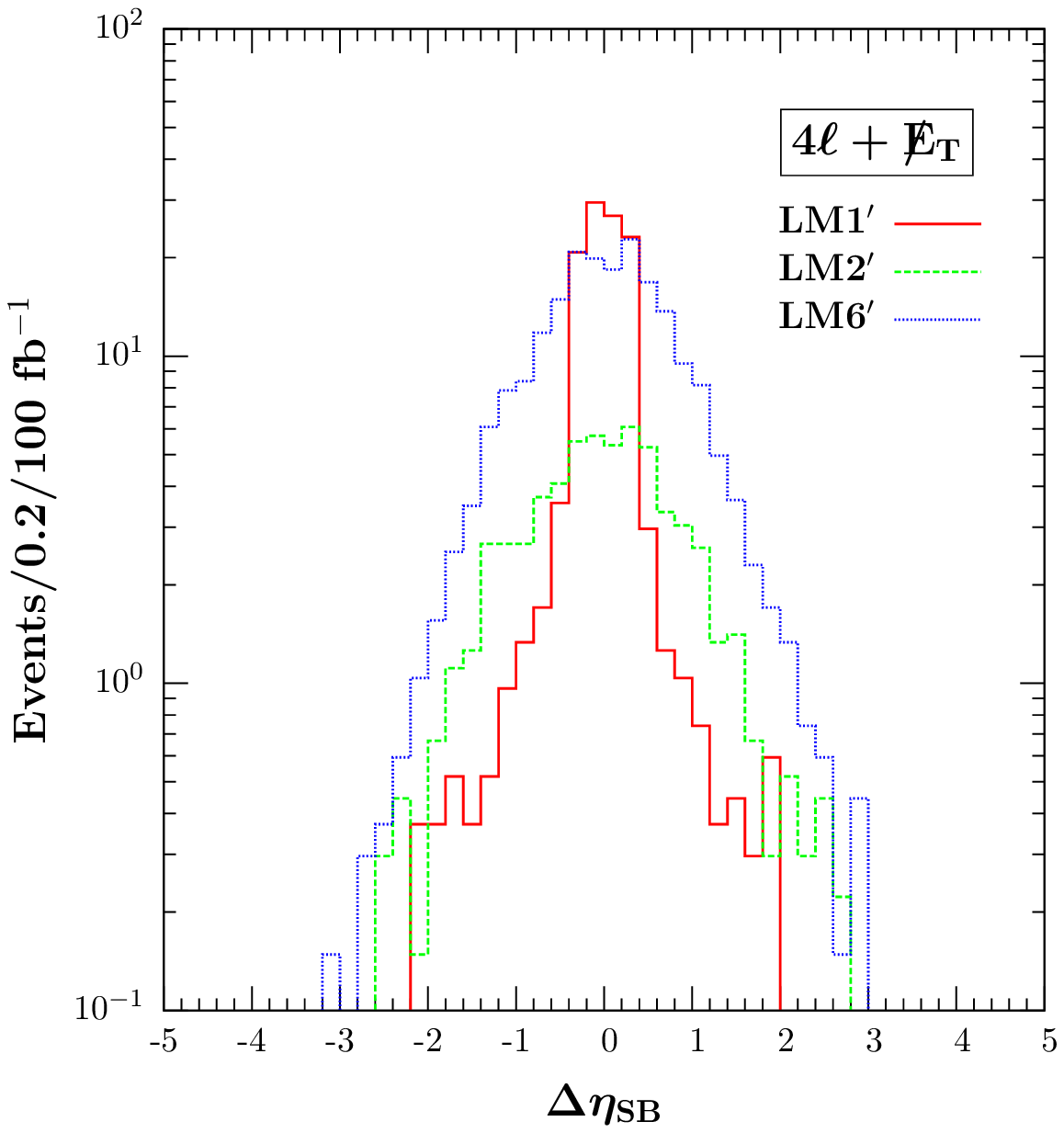}
&\hspace*{-1.5cm}
	\includegraphics[width=3.8in,height=3.2in]{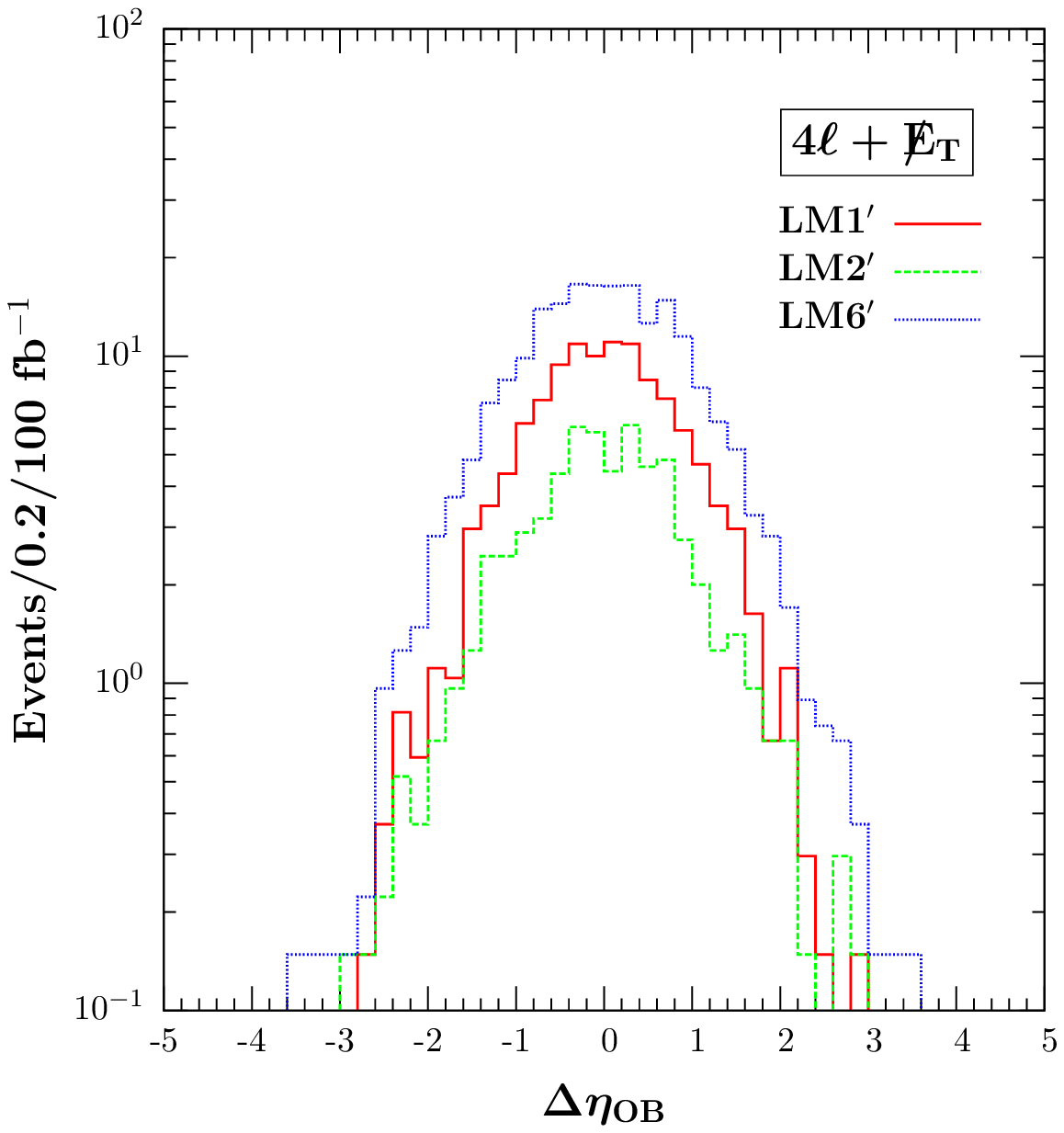}
	\end{array}$
\end{center}
\vskip -0.1in
      \caption{\sl\small The $\Delta R_{\ell^+\ell^-}$ and $\Delta
\eta_{\ell^+\ell^-}$ distributions of the $\fourlep$ signal at $14 \tev$ with
integrated luminosity ${\cal L}=100\xfb^{-1}$ for all three scenarios in the
secluded $U(1)^\prime$ model. Here and in what follows, `SB' is short-hand for `Same Branch' and `OB' for
'Opposite Branch'.}
 \label{fig:4lep_4m}
\end{figure}
%%%%%%%%%%%%%%%%%%%%%%%%%%%%%%%%%%%%%%%%%%%%%%%%%%%%%%%%%%%%%%%%%%%%%%%%%%%%%%%
%%%%%%%%%%%%%%%%%%%%%%%%%%%%%%%%%%%%%%%%%%%%%%%%%%%%%%%%%%%%%%%%%%%%%%%%%%%%%%%
\begin{figure}[htb]
%\vskip -0.3in
\begin{center}$
	\begin{array}{cc}
\hspace*{-1.7cm}
	\includegraphics[width=3.8in,height=3.2in]{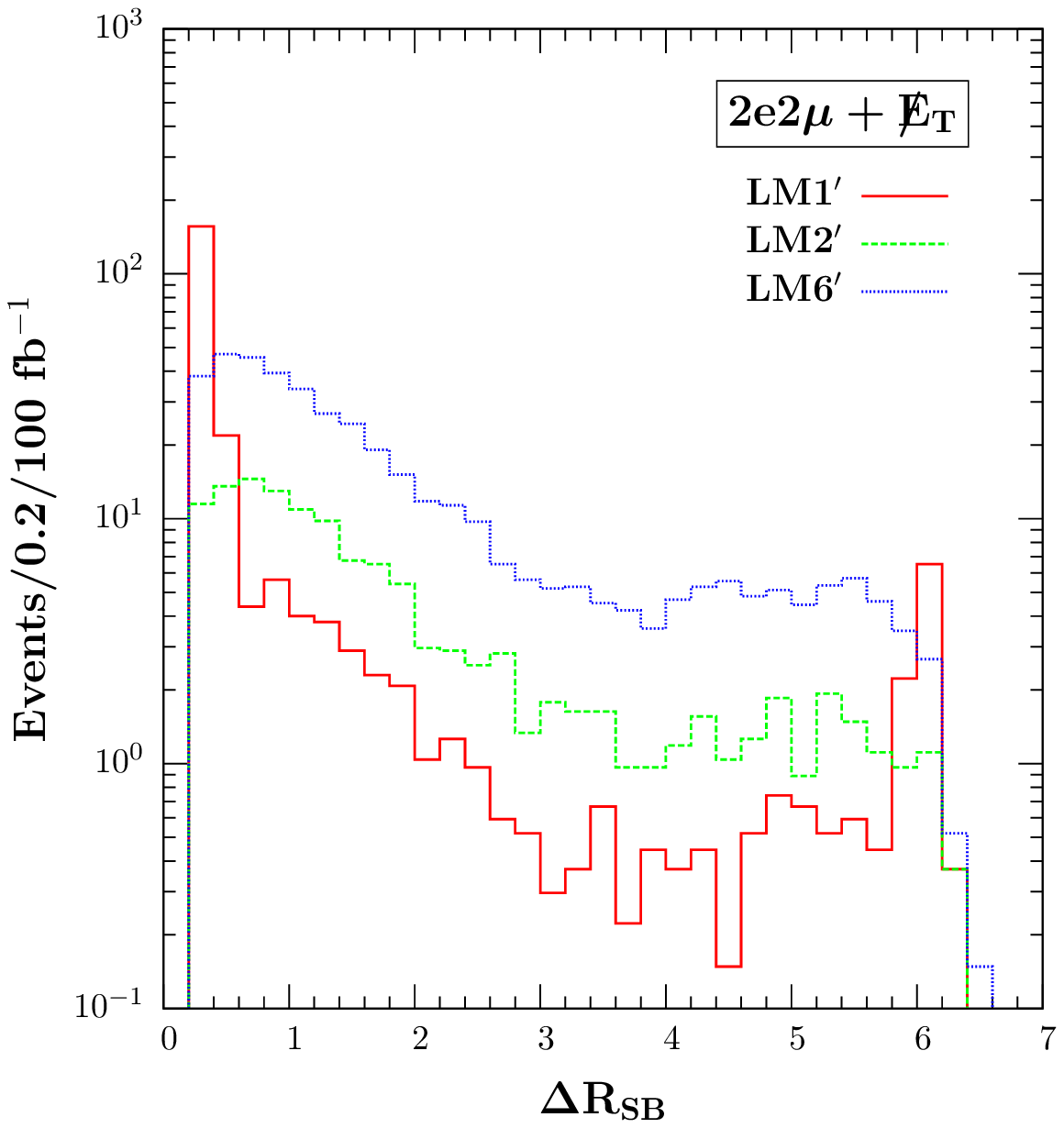}
&\hspace*{-1.5cm}
	\includegraphics[width=3.8in,height=3.2in]{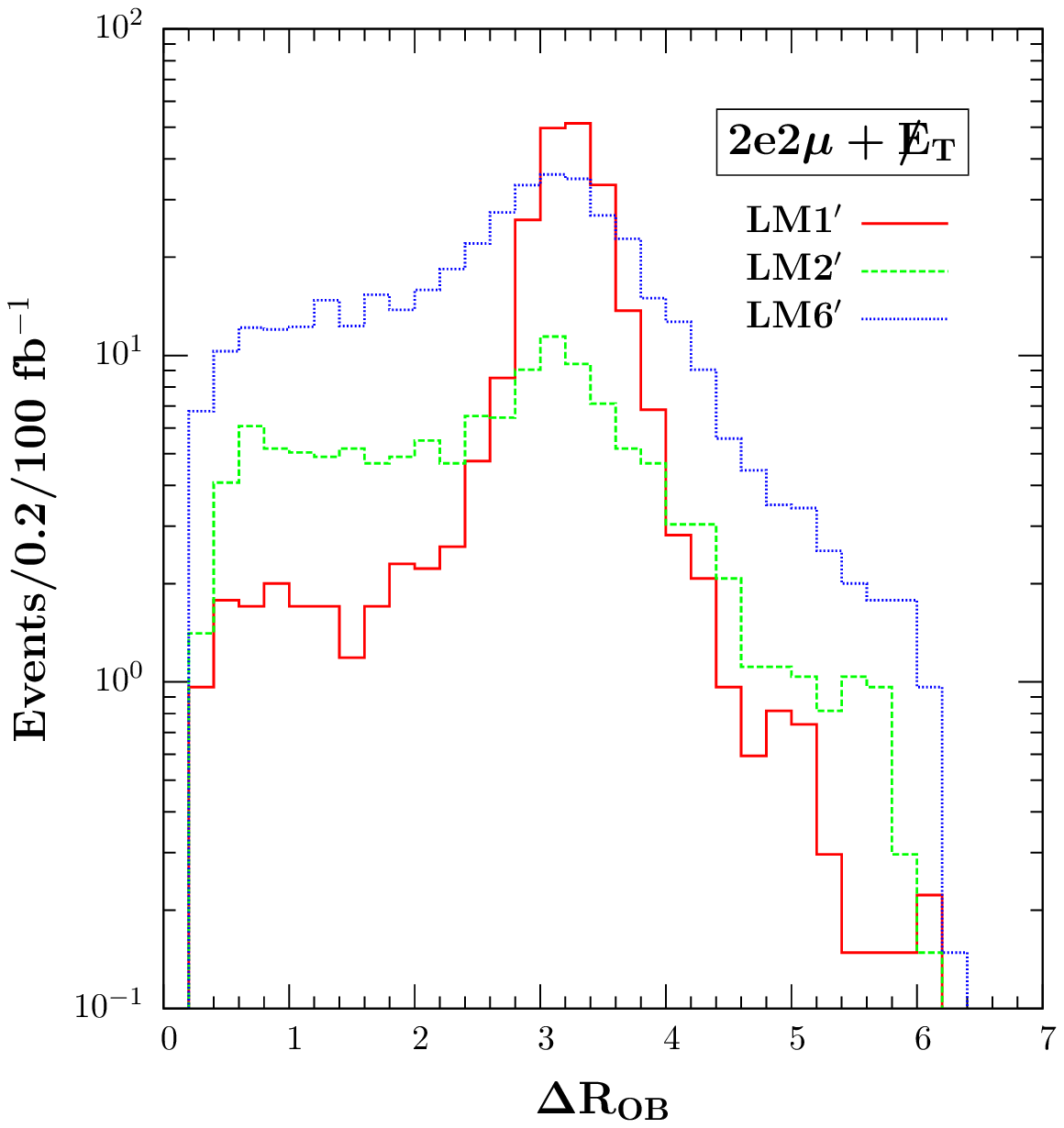}\\
\hspace*{-1.7cm}
	\includegraphics[width=3.8in,height=3.2in]{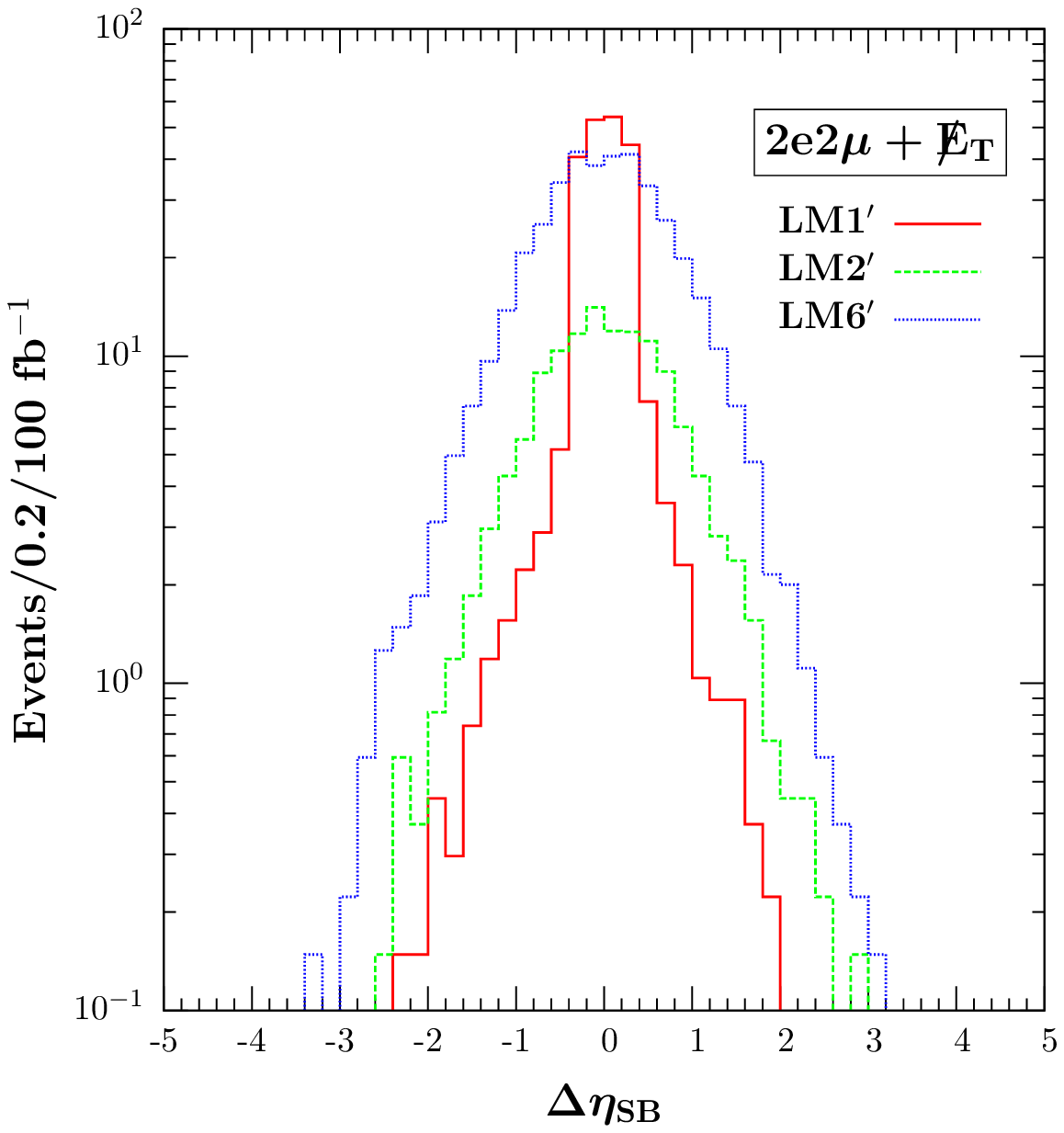}
&\hspace*{-1.5cm}
	\includegraphics[width=3.8in,height=3.2in]{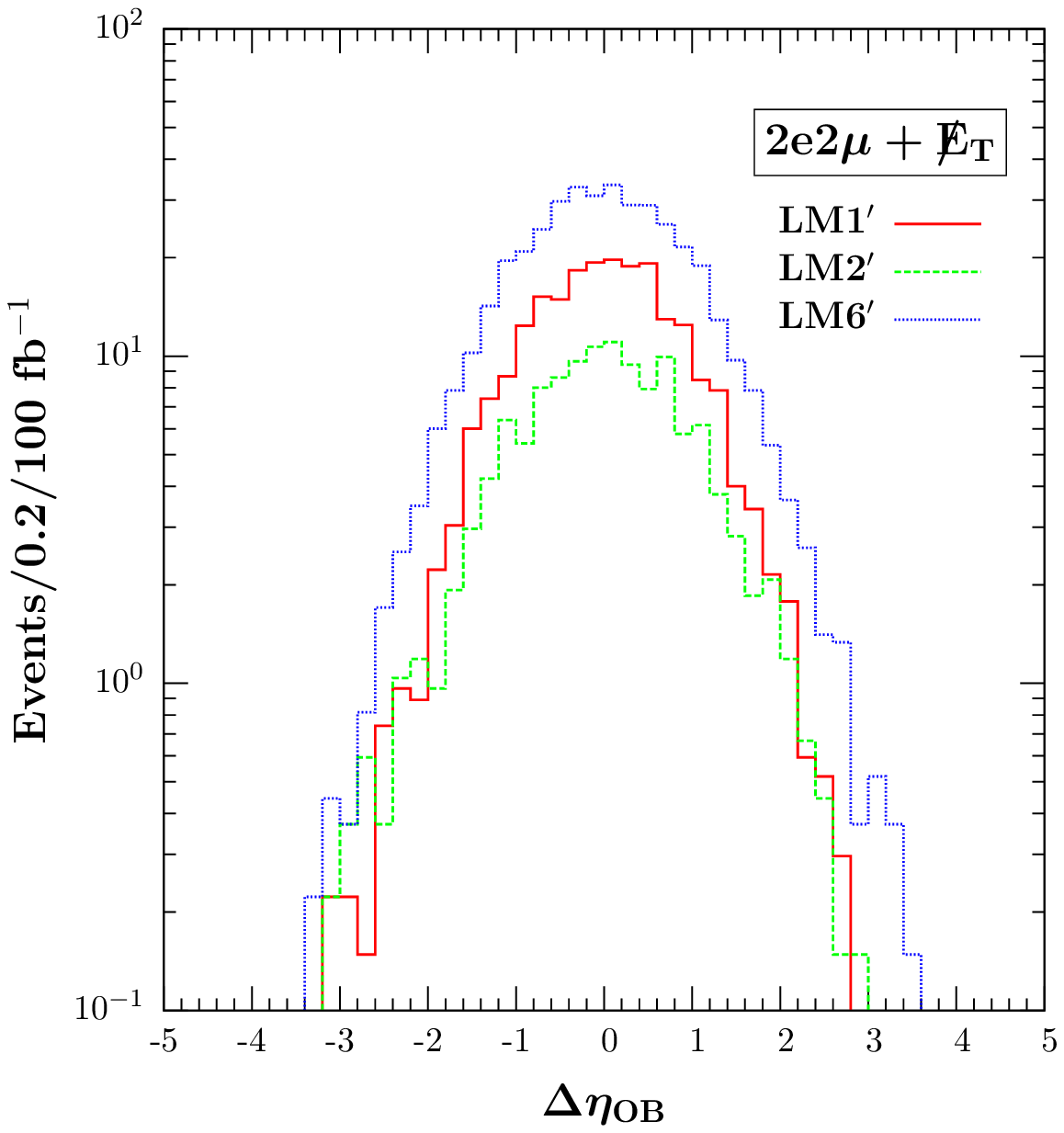}
	\end{array}$
\end{center}
\vskip -0.1in
      \caption{\sl\small The $\Delta R_{\ell^+\ell^-}$ and $\Delta
\eta_{\ell^+\ell^-}$ distributions of the $\tetmulep$ signal at $14 \tev$ with
integrated luminosity ${\cal L}=100\xfb^{-1}$ for all three scenarios in the
secluded $U(1)^\prime$ model.}
\label{fig:4lep_2em}
\end{figure}
%%%%%%%%%%%%%%%%%%%%%%%%%%%%%%%%%%%%%%%%%%%%%%%%%%%%%%%%%%%%%%%%%%%%%%%%%%%%%%%

The $p_T$ distributions of the $4e+\EmissT$ and $2e2\mu+\EmissT$ for $\lmsp$ are
given in Fig.~\ref{fig:4lep_pt}. The $4\mu$ case is very similar to $4e$. The
leptons seem slightly more energetic for the $2e2\mu$ case than in the other
cases. $\lmop$ and $\lmtp$ have less energetic leptons and we do not include
them here. Fig.~\ref{fig:4lep_minv} displays two-lepton invariant mass
distributions for various possibilities. As expected only Opposite Sign Same
Flavor (OSSF) distributions have peaks at the expected locations since both
leptons originate from the same parent unlike the other cases, Same Sign Same
Flavor (SSSF), Same Sign Opposite Flavor (SSOF) or Opposite Sign Opposite Flavor
(OSOF). The next figure, Fig.~\ref{fig:4lep_minv4l}, has four-lepton invariant
mass distributions for $4e$ and $2e2\mu$ cases. The last two figures,
Fig.~\ref{fig:4lep_4m} and Fig.~\ref{fig:4lep_2em}, are devoted to the  $\Delta
R_{\rm SB(OB)}$ and $\Delta \eta_{\rm SB(OB)}$ distributions of the
$4\ell+\EmissT$ and $\tetmulep$ signals. The subscript `SB(OB)' stands for the
Same Branch (Opposite Branch) and indicates where the leptons are coming from.
We see that the distributions are very similar for $4\ell$ and $2e2\mu$. If we
compare $\Delta R_{\rm SB}$ and $\Delta R_{\rm OB}$, the former peaks at small
$\Delta R$ while the latter peaks larger distances.
%This shouldn't come as surprise since when the leptons come from OB, they fall
% apart then when they are from the SB.
For the pseudorapidity, even though the shape of the distributions changes, they
both peak when the leptons have the same pseudorapidity.

%%%%%%%%%%%%%%%%%%%%%%%%%%%%%%%%%%%%%%%%%%%%%%%%%%%%%%%
\section{Conclusion \label{sec:conc}}
%%%%%%%%%%%%%%%%%%%%%%%%%%%%%%%%%%%%%%%%%%%%%%%%%%%%%%%

 We presented a thorough and complete analysis of the scalar neutrino production
and decays in a $U(1)^{\prime}$ model endowed with a secluded sector. This model
has several attractive features as compared to the MSSM. First, it extends the
gauge symmetry to include an extra neutral gauge boson, allowing for the
presence of right handed neutrinos. Neutrinos are Dirac particles in this model,
and masses are provided through an effective neutrino Yukawa coupling which is
naturally suppressed by the $U(1)^{\prime}$ invariance. This model generates the
$\mu$ term dynamically, through the VEV of a singlet scalar field. The secluded
sector consists of three chiral superfields in addition to $\widehat{S}$, and
generates correct $Z^\prime/Z$ mass hierarchy without affecting the $\mu$
parameter. Previous studies have provided extensive phenomenological analyses of
this model, and notably, have provided a novel way to explain the excess
positron flux in cosmic rays.

 The model has three right-handed scalar neutrinos, in addition to the three
left-handed states from the SM/MSSM spectrum.  Cross sections are considerably
enhanced compared to the ones estimated in MSSM, even though for most
of the parameter space studied, the signal is dominated by production of left
handed sneutrinos,  predicted to be lighter. To perform a through analysis, we
concentrate on three MSSM benchmark parameter points, denoted by LM1, LM2 and
LM6 and define correspondingly three $U(1)^\prime$ parameter points, denoted
by $\lmop,\,\lmtp, \,\lmsp$, specified in  such that the common parameters with
MSSM are identical. At this point, it is convenient to give a couple of remarks
on the scenarios adopted here. As we mentioned earlier, a $U(1)^\prime$ model
with one singlet and right handed sneutrinos \cite{Demir:2009kc} can explain the
excess positron flux observed by various satellite experiments. However, this
requires a rather special mass spectrum. Indeed, it turns out that the LSP must
one of the right handed scalar neutrinos with a mass around $100\gev$, and the
next-to-LSP must also be a right handed scalar neutrino weighing at the TeV
scale. All the other SUSY particles have to be heavier. One might ask why we did
not consider such a scenario here. There are a couple of reasons. First of all,
excess positron flux observation doesn't need to have an explanation coming from
particle physics, only. Secondly, the signal for such a scenario would be mainly
just missing transverse energy since all the SUSY particles other than the LSP
are above the TeV scale so that the cross section for left-handed sneutrino
production would be much smaller. Practically, missing energy signal with no
visible particle is not useful experimentally. Finally, in this work, we focused
on  only low-scale SUSY scenarios which would be discovered with the early LHC
data at 14 TeV.

After producing and decaying the sneutrinos, we identify three final-state
signals: $\nolep,$ $\twolep$ and $\fourlep$ and proceed to analyze them at LHC,
for  $14 \tev$ center-of-mass energy and with integrated luminosity ${\cal
L}=100\xfb^{-1}$.  We compare these signals with the $\nolep$ and $\twolep$
signals in MSSM, and discuss the SM background (coming from Drell Yan, $ZZ$ and
$WW$ production) for each. While $\nolep$ is the strongest, it has to compete
with MSSM and suffers from considerable background suppression, while the
$\fourlep$ signal has no MSSM equivalent, is practically background free, but
has few events.

We analyze the signals and suggest cuts to distinguish it from the background.
In particular $E_T^{\rm sum} \equiv m_{\rm eff}$, the scalar sum of the lepton
transverse momenta and the missing energy is found to be high for the signal,
thus a cut on $ m_{\rm eff}$ will likely reduce the background. Additionally a
new parameter ${\hat s}^{1/2}_{\rm min}$ is found to be useful for estimating
the mass of parent particles in hard scattering. (The peak in ${\hat
s}^{1/2}_{\rm min}$ gives the mass threshold of left and right handed sneutrinos
in the decay process). Using these considerations, we can estimate  the
production cross section, the products of decay and estimate the sneutrino
masses. The MSSM production differs both in the number of events expected, cross
section, $E_T^{\rm sum}$, ${\hat s}^{1/2}_{\rm min}$, in the $\nolep$ case;  and
additionally in the $p_T$ spectra of leptons (for $\twolep$ case). The
$\fourlep$ case has no MSSM equivalent and little, if any, background, so the
$U(1)^\prime$ is clear there; however the number of events, especially after
passing detector cuts, is small.

In conclusion, our extensive analysis  shows  significant enhancement of
$U(1)^\prime$ signal over the MSSM signal in sneutrino production and decays, and indicates
how the two models can be distinguished from each other and the background. This
provides a distinct collider signal for the secluded $U(1)^\prime$ model at the LHC.

\section{Acknowledgments}
This work is partially supported by Turkish Atomic Energy Agency (TAEK) through
the project CERN-A5.H2.P1.01-13. The work of D. A. D. is supported by Scientific
and Technological Research Council of Turkey (T{\"U}B{\.I}TAK) through the
project 109T718. D.A.D. is grateful to Levent Solmaz for fruitful discussions on
anomaly cancellation in $U(1)^{\prime}$ models. He also thanks to Oktay
Do{\~g}ang{\"u}n, Hale Sert, Koray Sevim and Onur Tosun for discussions on Higgs
sector of the model and $Z^{\prime}$ mass range. M.F. acknowledges NSERC of
Canada for partial financial support under Grant No. SAP01105354. L.S. is
grateful to the Department of Physics, {\.I}zmir Institute of Technology for its
generous hospitality, where part of this work was done, and T{\"U}B{\.I}TAK for
post-doctoral fellowship.

%%%%%%%%%%%%%%%%%%%%%%%%%%%%%%%%%%%%%%%%%%%%%%%%%%%
%%%%%%%%%%%%%%%%%%%%%%%%%%%%%%%%%%%%%%%%%%%%%%%%%%
%\clearpage
%\section*{Appendix A: The Lagrangian}
%\setcounter{equation}{0}
\appendix
\section{The Lagrangian}
In this Appendix, we present the complete Lagrangian of the $U(1)^{\prime}$
model and highlight the differences between this and   the MSSM
Lagrangian. Although parts of this formulation have appeared elsewhere
\cite{Erler:2002pr, Ali:2009md, cvetic}, we include the complete model
information for consistency, and to help future studies. The total Lagrangian
incorporates kinetic terms and various interaction terms among the fields. We
discuss below the
distinct pieces separately.

The kinetic terms of the Lagrangian are given by
\begin{eqnarray}\label{kinetic}
{\cal L}_{U(1)^\prime}^{Kinetic }&=&{\cal
L}_{MSSM}^{Kinetic}-\frac{1}{4}Z^{\prime\mu\nu}Z^\prime_{\mu\nu}+({\cal
D}_{\mu}S)^{\dagger}({\cal D}^{\mu}S)+\tilde Z^{\prime
\dagger}i\sigma^{\mu}\partial_{\mu}\tilde Z^{\prime}+\tilde
S^{\dagger}i\sigma^{\mu}{\cal D}_{\mu}\tilde S \nonumber \\
&+&({\cal D}_{\mu}S_j)^{\dagger}({\cal D}^{\mu}S_j)+\tilde
S^{\dagger}_ji\sigma^{\mu}{\cal D}_{\mu}\tilde S_j+({\cal
D}_{\mu}\tilde N)^{\dagger}({\cal D}^{\mu}\tilde N)
\end{eqnarray}
where $j=1,2,3$. The interactions of the gauge fields with the rest
(fermions,
sfermions, gauginos, Higgs and Higgsino fields) are contained in the
piece
\begin{eqnarray}
{\cal{L}}^{gauge}_{U(1)^{\prime}} = {\cal{L}}^{gauge}_{MSSM}\left(g_Y
\frac{Y_{X}}{2} B_{\mu} \rightarrow g_Y \frac{Y_{X}}{2}
B_{\mu} + g_{Y^{\prime}} Q_X^{\prime} Z^{\prime}_{\mu} \right)\,,
\end{eqnarray}
where $X$ runs over the fields charged under $U(1)^{\prime}$. In
(\ref{kinetic}), $Z^{\prime\, \mu \nu}$ is the field strength tensor
of $Z^{\prime}_{\mu}$, and ${\cal
D}_{\mu}S_j=(\partial_{\mu}+ig_{Y^{\prime}}Q^{\prime}_{S_{j}}Z_{\mu}^{\prime}
)S_j$ for $j=1,2,3$.

The part of the $U(1)^{\prime}$ Lagrangian spanned by the $F$--terms is given
by
\begin{eqnarray}\label{Fterm}
{\cal L}_{U(1)^\prime}^{F-term }&=&-\sum_i\left|\frac{\partial
W}{\partial\phi_i}\right|^2={\cal L}_{MSSM}^{F-term }(\mu\rightarrow
h_sS)-h_s^2|H_u\cdot H_d|^2\nonumber \\
&-&\left(h_u\tilde Q^*\tilde
U^*+h_s^*S^*H_d^*\right)\frac{h_{\nu}}{M_R}S_1\tilde L\tilde N\nonumber \\
&-&\frac{h_{\nu}}{M_R}S_1^*\tilde L^*\tilde N^*\left(h_u\tilde Q\tilde
U+h_sSH_d+\frac{h_{\nu}}{M_R}S_1\tilde L\tilde N\right)\nonumber \\
&-&\left(h_eH_d^*\tilde E^*\right)\frac{h_{\nu}}{M_R}S_1H_u\tilde N\nonumber \\
&-&\frac{h_{\nu}}{M_R}S_1^*H_u^*\tilde N^*\left(h_eH_d\tilde
E+\frac{h_{\nu}}{M_R}S_1H_u\tilde N\right)\nonumber \\
&-&\frac{h_{\nu}^2}{M_R^2}S_1^2|\tilde L\cdot H_u|^2\nonumber \\
&-&\frac{h_{\nu}^2}{M_R^2}|\tilde L\cdot H_u|^2\tilde
N^2-\bar{h}_s^2S_2^2S_3^2\nonumber \\
&-&\frac{h_{\nu}}{M_R}\tilde L^*\cdot H_u^*\tilde
N^*\bar{h}_sS_2S_3-\bar{h}_sS_2^*S_3^*\frac{h_{\nu}}{M_R}\tilde
L\cdot
H_u\tilde N\nonumber \\
&-&\bar{h}_s^2S_1^2S_3^2-\bar{h}_s^2S_1^2S_2^2
\end{eqnarray}
where $\phi_i$ is the scalar component of the $i$--the chiral superfield in the
superpotential.

The $D$--term contributions to the Lagrangian are given by
\begin{eqnarray}\label{Dterm}
{\cal L}_{U(1)^\prime}^{D-term }&=&-\frac{1}{2}\sum_a D^a D^a={\cal
L}_{MSSM}^{D-term }\nonumber \\
&-&\frac{g_{Y^\prime}^2}{2}\biggl(Q^{\prime}_Q\tilde Q^{*}\tilde
Q+Q^{\prime}_U\tilde U^{*}\tilde U+Q^{\prime}_D\tilde D^{*}\tilde
D+Q^{\prime}_L\tilde L^{*}\tilde L+Q^{\prime}_E\tilde E^{*}\tilde
E\nonumber \\
&+&Q^{\prime}_{H_d}H_d^{*}H_d+Q^{\prime}_{H_u}H_u^{*}
H_u+Q^{\prime}_N\tilde
N^*\tilde N+Q^{\prime}_SS^*S+Q^{\prime}_{S_1} S_1^* S_1\nonumber \\
&+&Q^{\prime}_{S_2}S_2^*S_2+Q^{\prime}_{S_3}S_3^* S_3\biggr)^2
\end{eqnarray}

The soft-breaking sector of the $U(1)^{\prime}$ Lagrangian is
\begin{eqnarray}\label{soft}
{\cal L}_{U(1)^\prime}^{Soft}&=&{\cal
L}_{MSSM}^{Soft
}(\mu\rightarrow0)-m^2_SS^*S-m^2_{S_1}S_1^*S_1-m^2_{S_2}S_2^*S_2-m^2_{S_3}
S_3^*S_3-m^2_N\tilde N^*\tilde N\nonumber \\
&-&[h_sA_sSH_u\cdot H_d+\frac{h_{\nu}}{M_R}A_{\nu}S_1\tilde L\cdot
H_u\tilde N+A_{\bar{h}_s}\bar{h}_sS_1S_2S_3+h.c.]\nonumber \\
&+&\frac{1}{2}\biggl(M_{\tilde Z^\prime}\tilde Z^\prime\tilde
Z^\prime+h.c.\biggr)\nonumber \\
&+&(m_{SS_1}^2SS_1+m_{SS_2}^2SS_2+m_{S_1S_2}^2S_1^*S_2+h.c.)
\end{eqnarray}
where $M_{\widetilde{Z}^{\prime}}$ is $U(1)^{\prime}$ gaugino mass
defined below  in (\ref{mneut}), and $A_s$
is the extra trilinear soft coupling.

Finally, the part of the Lagrangian describing  the fermion-sfermion-ino
interactions, as well as
the Higgs-Higgsino-Higgsino interactions, is given by
\begin{eqnarray}\label{mkl767}
{\cal L}_{U(1)^\prime}^{ino-f-\phi}&=&{\cal L}_{MSSM}^{ino-f-\phi
}(\mu\rightarrow0)+i\sqrt{2}g_{Y^\prime}[Q^\prime_QQ^{\dagger}\tilde
Z^\prime\tilde Q+Q^\prime_Uu^{\dagger}_R\tilde Z^\prime\tilde u_R\nonumber \\
&+&Q^\prime_Dd^{\dagger}_R\tilde Z^\prime\tilde
d_R+Q^\prime_LL^{\dagger}\tilde Z^\prime\tilde
L+Q^\prime_E\ell^{\dagger}_R\tilde
Z^\prime\tilde\ell_R+Q^\prime_{H_d}\tilde H_d^{\dagger}\tilde
Z^\prime H_d\nonumber \\
&+&Q^\prime_{H_u}\tilde H_u^{\dagger}\tilde Z^\prime
H_u+Q^\prime_{S}\tilde S^{\dagger}\tilde Z^\prime
S+Q^\prime_{S_j}\tilde S_j^{\dagger}\tilde Z^\prime
S_j+Q^\prime_N\nu^{\dagger}_R\tilde
Z^\prime\tilde\nu_R+h.c.]\nonumber \\
&+&[h_sS\tilde H_u\cdot \tilde H_d+h_s\tilde S H_u\cdot \tilde
H_d+h_s\tilde S\tilde H_u\cdot H_d+h.c.].
\end{eqnarray}

All parts of the the $U(1)^{\prime}$ model Lagrangian listed above
are described in the current basis. Eventually, the fields must be
transformed into the physical basis where each field obtains a
definite mass. The neutral gauginos and Higgsinos form the
neutralino sector whose physical states are expressed as in
(\ref{neutralino-def}), after diagonalizing the mass matrix
(\ref{mneut}). Unlike the neutralino sector, the structure of the
chargino sector is essentially the same as in the MSSM with the
 replacement $\mu \rightarrow h_s v_s /\sqrt{2}$. A detailed
analysis of the Higgs and chargino sectors of the $U(1)^{\prime}$
model  has been given in \cite{Erler:2002pr}.

In the gauge boson sector, spontaneous breakdown of the product
group $SU(2)_L\otimes U(1)_Y\otimes U(1)^{\prime}$ via the Higgs
VEVs
\begin{eqnarray}
\langle H_u \rangle = \frac{1}{\sqrt{2}}\left(\begin{array}{c} 0\\
v_u\end{array}\right)\,,\;\; \langle H_d \rangle =
\frac{1}{\sqrt{2}}\left(\begin{array}{c} v_d\\
0\end{array}\right)\,,\;\; \langle S \rangle =
\frac{v_s}{\sqrt{2}}\,,\;\;  \langle S_i \rangle =
\frac{v_{s_i}}{\sqrt{2}}
\end{eqnarray}
generates one massless state (the photon) and two massive states (the
$Z,~Z^\prime$ bosons) via orthonormal combinations of $W^{3}_{\mu}$,
$B^{\prime}_\mu$ and
$B_{\mu}$ gauge bosons. The $W^{1}_{\mu}$ and $W^{2}_{\mu}$ linearly
combine to give $W^{\pm}_{\mu}$, as the only charged vector bosons
in the model. In contrast to the MSSM, the $Z$ boson is not a
physical state by itself since it mixes with the $Z^{\prime}$ boson.
This mass mixing arises from the fact that the Higgs doublets
$H_{u,d}$ are charged under each factor of $SU(2)_L\otimes
U(1)_Y\otimes U(1)^{\prime}$, and the associated mass-squared matrix
is given by \cite{cvetic,langacker-review}
\begin{eqnarray}
\label{mzzp}
M^{2}_{Z-Z^{\prime}} = \left(\begin{array}{cc} M_Z^2 & \Delta^2 \\
\Delta^2 & M_{Z^{\prime}}^2\end{array}\right)\,,
\end{eqnarray}
in the $\left(Z_{\mu}, Z^{\prime}_{\mu}\right)$ basis. Its entries are
\begin{eqnarray}
M_Z^2 &=& \frac{1}{4} G_Z^2 \left(v_u^2 + v_d^2\right),\nonumber\\
M_{Z^{\prime}}^2 &=& g_{Y^{\prime}}^2 \left( Q^{\prime\ 2}_{H_u}
v_u^2 + Q^{\prime\ 2}_{H_d} v_d^2 + Q^{\prime\ 2}_{S}
v_s^2+\sum^3_{i=1}Q^{\prime\ 2}_{S_i} v_{s_i}^2
\right)\,,\nonumber\\
\Delta^2 &=& \frac{1}{2} G_Z g_{Y^{\prime}} \left(
Q^{\prime}_{H_u} v_u^2 - Q^{\prime}_{H_d} v_d^2\right)\,,
\end{eqnarray}
where $G_Z^2 = g_2^2 + g_Y^2$. The physical neutral vector bosons, $Z_{1,2}$,
are obtained by diagonalizing $M^{2}_{Z-Z^{\prime}}$:
\begin{eqnarray}
\label{mzzp-angle}
\left(\begin{array}{c} Z_1\\Z_2 \end{array}\right) =
\left(\begin{array}{cc} \cos\theta_{Z-Z^{\prime}} & \sin\theta_{Z-Z^{\prime}} \\
-\sin\theta_{Z-Z^{\prime}} &
\cos\theta_{Z-Z^{\prime}}\end{array}\right) \left(\begin{array}{c} Z
\\Z^{\prime}
\end{array}\right)\,,
\end{eqnarray}
where
\begin{eqnarray}
\theta_{Z-Z^{\prime}} = - \frac{1}{2} \arctan \left( \frac{ 2
\Delta^2}{M_{Z^{\prime}}^2 - M_Z^2}\right)\,,
\end{eqnarray}
is their mass mixing angle, and
\begin{eqnarray}
M^{2}_{Z_{1(2)}}= \frac{1}{2} \left[ M_{Z^{\prime}}^2 + M_Z^2 -
(+) \sqrt{\left(M_{Z^{\prime}}^2 - M_Z^2\right)^2 + 4
\Delta^4}\right]\,,
\end{eqnarray}
are their masses-squared . The collider searches at LEP and
Tevatron plus various indirect observations require
$Z$--$Z^{\prime}$ mixing angle $\theta_{Z-Z^{\prime}}$ to be at
most a few $10^{-3}$ with an unavoidable model dependence
coming from the $Z^{\prime}$ couplings
\cite{langacker-review,LEP,indirect,collid,spin,langacker-kang}.
This bound requires either $M_{Z_2}$ to be large enough (well
in the ${\rm TeV}$ range) or $\Delta^2$ to be sufficiently
suppressed by the vacuum configuration, that is,
$\tan^2\beta\equiv v_u^2/v_d^2 \sim
Q^{\prime}_{H_d}/Q^{\prime}_{H_u}$. Which of these options is
realized depends on the $U(1)^{\prime}$ charge assignments and
the soft-breaking masses in the Higgs sector ( see \cite{Erler:2002pr}
for a variant for reducing the $Z$--$Z^{\prime}$ mixing).

\section{The Scalar Fermions}
%\section*{Appendix B: The Scalar Fermions}
%\setcounter{equation}{0}
%\def\theequation{B.\arabic{equation}}

Given rather tight FCNC bounds, we neglect all the
inter-generational mixings, and consider only intra-generational
left-right mixings, though these turn out to be totally
negligible for the sfermions in the first and second
generations. The $2\times 2$ scalar fermion mixing matrix can
be written as
\begin{eqnarray}
{\cal M}^{2}_{\widetilde{f}^a}=
\left(
\begin{array}{cc}
{\cal M}^2_{\widetilde{f}^a_{LL}} & {\cal M}^2_{\widetilde{f}^{a,b}_{LR}}\\
\\
{\cal M}^{2 \dagger}_{\widetilde{f}^{a,b}_{LR}} & {\cal
M}^2_{\widetilde{f}^a_{RR}}
\end{array}
\right), ~~~~~~a\ne b=u,d\,,
\end{eqnarray}
where
\begin{eqnarray}\label{sferm-mass}
{\cal M}_{\tilde f_{L L}^\alpha}^{2}&=& \tilde M^2_{\tilde
f_L}+\frac{1}{2}h_{f^\alpha}^2v_\alpha^2 \kappa_s^2+ \frac{1}{4}
\biggl[g_Y^2Y_{f_L^\alpha} -(+) \frac{g^2}{2}\biggr](v_u^2-v_d^2)\nonumber \\
&&+\frac{1}{2}g_{Y'}^2Q'_{f_L^\alpha}(Q'_{H_u}v_u^2+Q'_{H_d}v_d^2+Q'_{S}
v_s^2\rho_s)
\\
{\cal M}_{\tilde f_{R R}^\alpha}^{2}&=&\tilde M^2_{\tilde
f_R}+\frac{1}{2}h_{f^\alpha}^2v_\alpha^2 \kappa_s^2+ \frac{1}{4}
\biggl[g_Y^2Y_{f_R^\alpha}\biggr](v_u^2-v_d^2)\nonumber \\
&&+\frac{1}{2}g_{Y'}^2Q'_{f_R^\alpha}(Q'_{H_u}v_u^2+Q'_{H_d}v_d^2+Q'_{S}
v_s^2\rho_s)
\\
\label{mkl7298} {\cal M}_{\tilde f_{L
R}^{\alpha,\beta}}^{2}&=&({\cal M}_{\tilde f_{R L}^{\alpha,
\beta}}^{2})^*=\frac{h_{f^\alpha}\kappa_s}{2\sqrt{2}}(\pm
2A_{f^\alpha}^*v_{\alpha}+\sqrt{2}h_sv_{\beta}v_s+2\sqrt{2}\xi_s)
\end{eqnarray}
where $\kappa_s=\frac{v_{s_1}}{\sqrt{2}M_R}$ and
$\xi_s=\frac{\bar{h}_sv_{s_2}v_{s_3}v_u}{2v_{s_1}}$ for sneutrinos
and $\kappa_s=1$ and $\xi_s=0$  for the others.
Here $\tilde{M}_{\widetilde{f}_{L,R}}^2$ are the soft mass-squared of the
sfermions, $v_{u,d,s,s_1,s_2,s_3}$ are the VEVs of the Higgs fields,
$Y_{f^a}(T_{3L})$ is the $U(1)_{Y}$ ($SU(2)_L$) quantum number,
$Q^\prime_{f^a}$ is the $U(1)^\prime$ charge, and $A_{f^a}$ are the
trilinear couplings. The mixing matrix can be diagonalized, in
general, by a unitary matrix $ \Gamma^f$ such that
$\Gamma^{f^a\dagger}\cdot{\cal M}^2_{\widetilde{f}^a}\cdot
\Gamma^{f^a} \equiv {\rm
Diag}(M_{\widetilde{f}^a_1}^2,M_{\widetilde{f}^a_2}^2)$.\footnote{We
note that unlike mixings in other sectors, $\Gamma^{f^a}$ is defined
differently, that is,
$(\widetilde{f}^a_{L,R})_i=\Gamma^{f^a}_{ij}\widetilde{f}^a_j$,
where $\widetilde{f}^a_j$ represent the mass eigenstates.} The
rotation matrix $\Gamma^{f^a}$ can be written for quarks and charged
leptons in the $2\times 2$ $\{\widetilde{f}^a_{L},
\widetilde{f}^a_R\}$ basis as
\begin{eqnarray}
\Gamma^{f^a}=
\left(
\begin{array}{cc}
\cos\theta_{\widetilde{f}^a} & -\sin\theta_{\widetilde{f}^a}\\
\sin\theta_{\widetilde{f}^a} & \cos\theta_{\widetilde{f}^a}
\end{array}
\right),
\end{eqnarray}
where $\displaystyle
\theta_{\widetilde{f}^a}=\frac{1}{2}\arctan 2(-2{\cal
M}^2_{\widetilde{f}^a_{LR}}, {\cal
M}^2_{\widetilde{f}^a_{RR}}-{\cal M}^2_{\widetilde{f}^a_{LL}})$
and $\arctan 2(y,x)$ is defined as
\begin{eqnarray}
{\rm \arctan 2}(y,x) = \left\{\begin{array}{ll}
                      \phi\ {\rm sign}(y),           ~~~& x>0 \\
              \frac{\pi}{2}\ {\rm sign}(y),  ~~~& x=0 \\
                      (\pi-\phi)\ {\rm sign}(y),     ~~~& x<0
                     \end{array}\right.
\end{eqnarray}
with $y$ being non-zero, and $\phi$ taken in the first quadrant
such that $\tan\phi=|y/x|$.

For the sfermions in the first and second generations, the
left-right mixings are exceedingly small as they are proportional to
the corresponding fermion mass. Therefore, the sfermion mass matrix
(\ref{sferm-mass}) is automatically diagonal. However, one has to
remember  that the sfermion masses, for fixed values of
$m_{\widetilde{f}_{L,R}^2}$, are different in the MSSM than in the
$U(1)^{\prime}$ models due to the additional $D$-term contribution
in the latter.

%%%%%%%%%%%%%%%%%%%%%%%%%%%%%%%%%%%%%%%%%%%%%%%%%%%%%%%%%%%%%%%%%%%%%%%%
\section{Gauge and Higgs Fermions}\label{app:C}
%\section*{Appendix C: Gauge and Higgs Fermions \label{app:C}}
%\setcounter{equation}{0}
%\def\theequation{C.\arabic{equation}}
Although the $U(1)^{\prime}$ model possesses no new charged
Higgsinos and gauginos it possesses five new fermion fields in the
neutral sector: the $U(1)^{\prime}$ gauge fermion
$\widetilde{Z}^{\prime}$ and four singlinos $\widetilde{S}$,
$\widetilde{S_1}$, $\widetilde{S_2}$, $\widetilde{S_3}$. In total,
there are 9 neutralino states $\widetilde{\chi}_i^0$ ($i=1,\dots,9$)
\cite{Erler:2002pr}:
\begin{eqnarray}
\label{neutralino-def} \widetilde{\chi}_i^0 = \sum_{a} N^0_{i a}
\widetilde{G}_a\,,
\end{eqnarray}
where the mixing matrix $N^0_{i a}$ connects the gauge-basis neutral
fermion states $\widetilde{G}_a \in \Big\{\widetilde{B},$
$\widetilde{W}^3,$ $\widetilde{H}^0_d,$ $\widetilde{H}^0_u,$
$\widetilde{S},$ $\widetilde{Z}^{\prime},$ $\widetilde{S_1},$
$\widetilde{S_2},$ $\widetilde{S_3} \Big\}$ to the physical
neutralinos $\widetilde{\chi}_i^0$. The neutralino masses
$M_{\widetilde{\chi}_i^0}$ and the mixing matrix $N^0_{i a}$ are
determined via the diagonalization condition $N^0 {\cal{M}} N^{0\ T}
= \mbox{Diag}$ $\Big\{M_{\widetilde{\chi}_1^0},$ $\dots,$
$M_{\widetilde{\chi}_9^0}\Big\}$ for the neutral fermion mass matrix

\begin{eqnarray}\label{mneut}
\left(
 \begin{array}{ccccccccc}
 M_{\tilde Y}&0&-M_{\tilde Y \tilde H_d}&M_{\tilde Y \tilde H_u}&0&M_{\tilde Y
\tilde Z'}&0&0&0 \\[1.ex]
  0&M_{\tilde W}&M_{\tilde W \tilde H_d}&-M_{\tilde W \tilde
H_u}&0&0&0&0&0\\[1.ex]
  -M_{\tilde Y \tilde H_d}&M_{\tilde W \tilde
H_d}&0&-\mu&-\mu_{H_u}&\mu'_{H_d}&0&0&0\\[1.ex]
  M_{\tilde Y \tilde H_u}&-M_{\tilde W \tilde
H_d}&-\mu&0&-\mu_{H_d}&\mu'_{H_u}&0&0&0\\[1.ex]
  0&0&-\mu_{H_u}&-\mu_{H_d}&0&\mu'_S&0&0&0\\[1.ex]
  M_{\tilde Y \tilde Z'}&0&\mu'_{H_d}&\mu'_{H_u}&\mu'_S&M_{\tilde
Z'}&\mu'_{S_1}&\mu'_{S_2}&\mu'_{S_3}\\[1.ex]
  0&0&0&0&0&\mu'_{S_1}&0&-\frac{\bar{h}_s v_{s_3}}{\sqrt{2}}&-\frac{\bar{h}_s
v_{s_2}}{\sqrt{2}}\\[1.ex]
  0&0&0&0&0&\mu'_{S_2}&-\frac{\bar{h}_s v_{s_3}}{\sqrt{2}}&0&-\frac{\bar{h}_s
v_{s_1}}{\sqrt{2}}\\[1.ex]
  0&0&0&0&0&\mu'_{S_3}&-\frac{\bar{h}_s v_{s_2}}{\sqrt{2}}&-\frac{\bar{h}_s
v_{s_1}}{\sqrt{2}}&0\\[1.ex]
 \end{array}
 \right)
\end{eqnarray}
where certain entries are generated by the soft-breaking sector
while others follow from the $SU(3)_c \otimes SU(2)_L\otimes
U(1)_Y\otimes U(1)^{\prime}$ breaking. The $U(1)_Y$ gaugino mass
$M_{\widetilde{Y}}$, the $SU(2)_L$ gaugino mass $M_{\widetilde{W}}$,
and the $U(1)^{\prime}$ gaugino mass
\begin{eqnarray}
M_{\widetilde{Z}^{\prime}} &=&
\frac{M_{\widetilde{Y}^{\prime}}}{\cos^2\chi}   - 2
\frac{\tan\chi}{\cos\chi} M_{\widetilde{Y} \widetilde{Y}^{\prime}} +
M_{\widetilde{Y}} \tan^2\chi\,,
\end{eqnarray}
as well as the mixing mass parameter between $U(1)_Y$ and
$U(1)^{\prime}$ gauginos
\begin{eqnarray}
M_{\widetilde{Y} \widetilde{Z}^{\prime}} &=& \frac{M_{\widetilde{Y}
\widetilde{Y}^{\prime}}}{\cos\chi} - M_{\widetilde{Y}} \tan\chi\,,
\end{eqnarray}
all follow from the soft-breaking sector. Through the mixing of the
gauge bosons, $M_{\widetilde{Z}^{\prime}}$ and $M_{\widetilde{Y}
\widetilde{Z}^{\prime}}$ exhibit an explicit dependence on the
masses of the $U(1)_Y$ and $U(1)^{\prime}$ gauginos, and their mass
mixing.  $M_{\widetilde{Y} \widetilde{Y}^{\prime}}$ is the
soft-breaking mass that mixes the $U(1)_Y$ and $U(1)^{\prime}$
gauginos. In the numerical analysis, we set the mixing mass
parameter $M_{\widetilde{Y}\widetilde{Z}^{\prime}}=0$ since we neglect
the kinetic mixing ($\tan\chi\to 0$) thus $M_{\widetilde{Y}
\widetilde{Y}^{\prime}}\to 0$. For convenience we also define
$R_{Y^\prime}\equiv M_{\widetilde{Y}^{\prime}}/M_{\widetilde{Y}}$.

The remaining entries in (\ref{mneut}) are generated by the
soft-breaking masses in the Higgs sector via the $SU(3)_c\otimes
SU(2)_L\otimes U(1)_Y\otimes U(1)^{\prime}$ breaking. Their explicit
expressions are given by
\begin{eqnarray}
M_{\widetilde{Y}\, \widetilde{H}_d} &=& M_Z \sin\theta_W \cos\beta\,,\;
M_{\widetilde{Y}\, \widetilde{H}_u} = M_Z \sin\theta_W \sin\beta\,,\nonumber\\
M_{\widetilde{W}\, \widetilde{H}_d} &=& M_Z \cos\theta_W \cos\beta\,,\;
M_{\widetilde{W}\, \widetilde{H}_u} = M_Z \cos\theta_W \sin\beta\,,\nonumber\\
\mu_{H_d} &=& h_s \frac{v_d}{\sqrt{2}}\,,\;
\mu_{H_u} = h_s \frac{v_u}{\sqrt{2}}\,,\; \mu^{\prime}_{H_d}=g_{Y^{\prime}}
Q_{H_d}^{\prime} v_d,\nonumber\\
\mu^{\prime}_{H_u} &=& g_{Y^{\prime}} Q_{H_u}^{\prime} v_u\,,\;
\mu^{\prime}_{S} = g_{Y^{\prime}} Q_{S}^{\prime} v_s\,,\;
\mu^{\prime}_{S_i} = g_{Y^{\prime}} Q_{S_i}^{\prime} v_{s_i}\,,
\end{eqnarray}
where $g_{Y'}$ is the coupling constant of $U(1)^{\prime}$. For
numerical analysis we choose the standard GUT value for it
$g_{Y'}=\sqrt{\frac{5}{3}}g\tan\theta_W$.

%%%%%%%%%%%%%%%%%%%%%%%%%%%%%%%%%%%%%%%%%%%%%%%%%%%%%%%%%%%%%%%%%%%%%%%%
%g_{Y'}=\sqrt{\frac{5}{3}}g\tan\theta_W

%%%%%%%%%%%%%%%%%%%%%%%%%%%%%%%%%%%%%%%%%%%%%%%%%%%%%%%%%%%%%%%%%%%%%%%%
\section{The Compositions of the Neutralinos}\label{app:D}
%\section*{Appendix D: The Compositions of the Neutralinos}
%\setcounter{equation}{0}
%\def\theequation{D.\arabic{equation}}

In this Appendix we give the Bino, Wino, Higgsino and Singlino compositions
of the physical neutralinos $\widetilde{\chi}_i^0, i=1,2,...,9$ for the three
scenarios $\lmop, \lmtp$ and $\lmsp$. They are listed in
Table~\ref{tab:neut_comps}.

\begin{table}[htbp]
  \begin{center}
    \setlength{\extrarowheight}{-2.0pt}
 \small
 \begin{tabular*}{0.99\textwidth}{@{\extracolsep{\fill}} cccccccccc}
 \hline\hline
  $LM1^\prime$&$\tilde\chi^0_1$&$\tilde\chi^0_2$
&$\tilde\chi^0_3$&$\tilde\chi^0_4$&$\tilde\chi^0_5$&$\tilde\chi^0_6$&
$\tilde\chi^0_7$&$\tilde\chi^0_8$&$\tilde\chi^0_9$\\
  \hline\hline
  $\tilde B$& -0.988 &0.046  &0.077 &0.043  &-0.056 &0.095 &-0.002 &0.0 &0.0 \\
 % \hline
  $\tilde W^3$&0.037   &-0.058  &0.955 &-0.122  & 0.086&-0.245 & 0.006&0.0 &
0.0\\
  %\hline
  $\tilde H_{d}^0$ &-0.126  &0.051  &-0.245 &-0.321  &0.692 &-0.581 &0.014 &0.0
&0.003 \\
  %\hline
  $ \tilde H_{u}^0$ &0.031  &-0.205  &0.115 &0.226  &0.698 &0.633 &-0.033 &0.002
&-0.040  \\
  %\hline
  $ \tilde S$ &0.057  &0.910  & 0.086&0.362  &0.140 &0.0 &0.048 &-0.004 &0.087
\\
  %\hline
  $ \tilde Z'$ &-0.010  &-0.180  &-0.010 &0.212  &0.022 &-0.092 &0.065 &0.013
&0.953   \\
  %\hline
  $\tilde S_{1}$ &-0.012  & -0.148 &-0.013 &0.492  & 0.012&-0.276 &-0.576 &0.555
&-0.133 \\
  %\hline
  $\tilde S_{2}$ &-0.006  &-0.089  &-0.006 &0.155  &0.008 &-0.060 &0.778 &0.586
&-0.120   \\
  %\hline
  $\tilde S_{3}$ & 0.018 &0.241  &0.020 &-0.621  &-0.023 &0.320 &-0.232 &0.589
&0.223  \\
  \hline
  $LM2^\prime$&& &&&&&&&\\
  \hline
  $\tilde B$&0.048  &-0.994  &0.032 &-0.009  &-0.044 &0.076 &-0.005 &0.0 &0.0 \\
  %\hline
  $\tilde W^3$ &-0.054  &0.0101  &0.974 & 0.029 &0.067 &-0.204 & 0.014&0.0 &0.0
\\
  %\hline
  $\tilde H_{d}^0$ &0.032  &-0.088  &-0.191 &0.135  &0.699 &-0.666 & 0.043&0.0 &
0.001\\
  %\hline
  $ \tilde H_{u}^0$ &-0.224  &0.013  &0.083 &-0.062  & 0.692&0.673 &-0.057
&0.002 & -0.033 \\
  %\hline
  $ \tilde S$ &0.942  &0.054  &0.075 &-0.238  &0.152 &0.130 &0.028 &-0.003
&0.069  \\
  %\hline
  $ \tilde Z'$ &-0.119  &-0.006  &-0.007 &-0.190  &0.015 &-0.019 &0.047 &0.011
&0.972   \\
  %\hline
  $\tilde S_{1}$ &-0.091  &-0.007  &-0.010 &-0.550  &0.006 &-0.143 & -0.585&
0.561&-0.100 \\
  %\hline
  $\tilde S_{2}$ &-0.069  &-0.005  &-0.006 &-0.212  &0.005 &0.013 &0.778 &0.578
&-0.094   \\
  %\hline
  $\tilde S_{3}$ &0.165  &0.012  &0.016 &0.732  &-0.013 &0.123 &-0.206 &0.591
&0.169  \\
  \hline
  $LM6^\prime$&& &&&&&&&\\
  \hline
  $\tilde B$&-0.036  &-0.995  &-0.034 &-0.003  &0.035 &-0.072 & 0.006&0.0 &
0.0\\
  %\hline
  $\tilde W^3$ &0.041  &0.015  &-0.978 &0.014  &-0.054 &0.194 &-0.017 & 0.0&0.0
\\
  %\hline
  $\tilde H_{d}^0$ &-0.015  &-0.080  & 0.174&0.093  &-0.700 &0.677 &-0.060 &0.0
&-0.002 \\
  %\hline
  $ \tilde H_{u}^0$ & 0.196 &0.021  &-0.089 &-0.015  &-0.695 & -0.680&0.073
&0.001 &0.031  \\
  %\hline
  $ \tilde S$ &-0.936  &0.040  &-0.057 &-0.278  &-0.143 &-0.114 &-0.028 &-0.004
&-0.073  \\
  %\hline
  $ \tilde Z'$ &0.138  &-0.005  &-0.006 &-0.202  &-0.012 &0.007 & -0.046&0.012
&-0.967   \\
  %\hline
  $\tilde S_{1}$ &0.116  &-0.007  &0.010 &-0.541  &-0.006 &0.120 &0.593 &0.560
&0.109 \\
  %\hline
  $\tilde S_{2}$ &0.082  &-0.004  &0.005 &-0.215  &-0.004 &-0.044 & -0.772&0.581
&0.101   \\
  %\hline
  $\tilde S_{3}$ &-0.203  &0.011  &-0.015 &0.729  &0.012 &-0.070 &0.197 &0.589 &
-0.185 \\
  \hline\hline
   \end{tabular*}
\caption{\label{tab:neut_comps}\sl\small The Bino, Wino, Higgsino and Singlino
composition of the neutralinos $\widetilde{\chi}_i^0, i=1,2,...,9$ for the
scenarios $\lmop, \lmtp$ and $\lmsp$.}
\end{center}
 \end{table}

\clearpage
%%%%%%%%%%%%%%%%%%%%%%%%%%%%%%%%%%%%%%%%%%%%%%%%%%%%%%%%%%%%%%%%%%%%%%%%

\end{document}